\documentclass[a4paper,11pt]{article}
\pdfoutput=1 

\usepackage{jcappub}

\usepackage[T1]{fontenc} 

\usepackage{amsmath,amssymb,epsfig}
\usepackage{wrapfig}
\usepackage{gensymb}
\usepackage[english]{babel}
\usepackage{graphics}
\usepackage{subcaption}
\usepackage{hyperref}
\usepackage[utf8]{inputenc}

\usepackage{booktabs}

\newcommand{\be}{\begin{equation}}
\newcommand{\ee}{\end{equation}}

\newcommand{\diff}{\text{d}}

\title{Ultra-High-Energy Cosmic Rays from Radio Galaxies}

\author[a, 1]{B. Eichmann,\note{Corresponding author.}}
\author[b]{J.P. Rachen,}
\author[a]{L. Merten,} 
\author[b]{A. van Vliet,}
\author[a]{and J.~Becker Tjus,}

\affiliation[a]{Ruhr Astroparticle and Plasma Physics Center (RAPP Center), Ruhr-Universit\"at Bochum, Institut f\"ur Theoretische Physik IV/ Plasma-Astroteilchenphysik, 44780 Bochum, Germany}
\affiliation[b]{Department of Astrophysics/IMAPP, Radboud University, P.O. Box 9010,\\
6500 GL Nijmegen, the Netherlands}

\emailAdd{eiche@tp4.rub.de, j.rachen@astro.ru.nl, lukas.merten@rub.de, a.vanvliet@astro.ru.nl, julia.tjus@rub.de}

\abstract{
Radio galaxies are intensively discussed as the sources of cosmic rays observed above about $3\,{\times}\,10^{18}\,\text{eV}$, called ultra-high energy cosmic rays (UHECRs). 
We present a first, systematic approach that takes the individual characteristics of these sources into account, as well as the impact of the extragalactic magnetic-field structures up to a distance of 120\,\text{Mpc}. 
We use a mixed simulation setup, based on 3D simulations of UHECRs ejected by observed, individual radio galaxies taken out to a distance of 120\,Mpc, and on 1D simulations over a continuous source distribution contributing from beyond 120\,Mpc. Additionally, we include the ultra-luminous radio galaxy Cygnus A at a distance of about $250\,$Mpc, as its contribution is so strong that it must be considered as an individual point source. 
The implementation of the UHECR ejection in our simulation setup, both that of individual radio galaxies and the continuous source function, is based on a detailed consideration of the physics of radio jets and standard first-order Fermi acceleration. This allows to derive the spectrum of ejected UHECR as a function of radio luminosity, and at the same time provides an absolute normalization of the problem involving only a small set of parameters adjustable within narrow constraints.  

We show that the average contribution of radio galaxies taken over a very large volume cannot explain the observed features of UHECRs measured at Earth. However, we obtain excellent agreement with the spectrum, composition, and arrival-direction distribution of UHECRs measured by the Pierre Auger Observatory, if we assume that most UHECRs observed arise from only two sources: The ultra-luminous radio galaxy Cygnus A, providing a mostly light composition of nuclear species dominating up to about $6\,{\times}\,10^{19}\,$eV, and the nearest radio galaxy Centaurus A, providing a heavy composition dominating above $6\,{\times}\,10^{19}\,$eV. Here we have to assume that extragalactic magnetic fields out to $250\,$Mpc, which we did not include in the simulation, are able to isotropize the UHECR events at about $8\,$EeV arriving from Cygnus A. Even in this case, significant anisotropy correlated with Cygnus A and Centaurus A could be present at higher energies, and thus allow for differences in UHECR spectrum and composition between the northern and southern hemispheres. 
If this scenario can be confirmed, it would also imply that the UHECR flux in our local cosmic environment is significantly above the average throughout the universe. 
}


\begin{document}
\maketitle
\newpage
\flushbottom

\section{Introduction}
Fully ionized nuclei that penetrate Earth's atmosphere with an energy above $3\,\text{EeV}\simeq 10^{18.5}\,\text{eV}$ are commonly defined as Ultra-High-Energy Cosmic Rays (UHECRs). The energy of $3\,$EeV hereby stems from a long known feature in the cosmic-ray spectrum called ``the ankle'' at about this energy.\footnote{This denotation stems from the similarity of the cosmic-ray spectrum with a human leg, with a spectral steepening at about $10^{15}\,$eV being called the ``knee'', followed by a flattening at about $3\,$EeV called the ``ankle''. This anatomic analogy has been stretched by the recent discovery of a ``second knee'' \cite{2011PhRvL.107q1104A}, but is still commonly used.} Observed up to several hundred EeV, they belong to the most enigmatic phenomena of astrophysics, and for some time they were even assumed to be products of some new-physics processes \cite{1995Sci...270.1977S}.   

The situation has somewhat changed in recent years, where the Pierre Auger Observatory (Auger) in the southern hemisphere, and the High Resolution Fly's eye (HiRes) as well as the Telescope Array (TA) experiments in the northern hemisphere, have provided measurements with unprecedented statistics \cite{2015JCAP...08..049P, ObservatoryMichaelUngerforthePierreAuger:2017fhr, Abbasi:2009ix, Jui:2016amg}. While confirming the known spectral features with high precision, besides the ankle also a cutoff above about $10^{19.5}\,\text{eV}$ \cite{Abbasi:2007sv, 2010PhLB..685..239A}, the main new result was the discovery of an increase of the fraction of heavier elements above $10^{18.3}\,\text{eV}$ among observed UHECRs \cite{Abraham:2010yv,Aab:2014aea, ObservatoryMichaelUngerforthePierreAuger:2017fhr}. This conclusion is based on observations of the depth of the maximal energy release, $X_{\rm max}$, in the particle shower (commonly called ``air shower'') unleashed by the primary UHECR. The translation of $X_{\rm max}$ into the nuclear mass $A$ of the primary UHECR is impeded by large fluctuations among individual air showers and the incomplete knowledge of the hadronic interaction model, but in the light of the current data the formerly common assumption that UHECRs are mostly protons is hardly tenable. 

This discovery has significantly eased the explanation of the origin of UHECRs, and at the same time caused a new problem: while before there were no sources known in the universe which could accelerate particles to the highest observed energies, now there are so many different source types that it becomes a challenge to find out the real responsible ones. As noted by Hillas already in 1984 \cite{Hillas:1985is}, there are objects in the universe spreading over 20 orders of magnitude in size and magnetic field strength, which all fulfill the condition $Z e \beta B R \sim 100\,$EeV (hereafter referred to as the \textit{Hillas-condition}) for $Z\sim 10$, as typical for the most abundant metals in the universe. Hillas grouped these sources into a diagram known after his name, and indeed, for all sources in this diagram a theory has been presented that they can not only reach the required energies, but also the observed flux --- from compact sources like pulsar winds \cite{Lemoine:2014ala,Kotera:2015pya}, tidal disruption of neutron stars or white dwarfs \cite{Farrar:2014yla, Biehl:2017hnb, AlvesBatista:2017shr, Guepin:2017abw}, gamma-ray bursts \cite{Waxman:1995vg, Vietri:1995hs, 1998AIPC..428..776R, Globus:2014fka}, to the largest structures in the universe \cite{1995ApJ...454...60N, Kang:1995xw, Kang:1996rp}. 

The objects in the center of the Hillas diagram are dominated by structures related to active galactic nuclei (AGN). The most powerful of them sit in the centers of so-called FR-II radio galaxies, which have also been shown to be able to produce the energy, flux, and essentially also the spectrum observed in UHECRs \cite{Rachen:1992pg, Rachen:1993gf}. The problem with these sources is that they are very rare, and none of them is closer to Earth than $100\,$Mpc --- which poses a problem as UHECRs arriving from such distances are constrained to observed energies below $60\,$EeV due to the so-called Greisen-Zatsepin-Kuzmin (GZK) effect \cite{1966PhRvL..16..748G, 1966JETPL...4...78Z}, while UHECRs are observed at significantly higher energies. It has therefore been suggested to also include less luminous AGN, which are related to so-called FR-I radio galaxies containing less powerful and less extended jets. The consequences of this have been investigated in relation to multi-messenger signals of UHECRs \cite{Mannheim:1998wp} or a potential anisotropy arising from the nearest FR-I galaxy, Centaurus A, at about $3\,$Mpc \cite{Piran:2000wh, Rachen:2008zg, 2012ApJ...746...72B, Farrar:2012gm}.       

In fact, if the observed radio brightness of radio galaxies were a reliable tracer for their expected contribution to the UHECR flux, there should be a significant anisotropy in the UHECR arrival directions. At about $1\,$GHz, the brightest three radio galaxies, i.e., Virgo A (M87), Centaurus A (NGC5198), and Cygnus A, contribute about 75\% of the total flux, most of it coming from Cygnus A alone. A strong anisotropy in UHECR arrival directions, however, is not observed. Early claims for anisotropy correlated with extragalactic structure \cite{Stanev:1995my, 1996PhRvL..77.1000H} were never confirmed to a sufficient level of statistical significance, and a joined data analysis of $2130+8259$ events above $10^{19}\,\text{eV}$ from TA and Auger are on a 99\% confidence level compatible with isotropy \cite{2014ApJ...794..172A}. Recently, Auger reported a $5\sigma$ detection of a dipole with an amplitude of $\approx 6.5\%$ in the UHECR arrival directions, while higher-order multipoles are still consistent with isotropy \cite{Aab:2017tyv}. This discrepancy between expectation and observation is often used to disregard radio galaxies or other powerful, steady sources as an explanation for the origin of UHECRs, and instead numerous transient sources are proposed which would trivially explain the observed isotropy \cite{Loeb:2002ee,Farrar:2008ex}.

An important unknown in this conclusion is the deflection of UHECRs in magnetic fields pervading extragalactic space and our Galaxy. The effect of extragalactic magnetic fields (EGMFs) can be simply estimated as follows: If we assume that a magnetic field with average strength $B_{\rm rms}$ is homogeneous over a coherence length $\lambda_{\rm c}$, a particle passing with energy $E$ and charge $Z$ will be deflected on average by an angle $\theta_{\rm rms} \approx Z e B_{\rm rms} \lambda_{\rm c}/2 E$, and after a path of length $d$ undergoing random deflections the total deflection angle will be $\theta \sim \theta_{\rm rms} \sqrt{d/\lambda_{\rm c}}$. For a typical UHECR as derived form observations, i.e., $E\sim 30\,$EeV, $Z\sim 10$, and a coherence length $\lambda_{\rm c} \gtrsim 1\,$Mpc, i.e., comparable to or larger than the typical distance between galaxies, and setting $B_{\rm rms} \sim 1\,$nG to the the current upper limit on large-scale extragalactic magnetic fields \cite{2016A&A...594A..19P, Pshirkov:2015tua}, we find $\theta_{\rm rms} \gtrsim 3^\circ$. Thus, a particle traversing a distance of $100\,$Mpc would be deflected by more than $30^\circ$, a quite significant amount. In fact, if these magnetic deflections are combined with the various energy-loss processes of UHECRs during propagation, one finds that for this simple setup and reasonable values of $B_0$, no UHECRs can reach us from beyond a ``magnetic horizon'' of about $300{-}500\,$Mpc, which is nearly constant over the UHECR energy regime \cite{Stanev:2000fb}. That means, sources outside this horizon will not contribute to the observed UHECR flux, and UHECRs from point sources near to it will reach us nearly isotropized. On the other hand, if UHECR protons exist at the highest energies $E\sim 100\,$EeV, and considering that they would need to arrive from not more than $d\sim 30\,$Mpc due to the GZK effect, one would expect deflections of not more than a few degrees. 

This simple scenario demonstrates that deflections in EGMFs are expected to be significant. For a quantitative estimates, however, it is not sufficient, because EGMFs are expected to be correlated with the observed large-scale structure of matter. Various work has been done to estimate the effect of realistic EGMF structure on UHECR propagation \cite{Sigl:2003ay, Hackstein:2016pwa, AlvesBatista:2017vob}. One of the currently most sophisticated descriptions of the EGMF in the local universe has been given by Dolag et al.\ \cite{2005JCAP...01..009D}, where constrained simulations of large-scale structure formation are used to derive the EGMF from cosmological magnetic fields grown in a magneto-hydrodynamic amplification process. The results of this simulation are straightforward to implement in the publicly available code CRPropa3 \cite{1475-7516-2016-05-038}, and therefore easy to apply to UHECR propagation. 

Apart from EGMFs, another important impact on arrival directions comes from the Galactic magnetic field (GMF). Unlike EGMFs, which are supposed to just scatter UHECR events around the line of sight to their sources, the largely coherent structure of the GMF can deflect them systematically, in addition to a random scattering induced by a small-scale turbulent field of similar strength. Simulations of UHECR trajectories through one of the currently most advanced GMF models \cite{2012ApJ...757...14J, Jansson:2012rt} have shown that, for fiducial values of UHECRs used above, both systematic and random deflections in the GMF should be again of order $30^\circ$ or larger \cite{Keivani:2014zja,Farrar:2017lhm}. This effect depends on direction, but is obviously independent of the distance of an extragalactic source, thus affects very nearby sources in the same way. 

These considerations show that a model proposing dominant and powerful sources like radio galaxies cannot be trivially abandoned on the base of observed UHECR isotropy. Rather, it appears to be time to review radio galaxies as putative UHECR sources in the light of current data on the UHECR spectrum, composition and arrival-direction distribution. In this paper, we present results of UHECR propagation simulations comprising three components: (a) a 3D structure of radio galaxies and EGMF within a radius of 120 Mpc; (b) a continuous source function derived from a luminosity function of radio galaxies for the contributions from beyond 120 Mpc; (c) a contribution of the powerful radio galaxy Cygnus A, which is near the magnetic horizon and is expected to deliver the dominant contribution to UHECRs due to its extreme radio power and brightness. All simulations are carried out with CRPropa3.

The paper is organized as follows: In Sect.~2 we outline the physics of radio galaxies as cosmic-ray accelerators and the derivation of the continuous source function. In Sect.~3 we explain the simulation setup and describe various tests performed to ensure its consistency. In Sect.~4 we present the simulation results for various choices of best-fit parameters. In Sect.~5 we discuss the implications of our results. We conclude in Sect.~6, where we also give an outlook on future work on the subject.

\section{Radio galaxies as UHECR accelerators}
\label{Sec:RG_physics}

\subsection{Phenomenology of radio galaxies}
\label{Sec:RG_pheno} 

The term \textit{radio galaxy} (RG) is commonly used for a subclass of AGN, also called {\textit{radio-loud AGN} \cite{Urry:1995mg}, which form extended plasma jets most likely due to accretion onto a strongly rotating supermassive black hole \cite{Begelman:1984mw}. It has to be distinguished from the more general term of radio-emitting galaxies, to which also other objects, like starburst galaxies, belong \cite{Condon:1992rq}. 

According to a classification introduced in 1974 by Fanaroff and Riley \citep{1974MNRAS.167P..31F}, radio galaxies fall into two major classes: FR-I RGs, in which the jets are terminating within the galactic environment on scales of a few kiloparsec, and which are observed as center brightened radio objects; and FR-II RGs, where the jets extend on scales of $\gtrsim 100\,$kpc deep into extragalactic space, and terminate there in two extremely bright \textit{hot spots}, causing a double-source appearance in radio observations. This morphological distinction strongly correlates with radio power\footnote{Here and in the following we denote the spectral power of a radio source at a frequency $\nu$, measured in Watt/Hz or Watt/Hz/sr, with $P_\nu$, where a three digit number without decimal point in the subscript refers to a frequency in MHz, while a number with decimal point refers to a frequency in GHz. With $L_\nu = \nu P_\nu$ we denote a radio luminosity, measured in erg/s, with the same conventions for frequency notation.}:
Sources with $P_{178} \lesssim 2\,{\times}\,10^{25}\,\text{W}\,\text{Hz}^{-1}\,\text{sr}^{-1}$ tend to be FR-I galaxies, while sources with $P_{178} \gtrsim 2\,{\times}\,10^{25}\,\text{W}\,\text{Hz}^{-1}\,\text{sr}^{-1}$ usually have FR-II morphology --- although there are notable exceptions to this, like the very powerful FR-I galaxy Hydra A at about 230\,Mpc distance.   

Plasma jets from AGN are usually assumed to start with relativistic velocities (bulk Lorentz factor $\Gamma \sim 10$), which is required for the explanation of highly-variable blazar emission in either leptonic or hadronic scenarios \cite{2007Ap&SS.309...95B}. At larger distances from the AGN, however, jets become sub-relativistic, with estimates for the dimensionless velocity of the jet $\beta_{\rm jet} = v_{\rm jet}/c$ ranging from $\lesssim 0.1$ for FR-I RGs \cite{1993ApJ...414..510S}, to $\gtrsim 0.3$ for FR-II RGs \cite{1989A&A...219...63M}. Cosmic-ray acceleration may take place in the relativistic part of the jet, where it is usually limited by radiative losses leading to significant emission of secondaries like neutrinos and gamma rays \cite{Mannheim:1993jg, 1998PhRvD..58l3005R, 2000AIPC..515...41R, 2001APh....15..121M, Muecke:2002bi, Becker:2007sv}. This scenario also allows for highly anisotropic UHECR ejection via neutrons produced in photohadronic interactions \cite{Waxman:1998yy,Mannheim:1998wp,Atoyan:2002gu,Rachen:2008zg,Farrar:2008ex}. We will assume here that the main sites of UHECR acceleration are located in, or at the end of, the mildly relativistic extended part of the jets, which will cause a largely isotropic emission and avoid limitations by radiative losses \cite{2003PhRvE..67d5401M, 2010PhyU...53..691P} (see also Sect.~\ref{Sec:Lradio2Ljet2Rmax}). 

The distribution of radio galaxies in the universe is described by the radio-luminosity function (RLF) of radio-loud AGN. An expression for the local RLF has been given by Mauch and Sadler \cite{2007MNRAS.375..931M} --- hereafter referred to as MS07 --- as\footnote{We use the RLF of \cite[Eq.\ (6)]{2007MNRAS.375..931M}, but revise the incorrect notation $(P_\star/P)$ to $(P/P_\star)$.} $\phi(P_{1.4})=\diff N_{\rm RG} / \diff M_{K}\, \diff V$, which we translate to the luminosity at $L_{1.1} \equiv 1.1\,{\rm GHz}\,P_{1.1}$ via the spectral behavior of $P_\nu\propto \nu^{-0.6}$, i.e., $P_{1.4}=0.79\,L_{1.1}\,\text{GHz}^{-1}$. Since the absolute $K$-band magnitude depends on the radio luminosity $L$ according to $\diff M_{K} = 1.09\,\diff \ln L$, the total number $N_{\rm RG}$ of radio-loud AGN per volume element $\diff V$ and luminosity interval $\diff L$ can be written as 
\be
\frac{\diff N_{\rm RG}}{\diff V\,\diff L_{1.1}} = \frac{C_{\rm RG}}{L_{1.1}\,\left[ (L_{1.1}/L_\star)^{\alpha}+(L_{1.1}/L_\star)^{\beta} \right]}\;,
\label{RLF}
\ee
with the parameters $C_{\rm RG} \approx 2.9\times 10^{-6}\,\text{Mpc}^{-3}\,$ and $L_\star \approx 4.9\times 10^{40}\,\text{erg/s}$, both determined within a factor of $2$, and the power-law indices $\alpha = 1.27\pm0.18$ and $\beta = 0.49\pm0.04$. This is essentially a broken power law with a break at $L_{1.1} = L_\star$, where most of the radio power is emitted by RGs with $L_{1.1}\sim L_\star$, which also corresponds quite well to the radio luminosity separating FR-I and FR-II RGs. The small value of $C_{\rm RG}$ makes clear that powerful radio galaxies are very rare: For sources like Centaurus A, at a distance of $\sim 3\,$Mpc the closest radio galaxy with $L_{1.1}\sim 5\times 10^{39}\,{\rm erg}/{\rm s}$, we would expect on average 1 within a sphere of $30\,$Mpc radius, and for the brightest radio galaxy in the sky, Cygnus A ($L_{1.1} \sim 2\times 10^{44}\,{\rm erg}/{\rm s}$) we have a number density of ${\sim} 0.3\,{\rm Gpc}^{-3}$, thus roughly a few hundred similarly powerful RGs in the entire observable universe!

\subsection{From radio luminosity to cosmic-ray luminosity and maximal rigidity}
\label{Sec:Lradio2Ljet2Rmax}
Radio galaxies have been proposed UHECR accelerators because their jets are suitable sites for first-order Fermi acceleration \cite{1977DoSSR.234.1306K, 1978MNRAS.182..147B, 1978MNRAS.182..443B, 1977ICRC...11..132A, 1978ApJ...221L..29B} at their discontinuities, i.e., in internal or external shocks \cite{1981heaa.book.....L, Biermann:1987ep, Rachen:1992pg, Mannheim:1993jg}, or by shear at the jet boundary \cite{1998A&A...335..134O}. First-order Fermi acceleration is a self-similar process producing power-law spectra with a canonical spectral index of $a=2$ at ideal discontinuities (e.g., strong shocks with large Mach-numbers). Relativistic effects can modify this to indices in a range of $1.5 < a < 2.3$ \cite{Drury:1983zz, 1990ApJ...360..702E}. Acceleration takes place on a timescale 
\be
\tau_{\rm acc}=\frac{f_{\rm diff}\,r_{\rm L}}{c \beta_{\rm sh}^2}
\label{acc-timescale}
\ee
for cosmic-ray particles with a Larmor radius $r_{\rm L}=E/ZeB = R/B$, where $R$ is called the particle rigidity, $c\beta_{\rm sh}$ is the shock or shear velocity, and $f_{\rm diff}\sim 3$ encapsulates all details of the diffusion process in a strongly turbulent magnetic field for standard geometries \cite{Drury:1983zz}. 

As the Fermi acceleration process is only dependent on rigidity, abundances $f_i$ of different nuclear species with charge $Z_i$ measured at a given rigidity $R$ remain unchanged in the acceleration process. The abundances at injection, however, will depend on the environment or details of the injection process. Regarding the environment, we would expect that the interior of radio-galaxy jets contains a similar material mix as the accretion disk around the supermassive black hole, from which the jet likely feeds \cite{Blandford:1982di, Falcke:1994eb}, and which can be assumed to have abundances close to solar (denoted $f_\odot$). For FR-II galaxies where these jets end far outside their host galaxy, it should be just this material which is injected into the acceleration. For FR-I galaxies, however, the jets dissipating inside the galactic environment may pick up material from gaseous shells and galactic cosmic rays, which can lead to a much heavier composition \cite{2010ApJ...720L.155G}. Additionally, the injection of a particles into the Fermi process may depend on other parameters than just the rigidity of the fully ionized nucleus. As a working hypothesis, we may assume that particles are injected at some minimal rigidity $\check{R}$ with abundances $f_i = f_{\odot} Z_i^q$, which reduces the heaviness of the composition to a single parameter $q$. 

To estimate the maximal rigidity which can be obtained in radio galaxies, we need to estimate the strength of the magnetic field $B$. For this we need to derive the total kinetic power of the jets from observational quantities. A relation between the jet power $Q_{\rm jet}$ and the extended radio luminosity $P_{151}$ at 151 MHz has been given by Willott et al.\ \cite{1999MNRAS.309.1017W}, using the narrow emission-line data from the 7C Redshift Survey. In general, the jet power depends on the projected linear size $D$ of the source according to $Q_{\rm jet}\propto D^{-4/7+\gamma/2}$ in the case of an electron number density profile that scales with the distance to the power of $-\gamma$ at radii $\sim 100\,\text{kpc}$. Further, it is pointed out that a very weak $D$-dependence is obtained in the case of a realistic value of $\gamma\simeq 1.5$, so that the jet power can be estimated by
\be
Q_{\rm jet} = f^{3/2}\,Q_{\rm jet,0} \simeq 3\times 10^{45}\,f^{3/2}\, \left({P_{151} \over 10^{28}\,\text{W}\,\text{Hz}^{-1}\,\text{sr}^{-1}}\right)^{6/7}\,{\text{erg} \over \text{s}}\,.
\ee
with $1\lesssim f \lesssim 20$ resulting from the uncertain efficiency of converting jet internal energy into observable radio luminosity (see also \cite{2000AJ....119.1111B, 2004MNRAS.349.1419C} for additional discussion on the normalization factor). Furthermore, the minimal energy condition for $Q_{\rm jet}$ is given by \cite{1970ranp.book.....P}
\be
Q_B\simeq {3\over 4}\,(Q_e+Q_{\rm cr})\,,
\ee
where $Q_B$, $Q_e$ and $Q_{\rm cr}$ denote the total power in the magnetic field, relativistic electrons, and cosmic rays, respectively. Assuming that these components dominate the energy density of the jet, we then have $Q_{\rm jet} \simeq Q_B+Q_e+Q_{\rm cr}$, hence $Q_{\rm jet}\simeq \frac73  Q_B$. The magnetic field power of the jet with a bulk velocity of $c\beta_{\rm jet}$ and a size $r$ is determined by $Q_B=c\,\beta_{\rm jet}\,\pi r^2\,B^2/8\pi$, so that we obtain
\be
B\, r \simeq \sqrt{{24\,Q_{\rm jet} \over 7 c\,\beta_{\rm jet}}}\,.
\label{magneticField2jetPower}
\ee
We consider the RGs to be in a steady state, where a characteristic loss time scale constrains the acceleration process: 
On the one hand, the CRs escape the accelerator on a time scale
\be 
\tau_{\rm esc} = {f_{\rm loss} r \over \beta_{\rm sh} c}\,,
\label{escape-timescale}
\ee 
where $f_{\rm loss}\sim 1$ is a fudge factor adapting the speed of expansion or diffusive particle escape downstream of the shock to the shock velocity; essentially, we suppose that there is one dominant velocity scale in the system.
On the other hand, energy losses due to synchrotron radiation, limit the acceleration process on a time scale
\be
\tau_{\rm syn}={6\pi\,m_p^2\,c^3 \over \sigma_{\rm T}\,B^2}\,\left({m_p\over m_e}\right)^2\,{A^2 \over Z^4}\,E^{-1}\,.
\label{synLoss-timescale}
\ee
where $\sigma_{\rm T} = 8 \pi e^4/3 m_e^2 c^4$ is the Thomson cross section. The maximal cosmic-ray energy then yields
\be
E_{\rm max}\simeq 2\times 10^{20}\,{A^2 \over Z^{3/2}}\,{\beta_{\rm sh}\over f_{\rm diff}^{1/2}}\,\left( {B \over 1\,\text{G}} \right)^{-1/2}\,\text{eV}\,,
\ee 
so that a magnetic field strength $B<0.02\,\text{G}$ is needed to enable the acceleration of CR protons up to an energy of $100\,\text{EeV}$, if we suppose a typical shock velocity of $\beta_{\rm sh}\sim 0.1$ and $f_{\rm diff}>1$. 
However, the comparison of the loss time scales shows that escape already dominates for
\be
B<0.02\,\left({\beta_{\rm sh} \over f_{\rm loss}} \right)^{1/2}\,\left( {A \over Z} \right)^2\,\left({E \over 100\,\text{EeV}} \right)^{-1/2}\,\left({r \over 1\,\text{kpc}} \right)^{-1/2}\,\text{G}\,.
\ee
Hence, using Eq.\ (\ref{magneticField2jetPower}), a jet with a power $Q_{\rm jet}$ features negligible synchrotron losses in structures with an extension 
\be
r>0.3\,f_{\rm loss}\, \beta_{\rm jet} ^{-1}\, \beta_{\rm sh}^{-1}\,\left( {Z \over A} \right)^4\, \left({Q_{\rm jet} \over 10^{46}\,\text{erg/s}} \right) \, \left({E \over 100\,\text{EeV}} \right) \, \text{pc}\,.
\ee
In the large-scale structures $\gtrsim 1\,\text{pc}$ of a radio galaxy, e.g.\ extended jets, hot spots, and lobes, the maximal CR energy is therefore determined by equating the acceleration time (\ref{acc-timescale}) to the escape time (\ref{escape-timescale}). Here, we absorb the factor $f_{\rm loss}$ into $f_{\rm diff}\sim 3$, so that the maximal cosmic-ray rigidity yields
\be
\hat{R}\equiv {E_{\rm max} \over Z\,e} = {\beta_{\rm sh} \over f_{\rm diff}}\,B\, r = g_{\rm acc} \sqrt{\frac{g_{\rm cr} Q_{\rm jet, 0}}{c}} = {g_{\rm acc} \over e}\,\sqrt{\alpha_{\rm f} \hbar\,g_{\rm cr} Q_{\rm jet, 0}\phantom{\big|}}\,.
\label{Rmax}
\ee
The last writing is useful when calculating energies, $Z e R$, as all electromagnetic constants disappear in the dimensionless Sommerfeld constant $\alpha_{\rm f} = e^2/\hbar c \approx 1/137$ \citep{Rachen:2008zg}.
Here, we introduced the dimensionless coefficients 
\be
g_{\rm cr} \equiv \frac47 f^{3/2} \quad{\rm and}\quad 
g_{\rm acc}=\sqrt{\frac{6\,\beta_{\rm sh}^2}{f_{\rm diff}^2\,\beta_{\rm jet}}}\;.
\ee
The cosmic-ray luminosity of a source in steady state is equal to the cosmic-ray fraction of the jet power, $Q_{\rm cr}$, so that we can use these terms synonymous and write for minimal energy conditions with $Q_{\rm cr}\gg Q_e$, 
\be
Q_{\rm cr}\simeq 1.3\times 10^{42}\,g_{\rm cr}\, \left({L_{1.1} \over L_\star} \right)^{6/7}\,\frac{\text{erg}}{\text{s}}\,,
\label{CRluminosity}
\ee
where we converted to the quantities used to describe the RLF, again assuming a typical $\nu^{-0.6}$ spectrum; note that $Q_{\rm cr} = \frac47 Q_{\rm jet}$ and $\hat{R}(Q_{\rm cr}) = g_{\rm acc} \sqrt{Q_{\rm cr}/c}$. Supposing the typical shock and jet velocities $\beta_{\rm sh} \sim \beta_{\rm jet}\sim 0.1$ in extended jets of radio galaxies, we obtain $0.01 \lesssim g_{\rm acc} \lesssim 1$, and as we assume $Q_{\rm cr} \gg Q_e$, the jet power normalization factor $f$ will rather be at the high end of its range, hence we assume $5 \lesssim g_{\rm cr} \lesssim 50$. If we distinguish between species of cosmic ray nuclei of charge $Z_i$ and assume that they represent a number fraction $f_i$ of mixed composition cosmic rays accelerated all over the same rigidity regime, the fractional contribution of this species to the total cosmic-ray power will be $f_i Z_i$. 

Consequently, even for optimal conditions ($g_{\rm acc} = 1$ and $g_{\rm cr} = 50$), only sources with radio luminosity $L_{1.1} > 0.028\,L_\star\,Z^{-7/3}$ can contribute cosmic rays with charge number $Z$ to the UHECR regime, i.e., an energy of $3\,\text{EeV}$ or higher. To reach the highest energies, $E\gtrsim 100\,$EeV, this requirement becomes $L_{1.1} \gtrsim 100\,L_\star\,Z^{-7/3}$. 
Thus, only the most powerful radio galaxies can accelerate protons to the highest energies, and for those radio galaxies emitting the most power per volume ($L_{1.1} \approx L_\star$), we need to require $Z>7$ to obtain this. Interesting is also the relation to $Q_{\rm cr}$, which implies 
\be 
Q_{\rm cr} \sim 10^{44} \text{erg}/\text{s} \quad\text{for}\quad \hat R \sim 10^{20}\,\text{V} 
\ee
This relation has been first pointed out by Lovelace \cite{1976Natur.262..649L} for a dynamo model of radio galaxies, but as we see here, it follows in general from combining the Hillas limit with minimal energy conditions in radio galaxy jets. We will refer to the general correspondence between cosmic ray power and maximal rigidity expressed by Eq.\ (\ref{Rmax}) as the \textit{Lovelace-Hillas relation} henceforth. Combining this with the Eddington luminosity of the supermassive black hole at the center of the RG, one could link this to the black-hole mass to produce a generic distribution of UHECR sources \cite{2012PhRvD..86f3005K}, but we have to note that many AGN accrete only at a small fraction of their Eddington luminosity and the scatter of this fraction is large \cite{2004A&A...414..895F}. 

\subsection{Continuous source function of UHECR from radio galaxies}
\label{Sec:AverageNonLocalSource}

Lacking the possibility to do efficient Monte Carlo calculations, early work on UHECR origin, e.g.\ \cite{Rachen:1992pg,Waxman:1995vg}, used continuous source functions derived from average properties of the sources considered. In this approach, cosmic rays are injected with a homogeneous rate per unit volume throughout the universe, where potentially dependences cosmological evolution (i.e., redshift) are allowed. The flux is then obtained by integration over the comoving volume, using modification factors to consider propagation effects \cite{Hill:1983mk, 1988A&A...199....1B}. For the radio-galaxy source function, the Lovelace-Hillas relation introduces a major impact as it predicts a power-dependent maximal rigidity of radio galaxies \cite{2013AdSpR..51..315P}. We derive here an analytical approximation based on the MS07 radio luminosity function for AGN.

The local continuous source function, $\Psi_{i,0}(R)$, derives from the local luminosity function of radio galaxies and the spectrum of individual sources as 
\begin{equation}
\Psi_{i,0}(R) \equiv {\mathrm{d}N_{\rm cr}(Z_i) \over \mathrm{d}V \mathrm{d}R\,\mathrm{d}t} = \int_{\check Q_{\rm cr}}^{\hat Q_{\rm cr}} S_i\big(R,\hat R(Q_{\rm cr})\big)\,{\mathrm{d}N_{\rm RG} \over \mathrm{d}V\,\mathrm{d}Q_{\rm cr}}\,\mathrm{d}Q_{\rm cr} 
\label{CRsourceRateDensity}
\end{equation}
where $\diff N_{\rm cr}$ is the number of cosmic ray particles of charge number $Z$ emitted at rigidity $R$ per energy bin $Z \,e \, \diff R$, volume element $\diff V$ and time interval $\diff t$. \mbox{$S_i(R,\hat R(Q_{\rm cr})) \equiv \diff N_{\rm cr}(Z_i)/\diff R\,\diff t$} is the cosmic ray spectrum of element species $i$ with charge number $Z_i$, emitted by a source with total cosmic ray power per charge number, $Q_{{\rm cr},i}\equiv Q_{\rm cr}(Z_i)=f_i\,Z_i\,Q_{\rm cr}/\bar Z$, up to a maximal rigidity $\hat R(Q_{\rm cr})$, and $\diff N_{\rm RG}/\diff V\,d Q_{\rm cr}$ is the number of radio galaxies per volume per power bin. The limits of integration are the smallest, $\check Q_{\rm cr}$, respectively largest, $\hat Q_{\rm cr}$, CR powers we need to consider. To convert rigidity into energy we use the average cosmic ray charge number at a given rigidity, $\bar Z = \sum_i f_i\,Z_i$, where the sum goes over all elements we consider with modified solar injection abundances $f_i$, at a constant injection rigidity $\check R\simeq 10\,\text{GV}$ corresponding to the peak of the Galactic CR flux.

To solve this integral analytically, we make the simplifying assumption that individual source spectra are power laws with a sharp cutoff at $\hat R(Q_{\rm cr}) = g_{\rm acc} \sqrt{Q_{\rm cr}/c}$, i.e. 
\begin{equation}\label{eq:siglespec}
S_i(R,\hat R(Q_{\rm cr})) = \nu_i(a)\,Q_{\rm cr}\,\left({R \over \check R}\right)^{-a}\,\Theta\big(\hat R(Q_{\rm cr}) - R\big)\;,
\end{equation}
where $\Theta(x)$ is the Heaviside step function. 
From the requirement 
\be
Q_{\text{cr},i}={f_i\,Z_i\,Q_{\rm cr} \over \bar Z} =eZ_i\int_{\check{R}}^{\hat{R}(Q_{\rm cr})} \text{d}R\,R\;S_i(R,\hat R(Q_{\rm cr}))\,,
\ee 
we obtain the spectral normalization correction $\nu_i(a)$ as
\be
\nu_i(a) = {f_i \over e \bar Z \check{R}^2}\times
\begin{cases}
\;(2-a)\,\big/\,(\rho_{\rm cr}^{2-a}-1)\;,& \text{ for }\, a\neq 2\,\\
\;1/\ln \rho_{\rm cr} \phantom{\Big|}\;,& \text{ for }\, a=2
\label{nonlocalNorm0}
\end{cases}
\ee
where we introduced the cosmic ray dynamical range $\rho_{\rm cr} \equiv \hat{R}(Q_{\rm cr})/\check{R}$.
The radio luminosity function $\Phi_{\rm RG}(Q)$ provides the number of radio galaxies per volume per logarithmic power bin, so that $\diff N_{\rm RG}/\diff V\,\diff Q_{\rm cr} = Q_{\rm cr}^{-1}\,\Phi_{\rm RG}(Q_{\rm cr})$. 
Using the MS07 luminosity function dependent on the CR power where 
\be
P_{1.4}=3.31\times 10^{28}\,\left({ Q_{\rm cr} \over 3\times 10^{38}\,g_{\rm cr}\,\text{W} } \right)^{7/6}\,\text{W}\,\text{Hz}^{-1}\,,
\ee
for a fiducial radio spectrum $\propto \nu^{-0.6}$, we obtain 
\be
{\diff N_{\rm RG}\over \diff V\,\diff Q_{\rm cr} }= {7 \over 6}\, Q_{\rm cr}^{-1}\,\Phi_{\rm RG}\big(P_{1.4}(Q_{\rm cr})\big) = \frac76\,\frac{C_{\rm RG}}{g_{\rm cr} Q_\star}\,\left[\left({Q_{\rm cr}\over g_{\rm cr}\,Q_{*}} \right)^{\alpha'} + \left({Q_{\rm cr}\over g_{\rm cr}\,Q_{*}} \right)^{\beta'}\right]^{-1}\, , 
\ee
where $C_{\rm RG} \approx 3\times 10^{-6}\,{\rm Mpc}^{-3}$ is the source density parameter and $\alpha' = \frac 76 \alpha + 1 \simeq 2.48$ as well as $\beta' = \frac76 \beta + 1 \simeq 1.55$ are related to the indices for the radio luminosity function given below Eq.\ (\ref{RLF}). 
Further, we introduced $Q_\star \simeq 1.3\times 10^{42}\,$erg/s, according to Eq.\ (\ref{CRluminosity}), as the CR power related to the break in the radio luminosity function. 
We can then find an approximate analytical solution to Eq.\ (\ref{CRsourceRateDensity}) as 
\begin{equation}
\Psi_{i,0}(R) \;\approx\; {7 \over 6}\; \frac{C_{\rm RG}\,f_i\,\tilde\nu(a)\,c}{e\,\bar Z\,g_{\rm acc}^2}\,\left(\frac{R}{R_\star}\right)^{-a}\!\times\; 
\left\{ F_{\tilde{\alpha}\tilde{\beta}}\left(\frac{R^2}{R_\star^2}\right) - F_{\tilde{\alpha}\tilde{\beta}}\left(\frac{\hat Q_{\rm cr}}{g_{\rm cr}\,Q_{*}}\right) \right\}\,,
\label{ProdRateDens1}
\end{equation}
for the three simplifying cases
\begin{align*}
\tilde\nu(a) &= 2-a\;; & \tilde\alpha &= \alpha'-a/2\;; & \tilde\beta &= \beta'-a/2\;; &\qquad\text{for} &\;a<2,\,a\not\approx 2\\
\tilde\nu(a) &= 1/\ln \rho_\star\;; & \tilde\alpha &= \alpha'-1\;; & \tilde\beta &= \beta'-1\;; &\qquad\text{for} &\;a\simeq 2\\
\tilde\nu(a) &= (a-2)\rho_\star^{2-a}\;; & \tilde\alpha &= \alpha'-1\;; & \tilde\beta &= \beta'-1\;; &\qquad\text{for} &\;a>2,\,a\not\approx 2
\end{align*}
where we introduced $R_\star = g_{\rm acc}\sqrt{g_{\rm cr} Q_\star/c}$, as well as the corresponding dynamic range $\rho_\star = R_\star/\check{R}$, and
\begin{equation*}
F_{\alpha\beta}(x) \equiv \frac{x^{1-\alpha}}{\alpha-1}\;_2F_1\!\left[ 1, \frac{\alpha-1}{\alpha-\beta};\frac{\alpha-1}{\alpha-\beta}+1; -x^{\beta-\alpha}\right]
\end{equation*}
with $_2F_1$ denoting the hypergeometric function.\footnote{This solution is only valid for $0 < \tilde\beta < 1 < \tilde\alpha$, which is the case here.} The first respectively last case corresponds to $\rho_\star^{2-a}\gg 1$ for $a<2$ (and is thus included in the integration), and $\rho_\star^{2-a} \ll 1$ for $a>2$, which can both be assumed even for moderate deviations of the spectral index from $a=2$ as $\rho_\star \sim 10^9$. For $a\simeq 2$, we simply set $\hat{R}(Q_{\rm cr}) \simeq R_\star$ constant in the slowly varying logarithm. $\hat Q_{\rm cr}$ is the highest jet power considered, 
and should correspond to the most powerful FR-II galaxies, like Cygnus A; we set it to $10^{45}\,g_{\rm cr}\,$erg/s. Assuming higher values of $\hat Q_{\rm cr}$ would not significantly change the resulting production rate.

The function $\Psi_{i,0}(R)$ is the local continuous source function as it is derived from a radio luminosity function determined in the local universe ($z<0.3$). To extend it to larger redshifts, we use the approximation
\be
\Psi_i(R,z) \approx \Psi_{i,0}(R)\,(1+z)^{m-1}\,,
\label{eq:SourceFunctionEvolution}
\ee
where $m$ is called the source evolution index. This writing disregards the (likely) dependence on characteristic parameters like $L_\star$, $\alpha$ and $\beta$ on $z$. However, it has been shown that different redshift dependent RLFs with roughly the same mean index $m$ produce very similar UHECR spectra \cite{RachenDiplom1992,Rachen:1992pg}, so we adopt Eq. (\ref{eq:SourceFunctionEvolution}) for $m \in [2,4]$. 

With this, the free parameters of the model are $f_i$, $a$, $g_{\rm cr}$, $g_{\rm acc}$, and $m$, with quite limited plausibility ranges as discussed the previous sections (note that $R_\star$ and $\rho_\star$ depend on $g_{\rm cr}$ and $g_{\rm acc}$). This means, that there is little margin to adjust the total normalization of the model.  Setting canonical values of $a=2$, $g_{\rm cr}=50$, $g_{\rm acc}=0.5$ and using solar abundances $f_i=f_{\odot}$, we get a total UHECR production rate density of 
\be 
\langle\Psi\rangle_{\rm UHE} = \sum_i\,\int_{10^{19}eV}^{10^{20}eV} \diff E\,E\,{\Psi_{i,0}(E/Z_i e) \over Z_i\,e} \approx 4\times10^{44}\,{\rm erg}\,{\rm Mpc}^{-3}\,{\rm year^{-1}}
\ee
comparable to the value obtained by Waxman \cite{Waxman:1995dg} as required to explain the observed UHECR flux at Earth. As the total normalization at $R \gtrsim R_\star$ is approximately $\propto g_{\rm cr}^{1.5}\,g_{\rm acc}$ (see below), this estimate presents and upper limit regarding these parameters. The source evolution parameter $m$ is mostly relevant below the ankle and cannot significantly modify the UHECR normalization. Looking at $\tilde\nu(a)$, up to an order of magnitude increase is possible for $a<2$, while for $a>2$ every increase of $a$ by $0.1$ drops the normalization by almost one order of magnitude.  Thus, spectral indices significantly steeper than $a=2$ are strongly disfavored.

\begin{figure}[tbh]
  \centering
    \includegraphics[width=0.79\textwidth]{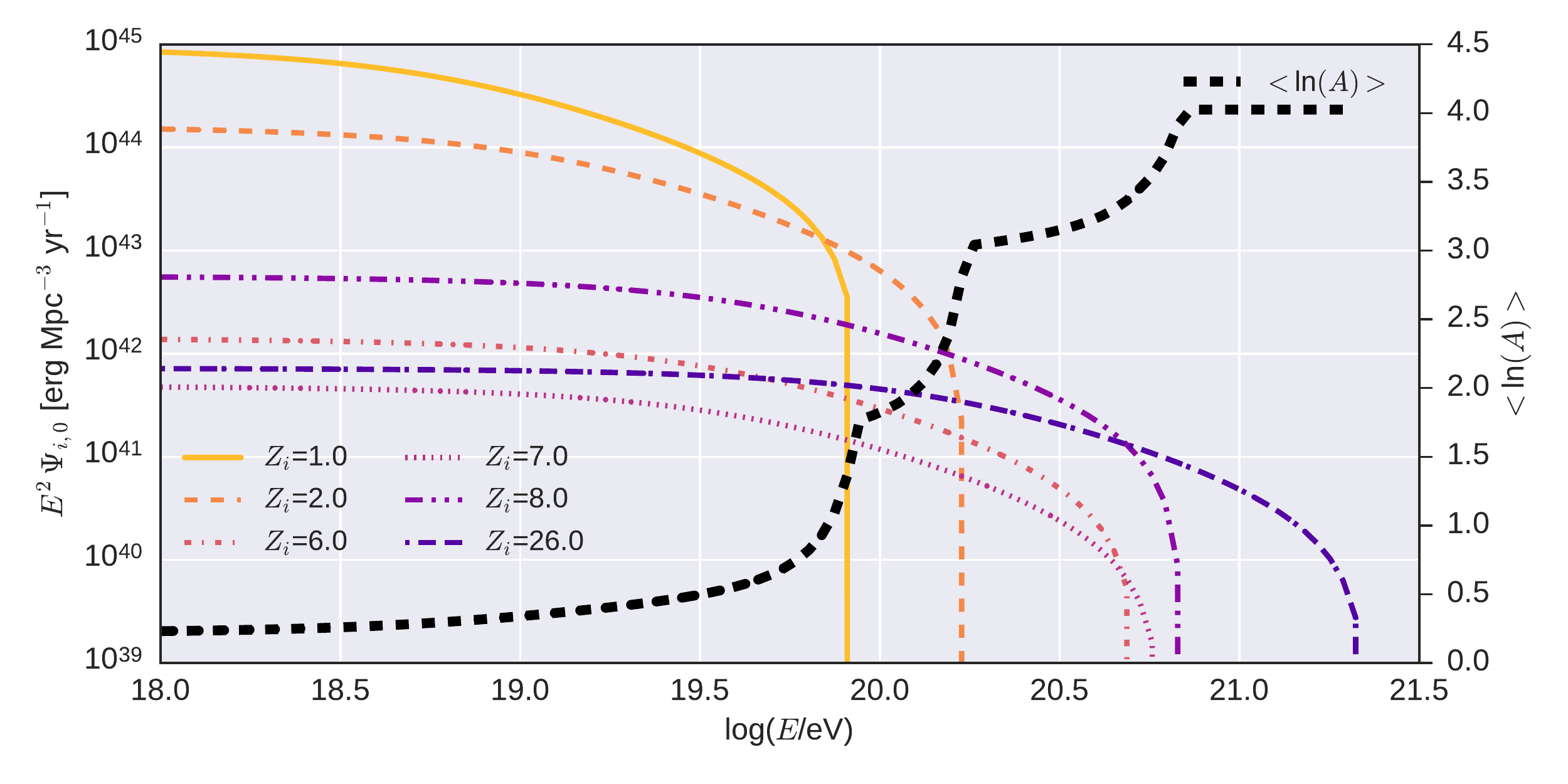}
\caption{UHECR production rate density (colored, solid lines) and corresponding $\langle \ln(A) \rangle$ (black, dashed line) using $a=2$, $g_{\rm cr}=50$, $g_{\rm acc}=0.5$, as well as an initial solar abundance $f_i=f_{\odot}$. Note that the sharp cutoffs in the spectra and also the ``stairs'' in the $\langle \ln(A) \rangle$ curve are artefacts from using a Heaviside cutoff in the derivation of the continuous source function.}
\label{UHECRprodRateDens}
\end{figure} 

Let us now analyze the shape and characteristic energies connected to this spectrum.  
The term in curly brackets of Eq.\ (\ref{ProdRateDens1}) consists of two modified hypergeometric functions, where the second is simply a (usually very small) constant. 
In the case of a small value of $R$, the first term runs asymptotically to a constant $\sim 1$, so at low rigidities we see an undistorted $R^{-a}$ power law. 
Towards large values of $R$, it approaches asymptotically a power law $R^{2-2\tilde{\alpha}}$, so that in total we obtain a production rate density $\propto R^{4-2\alpha'-a}$ for $a\gtrapprox 2$, which approaches a limit $\propto R^{2-2\alpha'}$ for $a < 2$; thus for flat spectra and rigidities above the break, the spectral index is only determined by the spectral slope of the RLF, and is close to $E^{-3}$. This is a result of the sliding-cutoff effect introduced by the Lovelace-Hillas relation, and a qualitatively new feature not contained in any source luminosity function determined so far.  The break occurs at a critical rigidity $R_* = g_{\rm acc} \sqrt{g_{\rm cr}\,Q_* / c} \approx 2\times 10^{18}\,g_{\rm acc}\,\sqrt{g_{\rm cr}}\,\text{V}$, thus below the UHECR regime for protons. Viewed in energy, we then get a succession of smooth breaks for all element species. Effectively, this could lead to a successive increase of the heaviness of the UHECR spectrum, as seen in the data, provided that abundances of heavy elements are not increased that much that they dominate already below the break. Fig.\ \ref{UHECRprodRateDens} shows this effect for a scenario of solar abundances and maximal parameters $g_{\rm cr}$ and $g_{\rm acc}$. It clearly shows an at first slow increase of $\langle A \rangle$ due to a succession of breaks, and then a fast increase due to a succession of cutoffs. If we had considered realistic cutoffs of the individual sources, we would expect that the spectra of different species just run out into exponentials, and the $\langle A \rangle$ curve would run smoothly below the dotted line shown. Lower values of $g_{\rm cr}$ and $g_{\rm acc}$ would shift that whole scenario to lower energies.   

Putting everything together, this means that for light nuclei the spectral index in the UHECR regime is about $E^{-3}$, thus too steep to explain the observed UHECR spectrum (as propagation will make the spectrum even steeper). For heavy nuclei, $e Z R_\star$ may be large enough to allow UHECR spectra consistent with the data, but this would mean that they have to dominate in the entire UHECR regime, which is inconsistent with the measured composition. We can therefore already get to the preliminary conclusion that the \textit{cosmic-ray spectrum averaged over the radio luminosity function cannot explain the known UHECR data}. This agrees with the earlier finding of Ptuskin et al. \cite{2013AdSpR..51..315P} using an approach similar to ours. We note, however, that their attempt to overcome this problem by distinguishing between \mbox{FR-I} and \mbox{FR-II} radio galaxies and assuming a significantly larger acceleration efficiency for the latter does not reproduce the UHECR spectrum up to the highest energies observed \cite[Figs.~2 \& 3]{2013AdSpR..51..315P}. Therefore, we need to consider the contribution of powerful individual sources in the local universe, where at least one has to be nearby to be unaffected by GZK losses. The estimation of their impact is the subject of our simulation setup, which is explained in the following.

\section{Simulation setup}
\label{sec:1}
The Monte Carlo based algorithm of CRPropa has been designed to describe the propagation of UHECR nuclei in extragalactic background light and cosmic magnetic fields \cite{2007APh....28..463A, 2013APh....42...41K}. 
The recently published version CRPropa3 has been redesigned for better performance, and includes improved methods and new functionalities \cite{1475-7516-2016-05-038}. It thus represents the ideal framework to explain the energy spectrum, composition and arrival directions of UHECRs (e.g.\ \cite{AlvesBatista:2017vob, Hackstein:2016pwa, 2017arXiv171005617W}). 

In CRPropa3, the trajectory of the particles is determined by the Lorentz force according to the Cash-Karp algorithm, an adaptive Runge-Kutta algorithm of the fourth order. In doing so, energy loss due to photonuclear interactions (e.g., pair production, photo-pion production, and photo-disintegration) with the cosmic microwave background (CMB) and the UV/optical/IR background (IRB) is taken into account. 
Furthermore, the decay of unstable nuclei is considered as well. 
The necessary photon fields of the CMB and the IRB, where the IRB model of Gilmore et al.\ \cite{2012MNRAS.422.3189G} is used, are already implemented. The uncertainties in the simulation results due to poorly known physical quantities as the IRB and the photo-disintegration cross sections should not affect the results in a significant way \cite{Batista:2015mea}.

In order to obtain suitable statistics, an observer sphere with a radius of $r_{\rm obs}=100\,\text{kpc}$ is chosen. This size also ensures, that no significant deformation of the higher multipole moments of the arrival directions occurs, as recently shown by Dundovi{\'c} and Sigl \cite{2017arXiv171005517D}. 
Thus, UHECR candidates are only propagated on extragalactic scales at first.
Afterwards, deflection of the candidates inside the Milky Way is easily applied using the Galactic lensing module.  
Here, the GMF model of Jansson \& Farrar \cite{2012ApJ...757...14J} --- hereafter referred to as the JF12 model --- is adopted to trace the particles from the observer sphere through the Galactic magnetic field to Earth. 
This treatment neglects energy losses inside the Galaxy, which is justified as Galactic trajectory lengths are small compared to the mean free path of interactions.
Further details and explanations of the simulation setup are given in Appendix \ref{detailsOfSim}.
Consequently, the physical constituents that still need to be applied to the simulation setup are (i) the EGMF and (ii) the source setup (e.g. spatial distribution, energy spectrum, composition).

\subsection{The extragalactic magnetic field}
\label{Sec:EGMF}
We apply the local EGMF model of Dolag et al.\ \cite{2005JCAP...01..009D} --- hereafter referred to as D05 --- to define the magnetic field up to a distance of about 120 Mpc from Earth. 
Here, a uniform magnetic seed field with $B_0=2\times 10^{-12}\,\text{G}$ is treated in the framework of ideal MHD in a constrained structure-formation simulation of the local universe (see \cite{1999A&A...348..351D} for more details). 
The arbitrarily chosen initial orientation of the uniform seed magnetic field gets randomized in dense environments, e.g. galaxy clusters, but in low density regions the initial orientation of the seed field is conserved. 
Nevertheless, it is shown by Dolag et al.\ \cite{2005JCAP...01..009D} that there is no statistical difference in the distribution of deflected UHE protons with respect to the chosen orientation of the seed field.
Magnetic-field strengths of the order of a few nG and higher are only found in dense environments like galaxy clusters, so significant deflections are predominantly expected from UHECRs that pass through dense environments. On large scales these environments are rare and not surrounding the Milky Way, so that the sky is dominated by areas showing small deflections \cite{2005JCAP...01..009D}. We note, however, that other simulations obtain higher cumulative filling factors of strong fields \cite{AlvesBatista:2017vob}.

The advantage of using the D05 field is that it represents the true structure of the local universe, which is essential when comparing with a source distribution based on astronomical catalogs.
Therefore, the D05 EGMF is sampled in this simulation with a high resolution of $14.6\,\text{kpc}$ and subsequently stored in a multi-resolution grid that has been developed by M{\"u}ller \cite{2016JCAP...08..025M} using a relative error of $0.4$ and an absolute error of $10^{-14}$.

\subsection{Characteristics of the source sample of UHECRs}
\label{Sec:SourceSample}
As our implemented EGMF structure extends only to 120\,Mpc, we need to distinguish two groups of sources: (1) Sources up to a distance of 120 Mpc from Earth, called \emph{local} RGs, which are treated as individual sources in 3D simulations making full use of the EGMF structure; (2) Sources beyond 120 Mpc, called \emph{non-local} RGs, for which the continuous source function derived in Sect.\ \ref{Sec:AverageNonLocalSource} is used as input for 1D simulations. Continuity requirements imply that the average source features of local and non-local RGs are the same in particular in the transition regime around 120\,Mpc, which is carefully checked (see Appendix \ref{TestNormNonLocal}).
For all sources we assume a temporally constant and isotropic outflow of UHECRs described by a single power law with the same spectral index $a$, limited to a range $1.7\leq a \leq 2.2$ according to the constraints discussed in sections \ref{Sec:Lradio2Ljet2Rmax} and \ref{Sec:AverageNonLocalSource}.  
We explicitly simulate the propagations of the nuclear species $^1$H, $^4$He, $^{12}$C, $^{14}$N, $^{16}$O, and $^{56}$Fe, i.e., $Z_i \in \{1,2,6,7,8,26\}$. As nuclei in the range $Z=3,4,5$ have low abundances and are quite unstable to spallation in high-energy propagation, this range can be considered complete up to oxygen. Reducing the heavy elements beyond CNO to iron, however, is a simplification chosen just for convenience, and we need to discuss potential artifacts introduced by it in our results.  

For the chemical composition at a given rigidity $R$ we distinguish two cases: (1) the \textit{light composition scenario} assuming solar abundances (in particle number, e.g.\ \cite{2009LanB...4B...44L}), given by
\be
f_{\odot} = \begin{cases}         
0.922\,,\text{ for } \text{H}\,,\\
0.078\,,\text{ for } \text{He}\,,\\
2.3\times 10^{-4}\,,\text{ for } \text{C}\,,\\
6.7\times 10^{-5}\,,\text{ for } \text{N}\,,\\
4.9\times 10^{-4}\,,\text{ for } \text{O}\,,\\
2.7\times 10^{-5}\,,\text{ for } \text{Fe}\,,
\end{cases}
\ee
with a strong dominance of ${}^{1}\text{H}$ and ${}^{4}\text{He}$ nuclei; 
(2) the \textit{heavy composition scenario}, where we assume modification of the abundances at injection to Fermi acceleration as $f_i \propto f_{\odot} Z^q$ with $1\lesssim q\le 2$. For the limiting case of $q=2$ we obtain abundances 
\be
f_{\bullet}\equiv Z^2\,f_{\odot} = \begin{cases}         
0.706\,,\text{ for } \text{H}\,,\\
0.238\,,\text{ for } \text{He}\,,\\
0.006\,,\text{ for } \text{C}\,,\\
0.003\,,\text{ for } \text{N}\,,\\
0.033\,,\text{ for } \text{O}\,,\\
0.014\,,\text{ for } \text{Fe}\,.
\end{cases}
\ee
In this case, the iron contribution to the UHECR source spectrum is comparable to that of hydrogen (note that the abundances compared at a given energy $E=ZeR$ are further enhanced by $Z^{a-1}$). We treat these two cases as extremes in many explanatory examples, but for the fit to the data we optimize the value of $q$ in its given range. We note that other composition enhancements would be possible depending on the unknown details of the injection process, but they are not considered here unless explicitly noted.

Based on the characteristics of the Monte Carlo method, most source characteristics are changeable after the simulation by re-weighting (e.g.\ \cite{2007APh....28..463A}). 
Further details on the weights we use are given in Sect.\ \ref{AbsoluteNormalization}.

\subsubsection{Local sources}
\label{Sec:LocalSources}
To obtain a complete sample of radio galaxies powerful enough to contribute to the UHECR spectrum we use the catalog of van Velzen et al.\ \cite{2012A&A...544A..18V} --- hereafter referred to as \emph{vV12 catalog}. In total it contains 575 radio-emitting galaxies with a flux greater than $213\,\text{mJy}$ at $1.4\,\text{GHz}$. 
We note, however, that the catalog is based 2MASS redshift survey \cite{2012ApJS..199...26H}, which introduces other selection criteria, in particular a limiting infrared $K$-band magnitude of $11.75$. This excludes preferentially distant powerful radio galaxies for which the radio-to-infrared luminosity ratio is larger than average. An example is the powerful FR-II galaxy Pictor A at 125\,Mpc distance, which is among the ten brightest radio galaxies in the sky but not contained in the vV12 catalog. Cross checks with searches in the NASA/IPAC extragalactic database\footnote{http://ned.ipac.caltech.edu/} showed an increasing incompleteness\footnote{We note for clarification that this refers to the secondary goal of the vV12 catalog to be a complete collection of radio galaxies potentially relevant as UHECR sources. There is no doubt that the vV12 catalog is a complete sample with respect to its astronomical selection criteria.} of the vV12 catalog at the bright end for distances ${>}\,120\,$Mpc. Up to $120\,$Mpc, however, we found no evidence for significant incompleteness in this respect. 

Excluding 52 starforming galaxies, we are finally left with 121 radio galaxies up to a distance of $120\,$Mpc, predominantly of FR-I type, which will subsequently be called the \emph{visible, local} RGs.  
In addition to the spatial coordinates $(x_0,y_0,z_0)$, the catalog also provides the radio luminosity $L$ at $1.1\,\text{GHz}$ of each RG and the angular distance $\delta_d$ between the galaxy and the radio matches. 
But still, the catalog only covers 88\% of the sky and misses a multitude of low-luminous, distant RGs due to its flux limit. 

To complete the source sample, we use the MS07 RLF (\ref{RLF}) and consider spherical shell elements with a thickness $\hat{d}-\check{d}$ at small redshift and a radio luminosity between $\check{L}$ and $\hat{L}$. 
Hence, we obtain the expected total number of radio-loud AGNs in the fraction $P_{V}$ of the volume by
\be
N_{RG} = C_{\rm RG}\,{4\pi \over 3}\,\left(\hat{d}^3-\check{d}^3 \right)\,P_{V}\, \int_{\check{L}_{1.1}}^{\hat{L}_{1.1}} \diff L\,\,L^{-1}\,\left[ (L/L_\star)^{\alpha}+(L/L_\star)^{\beta} \right]^{-1}\,.
\label{numberOfAGNfromRLF}
\ee
with the best-fit parameter $C_{\rm RG} \approx 3.162\times 10^{-6}\,\text{Mpc}^{-3}\,$, which has been determined by fitting the Eq.\ (\ref{numberOfAGNfromRLF}) with $P_{V}=0.88$ to the known number of radio-loud sources from the vV12 catalog. All other parameters are taken as the best-fit values given by MS07 \cite{2007MNRAS.375..931M}.
Since the integrand of Eq.\ (\ref{numberOfAGNfromRLF}) hardly depends on the upper integration limit $\hat{L}_{1.1}$, we set $\hat{L}_{1.1}=10^{45}\,\text{erg/s}$ in the following.\footnote{The most luminous RG in the vV12 catalog, Cygnus A, has a radio luminosity $L_{1.1}=1.58\times 10^{44}\,\text{erg/s}$.}
Due to the flux limitation of the catalog we had to increase the lower luminosity limit $\check{L}_{1.1}$ with distance according to $\check{L}_d^{\rm obs} = 3.241\times 10^{35}\,(d/\text{Mpc})^2\,\text{erg/s}$. 
Fig.\ \ref{FitRLF2catalog} shows the best-fit result with a $p$-value from a Kolmogorov-Smirnov test of about $0.9$. 
\begin{figure}[tbh]
  \centering
    \includegraphics[width=0.79\textwidth]{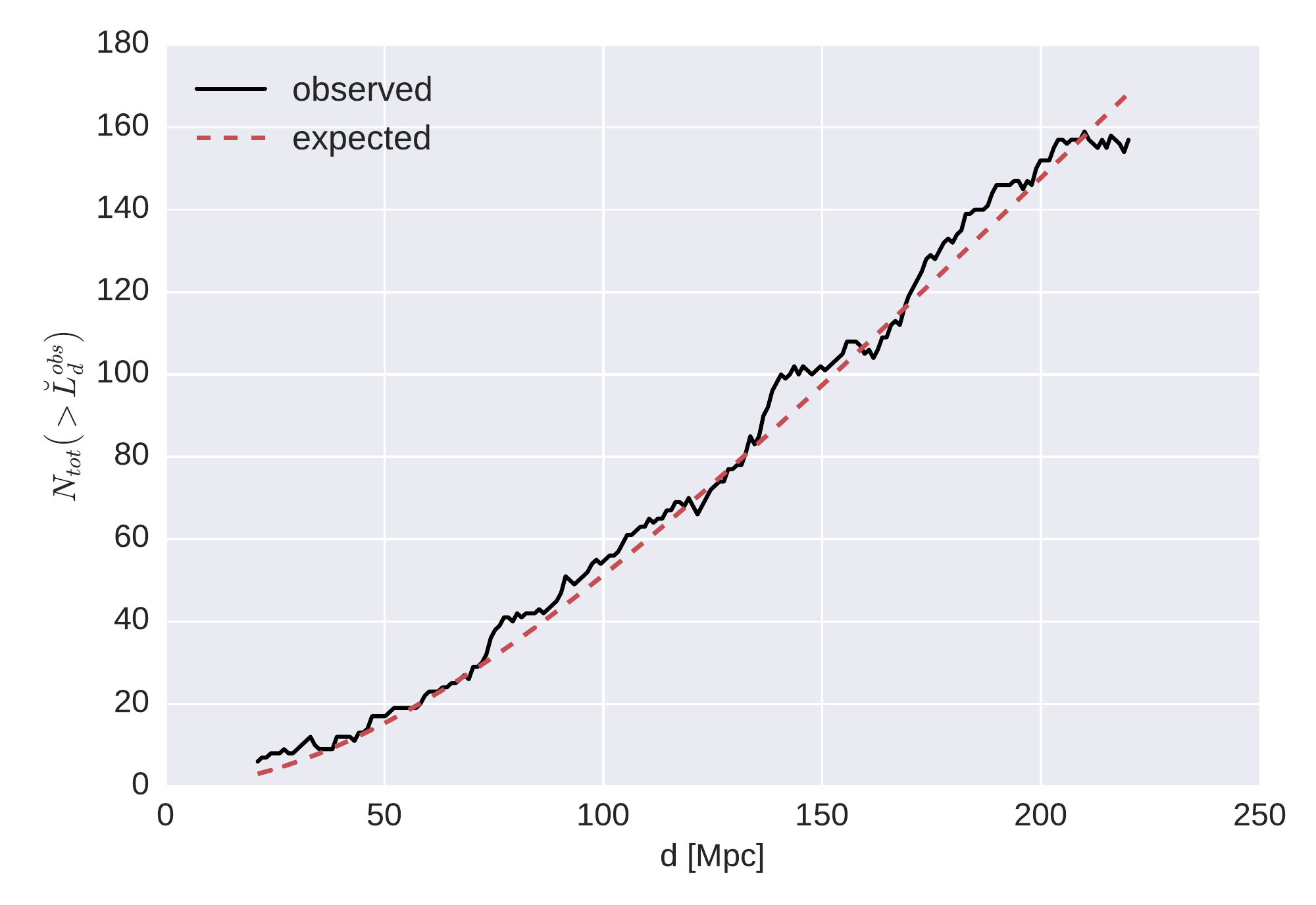}
\caption{Best-fit result of the expected number of radio-loud AGNs (green) according to Eq.\ (\ref{numberOfAGNfromRLF}) to the observed source number from the vV12 catalog (blue), where the constant of proportionality yields $C_{\rm RG} = 3.162\times 10^{-6}$. Here, a distance dependent minimal luminosity of $\check{L}_d^{obs} = 3.241\times 10^{35}\,(d/\text{Mpc})^2\,\text{erg/s}$ is used.}
\label{FitRLF2catalog}
\end{figure}
Note that the resulting fit slightly depends on the number $N_{\rm bins,d}$ of distance bins, but in order to obtain a sufficient number of known sources in each distance bin we suggest $100\lesssim N_{\text{bins},d} \lesssim 300$. 
More important, the constant of proportionality $C_{\rm RG}$ does not change significantly with $N_{\text{bins},d}$. 

To complete the catalog at the low luminosity end, we use a strategy as follows. 
According to Eq.\ (\ref{Rmax}) the minimal total luminosity at 1.1 GHz that enables an acceleration of UHECRs yields 
\be
\check{L}_{1.1}(Z) = 1.4\times 10^{39}\,Z^{-7/3}\,\text{erg/s}
\label{minL1.1}
\ee
Since the chemical composition around the ankle shows that $Z < 6$ (e.g.\ \cite{PhysRevD.90.122005}), there cannot be a significant contribution from sources with $L_{1.1}\leq 10^{38}\,\text{erg/s}$. 
Therefore, the luminosity of M49 at a distance of $15.8\,\text{Mpc}$, which is the least luminous source of the vV12 catalog, provides a reasonable lower luminosity limit of $\check{L}_{1.1} = 9.7\times 10^{37}\,\text{erg/s}$.
As an equally luminous radio galaxy at lower distances would likely be contained in the catalog, we distribute additional RGs only beyond a distance of $\check{d}=15.8\,\text{Mpc}$.
So, Eq.\ (\ref{numberOfAGNfromRLF}) is used to determine the unknown RGs up to a maximal distance of $\hat{d}=120\,\text{Mpc}$, with the following distinction: 
\begin{enumerate}
 \item[(i)] \emph{invisible, local} RGs with a luminosity $L_{\rm inv} \in [\check{L}_{1.1},\,\hat{L}_{1.1}]$ within the distance $d_{\rm un}=[\check{d},\,\hat{d}]$, that are located in the non-covered part of the sky\footnote{The vV12 catalog includes only sources with $|b|>5\degree$ for $30\degree <l<330\degree$, $|b|>8\degree$ for $l<30\degree$ or $l>330\degree$, and $|b|>10\degree$ for $\delta<-40\degree$, where $l$ and $b$ denote the Galactic longitude and latitude, respectively, and $\delta$ refers to the declination of the source.}, i.e. $P_V=0.12$. Here, we obtain a total number $N_{\rm RG}^{\rm inv}=100$ of invisible sources.
 \item[(ii)] \emph{low luminous, local} RGs with a luminosity $L_{\rm low} \in [\check{L}_{1.1},\,\check{L}_d^{\rm obs}]$ within the distance $d_{\rm un} \in [\check{d},\,\hat{d}]$, that are located in the covered part of the sky, i.e. $P_V=0.88$, but emit a radio flux below the observable flux limit. Here we obtain a total number $N_{\rm RG}^{\rm low}=617$ of low luminous sources.
\end{enumerate}
Next, we need to determine different realizations of a distribution $\Phi$ of these invisible and low luminous sources in space and luminosity. 
For this, each distribution has to comply with the following expectations.

First, as we expect the sources to be predominantly located within the supergalactic plane \cite{1953AJ.....58...30D, 2000MNRAS.312..166L} up to a distance of about $40\,h^{-1} \approx 59\,\text{Mpc}$, where $h=0.678$ \cite{2016A&A...594A...1P} is the dimensionless Hubble parameter, while being uniformly distributed in space beyond this distance, the radial source distribution within the distance interval $d_{\rm un}$ yields 
\be 
\phi_1(d) \propto \begin{cases} 
				  d\,,\quad&\text{ for }d\leq59\,\text{Mpc}\,,\\
                                  d^2\,,\quad&\text{ for }d>59\,\text{Mpc}\,.
                                 \end{cases}
\ee
\\ 
Second, the RLF
\be
\phi_2 (L_{1.1}) = {\diff N_{RG} \over \diff L_{1.1}\, \diff V} =  C_{\rm RG}\, L_{1.1}^{-1}\,\left[ (L_{1.1}/L_\star)^{\alpha}+(L_{1.1}/L_\star)^{\beta} \right]^{-1}
\ee
is applied to distribute the luminosity of the invisible and low-luminous sources in the corresponding luminosity interval $L_{\rm inv}$ and $L_{\rm low}$, respectively. 

Finally, the invisible and low-luminous RGs need to be placed at a certain position $(x_0,\,y_0,\,z_0)$ at the previously derived distance. 
Here, we use the magnetic-field strength $B$ from Dolag et al., since $B$ is to first order proportional to the gas-particle density \cite{2005JCAP...01..009D}. 
The unknown spatial distribution of RGs is also expected to comply with $\phi_3 (x_0,\,y_0,\,z_0)\propto B(x_0,\,y_0,\,z_0)$, so that the total RG distribution function is described by
\be
\Phi(d,\,L_{1.1},\,x_0,\,y_0,\,z_0) \propto \phi_1(d)\,\phi_2(L_{1.1})\,\phi_3 (x_0,\,y_0,\,z_0)\,,
\label{UnknownRGdistr}
\ee
where $d = \sqrt{x_0^2+y_0^2+z_0^2}$.
\begin{figure}[tbh]
  \centering
    \includegraphics[width=0.79\textwidth]{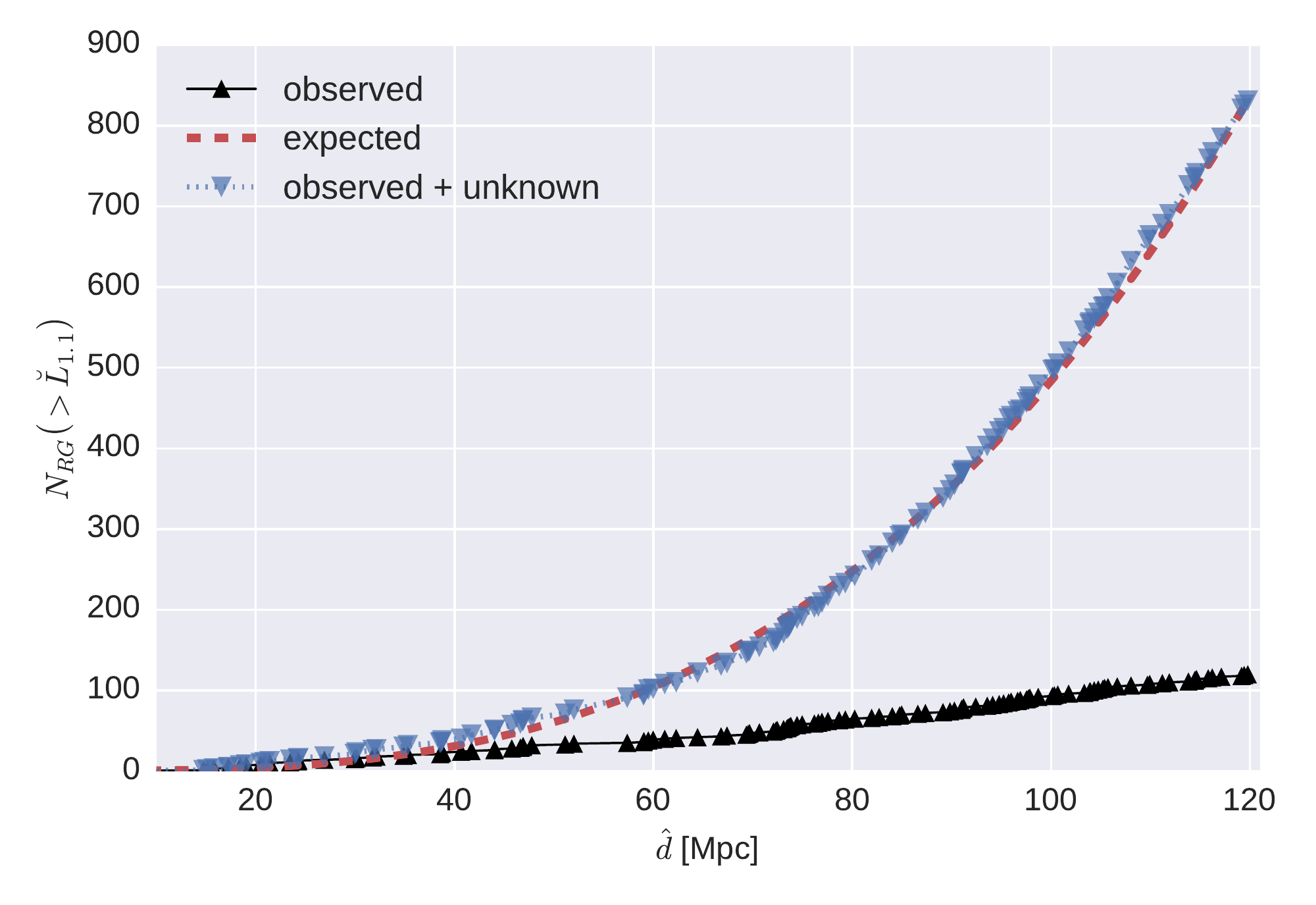}
\caption{The expected number of radio-loud AGNs (green) according to Eq.\ (\ref{numberOfAGNfromRLF}), the observed source number from the vV12 catalog (blue) and the extended source number (red) according to Eq.\ (\ref{UnknownRGdistr}). Here, a constant minimal luminosity of $\check{L}_{1.1} = 9.7\times 10^{37}\,\text{erg/s}$ is used.}
\label{CumRGnumber}
\end{figure} 
Fig.\ \ref{CumRGnumber} shows the resulting total number of RGs up to a distance $\hat{d}$ with a radio luminosity above $\check{L}_{1.1}$. 
Extending the source catalog of the vV12 catalog by the invisible and low luminous local RGs obviously leads to a very good agreement with the expected total number $N_{\rm RG}$ of radio-loud AGNs in the local Universe, i.e. $\hat{d}=120\,\text{Mpc}$ and $P_V=1$. 
\begin{figure}[tbh]
  \centering
    \includegraphics[width=0.79\textwidth]{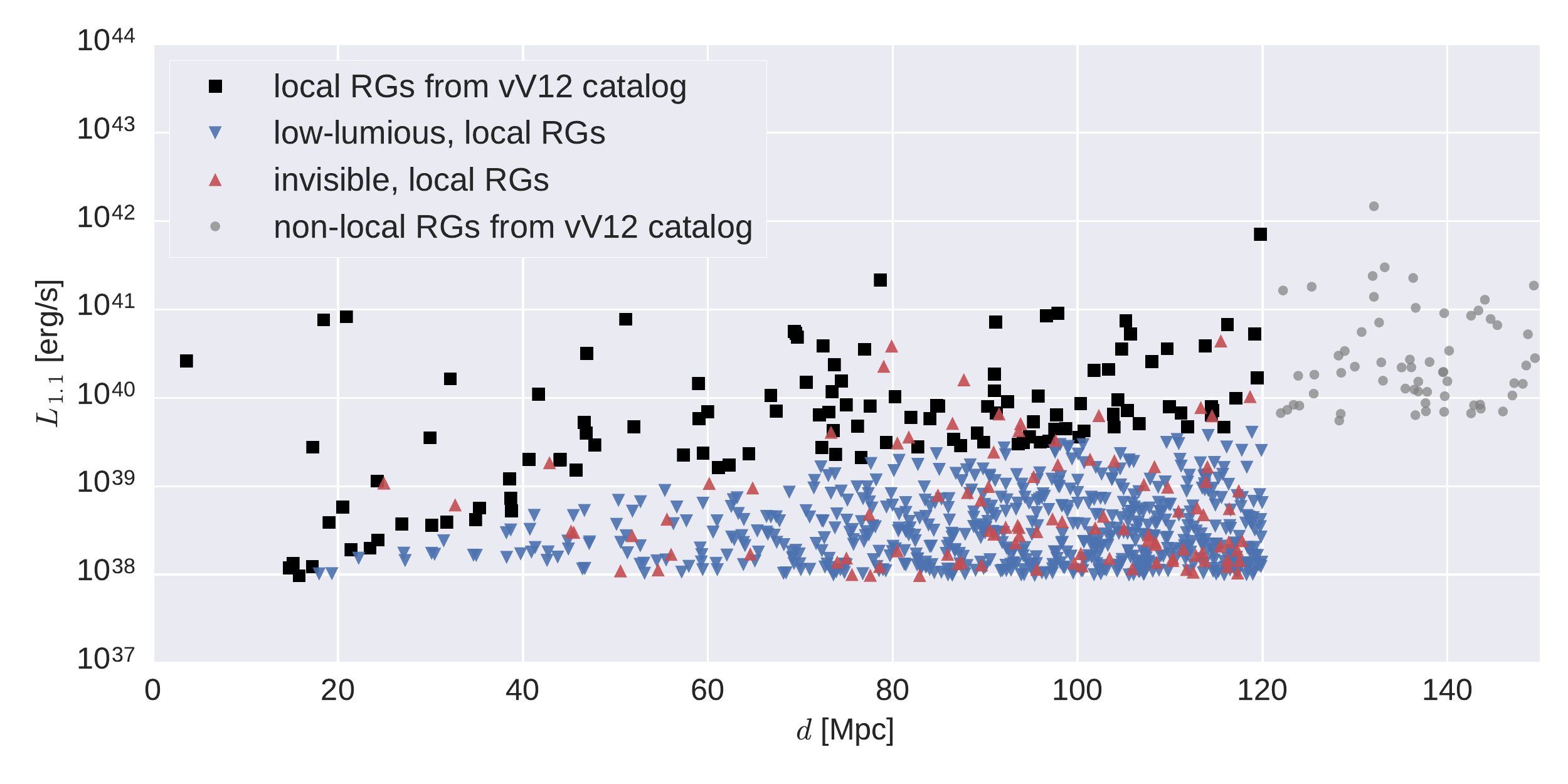}
\caption{Radio luminosity at 1.1 GHz of the local RGs dependent on their distance to the Earth.}
\label{RGdistr}
\end{figure} 

Eventually, the local RG distribution is composed of 121 RGs from the vV12 catalog, 100 invisible RGs and 617 low luminous RGs with individual positions and radio luminosities as displayed in Fig.\ \ref{RGdistr}.
Here, 'ESO 075-G 041' is the most distant RG in the sample with $d\simeq \hat{d}=120\,\text{Mpc}$ and also the most luminous one with $L_{1.1}=7.1\times 10^{41}\,\text{erg/s}$. 
The most luminous sources at a small distance are Centaurus A (with $L_{\rm CenA}=2.6\times 10^{40}\,\text{erg/s}$, $d_{\rm CenA}=3.6\,\text{Mpc}$), M87 (with $L_{\rm M87}=7.6\times 10^{40}\,\text{erg/s}$, $d_{\rm M87}=18.4\,\text{Mpc}$), and Fornax A (with $L_{\rm ForA}=8.3\times 10^{40}\,\text{erg/s}$, $d_{\rm ForA}=20.9\,\text{Mpc}$). 
Due to these three sources we obtain a local excess in luminosity up to a distance of about 30 Mpc. 
If the deflection of the UHECRs by the EGMF is rather small and the CR luminosity is roughly proportional to radio luminosity (which is both expected), we expect a dominant contribution to the observed UHECR flux by Centaurus A, M87 and Fornax A. 
As derived in Sect.\ \ref{Sec:RG_pheno}, the radio luminosity also provides an estimate of the corresponding maximal rigidity given by
\be
\hat{R}(L_{1.1}) = 0.139\,g_{\rm acc}\,\sqrt{g_{\rm cr}}\,\left({L_{1.1} \over 10^{38}\,\text{erg}\,\text{s}^{-1}}\right)^{3/7}\,\text{EV}\,.
\label{Rmax2}
\ee
The parameters $g_{\rm cr}$ and $g_{\rm acc}$ are used as free fit parameters with their value ranges $5\leq g_{\rm cr}\leq 50$ and $0.01\leq g_{\rm acc}\leq 0.5$, as discussed in Sect.\ \ref{Sec:Lradio2Ljet2Rmax}. 
Assuming maximal values for these parameters and a canonical spectral index $a=2$, Fig.\ \ref{ContribOfLocalRGs} shows that few of the local sources can contribute significantly to the UHECR flux observed at the highest energies, and only for a composition significantly heavier than solar. It further confirms our expectation that Centaurus A, M87, and/or Fornax A, are the prime candidates to explain the UHECR flux these energies, and we see that at least one of them must have enhanced metal abundances. 
Up to energies of about $10^{19.9}\,\text{eV}$, the contribution of all other local sources together is similar to that of M87 or Fornax A alone, while it is negligible for higher energies. At energies around the ankle, all local sources are able to contribute to the observed CR flux, even for solar composition, but still the flux is expected to be dominated by a few bright sources.  

\begin{figure}[tbh]
  \centering
    \includegraphics[width=0.49\textwidth]{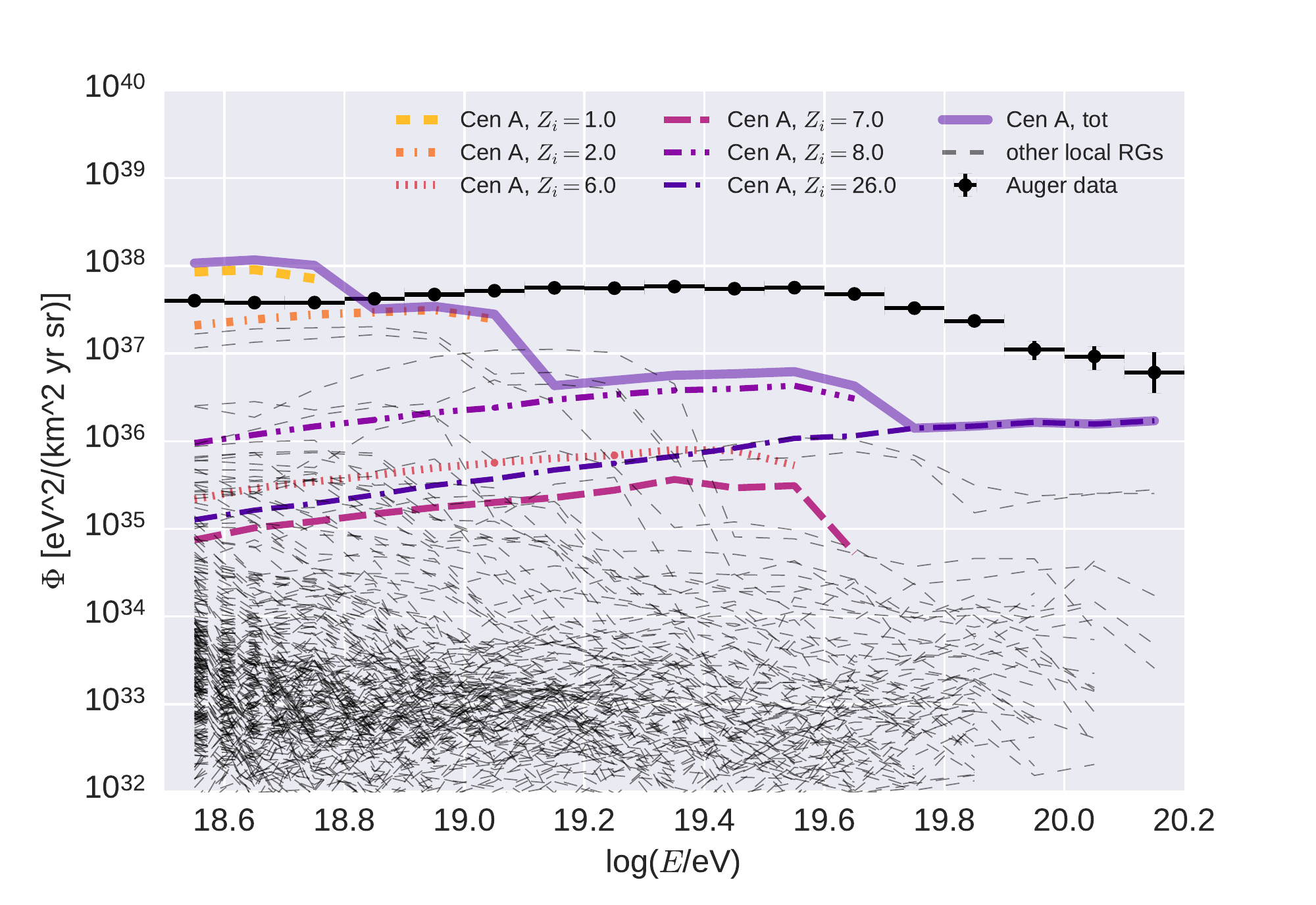}
    \includegraphics[width=0.49\textwidth]{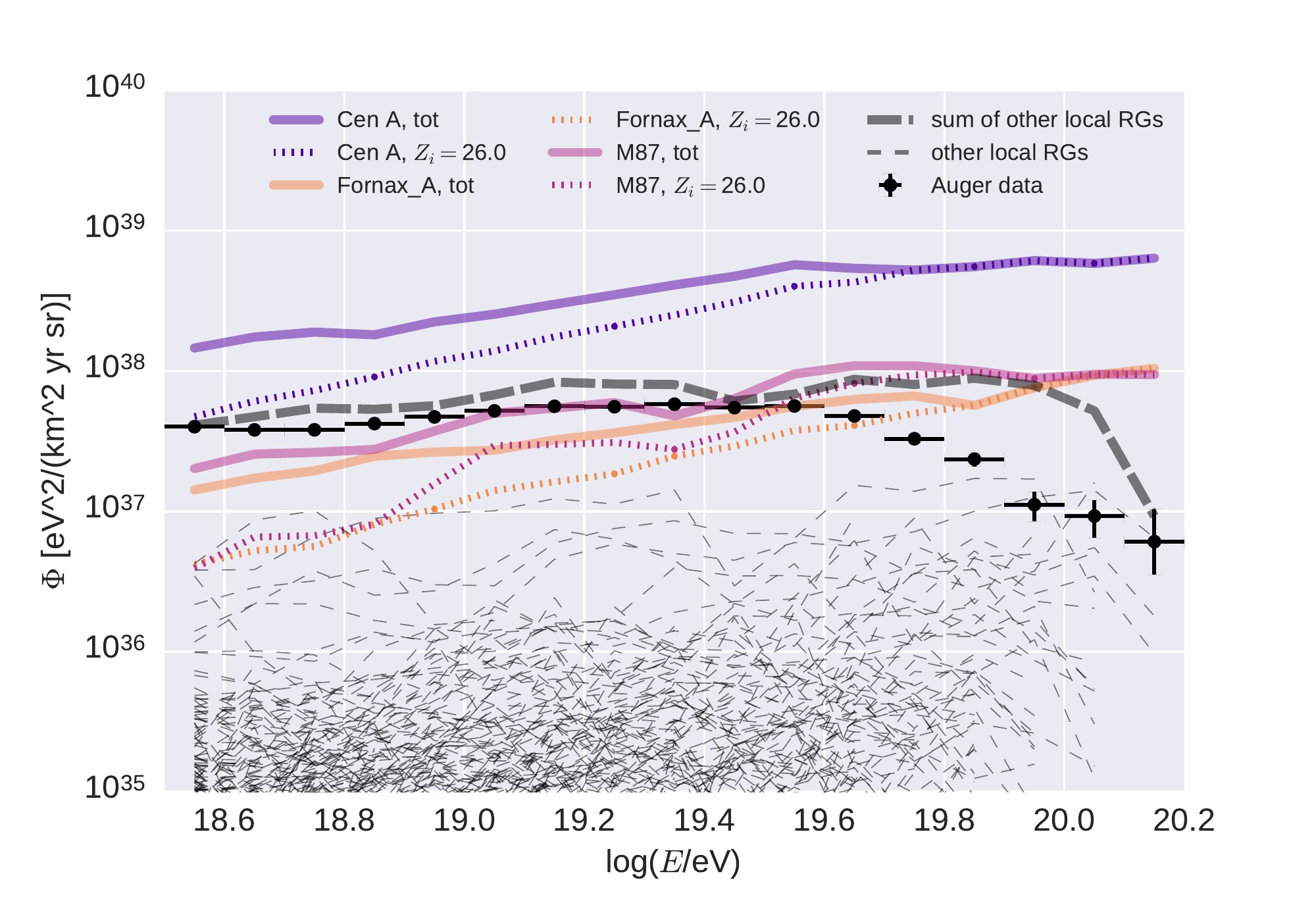}
\caption{Resulting flux of the local RGs using $a=2$, $g_{{\rm cr}}=50$, $g_{{\rm acc}}=0.5$ as well as $f_i=f_\odot$ (left plot) and $f_i=f_{\bullet}$, respectively.}
\label{ContribOfLocalRGs}
\end{figure} 

Further, the left panel of Fig.~\ref{ContribOfLocalRGs} shows that the dominant sources in the local regime cannot eject a solar like initial composition due to the resulting jumps in the energy spectrum. 
Note that most of these jumps are not mathematical artifacts, as there is no missing contribution between protons and oxygen. Individual adjustment of the initial abundances would be needed to fit the observed spectral behavior --- without significant jumps --- and the chemical composition.
However, we consider such an ad hoc fine-tuning as highly unlikely to be realized in nature, and thus implausible.

\subsubsection{The average non-local contribution}
\label{Sec:NonLocalContribution}

From simple geometrical considerations one can see that the flux provided by sources inside 120\,Mpc cannot give a complete picture of the UHECR flux: assuming a homogeneous source distribution and neglecting propagation effects, the contribution of every distance bin $\delta d$ to the flux would be equal. As UHECR around 10\,EeV can reach us in principle from 1 Gpc distance, the contribution from beyond 120\,Mpc should be an order of magnitude higher at this energy than from inside this boundary. This is somewhat reduced by propagation effects, but it is enhanced by positive source evolution which we need to assume for radio galaxies.  

In fact, the main reason for limiting our 3D simulations to a volume of 120\,Mpc radius is of rather technical nature:
\begin{enumerate}
\item The EGMF structure available for our local universe is limited to this distance. 
\item The vV12 catalog is found to be incomplete as a collection of relevant UHECR sources for distances larger than $120\,$Mpc, expressed best by the omission of the very powerful radio galaxy Pictor A close to this distance.
\item With the current setup, the necessary CPU time of 3D simulations with the same precision will become unmanageable in the case of a substantial increase of the source sample or the trajectory lengths. This argument becomes even more restrictive if we consider that for such propagation distances cosmological-evolution effects cannot be neglected, requiring 4D simulations which are even more CPU demanding \cite{1475-7516-2016-05-038, 2017arXiv171005617W}. 
\end{enumerate}
Therefore we follow a different approach to include distant sources: We perform 1D simulations based on the continuous source function for radio galaxies derived in Sect.\ \ref{Sec:AverageNonLocalSource}. This simulation is carried out from 120\,Mpc up to a distance corresponding to a redshift $z=2$. An advantage of this treatment is that it is fast, and cosmological evolution effects are easy to include. The disadvantage is that we neglect all potential anisotropy provided by distant, but extremely powerful sources. We return to this issue in the next section.  

\begin{figure}[tbh]
  \centering
    \includegraphics[width=0.49\textwidth]{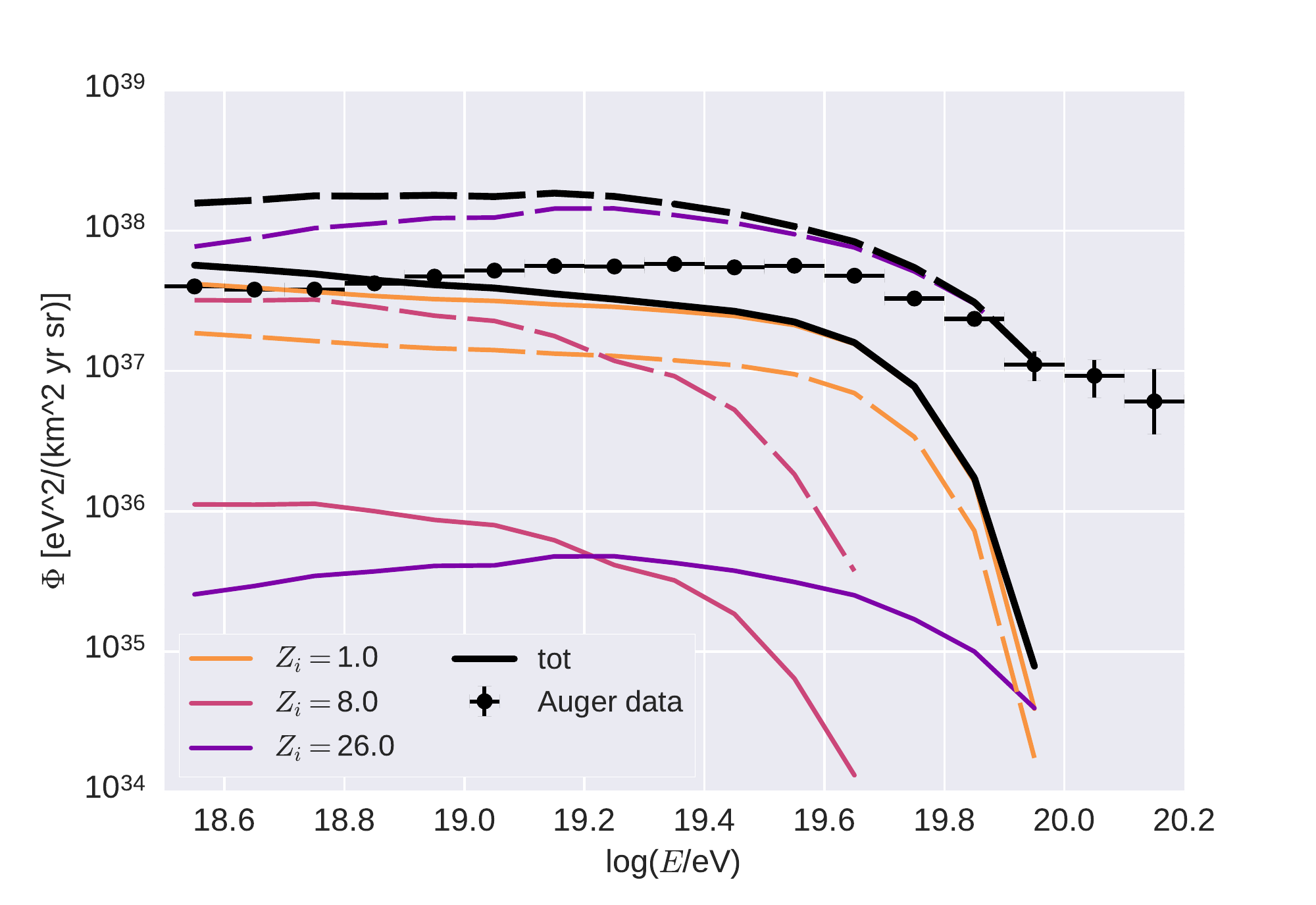}
    \includegraphics[width=0.49\textwidth]{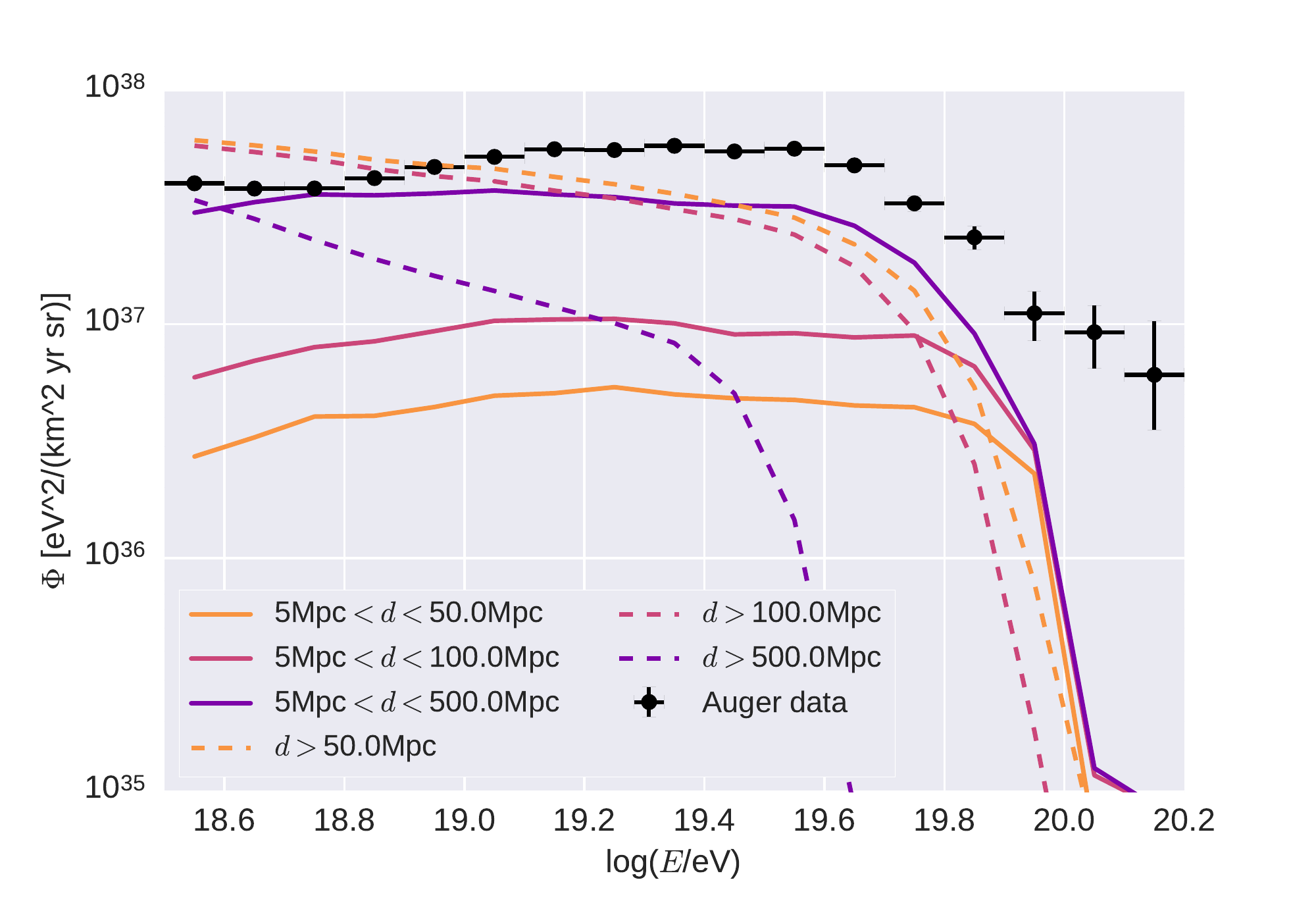}
\caption{UHECR flux from the non-local contribution at the observer using $a=2$, $g_{\rm cr}=50$, $g_{\rm acc}=0.5$. The left plot shows the influence of the different initial elements, where $f_i=f_{\odot}$ (solid line) and $f_i=f_{\bullet}$ (dashed line). The right plot exhibits the change of the spectrum dependent on the different distance ranges the continuous source function is adopted.}
\label{nonlocalContr}
\end{figure}  

Fig.\ \ref{nonlocalContr} demonstrates two important aspects of the non-local contribution. The left panel shows that, as for local sources, a heavy composition is needed to reach out to the highest energy data. However, as the corresponding mean logarithm of the mass number at around $10^{18.5}\,\text{eV}$ is already above 2.5 in this scenario, it is in disagreement with the observations of the chemical composition. It thus confirms our prior expectation outlined in Sect.\ \ref{Sec:AverageNonLocalSource}. The right panel shows the spectral steepening that results from adopting the continuous source function for an increasing distance range due to the increasing impact of interactions. 
Hence, the flux contribution from the average local sources is about a magnitude smaller and provides a significantly flatter spectral behavior compared to the one of the average non-local sources. 
We show in Appendix \ref{TestNormNonLocal} that indeed the continuous source function in the limited range of 5 to 120\,Mpc agrees with the sum of contributions from all local source described in the previous section.
\begin{figure}[tbh]
  \centering
    \includegraphics[width=0.49\textwidth]{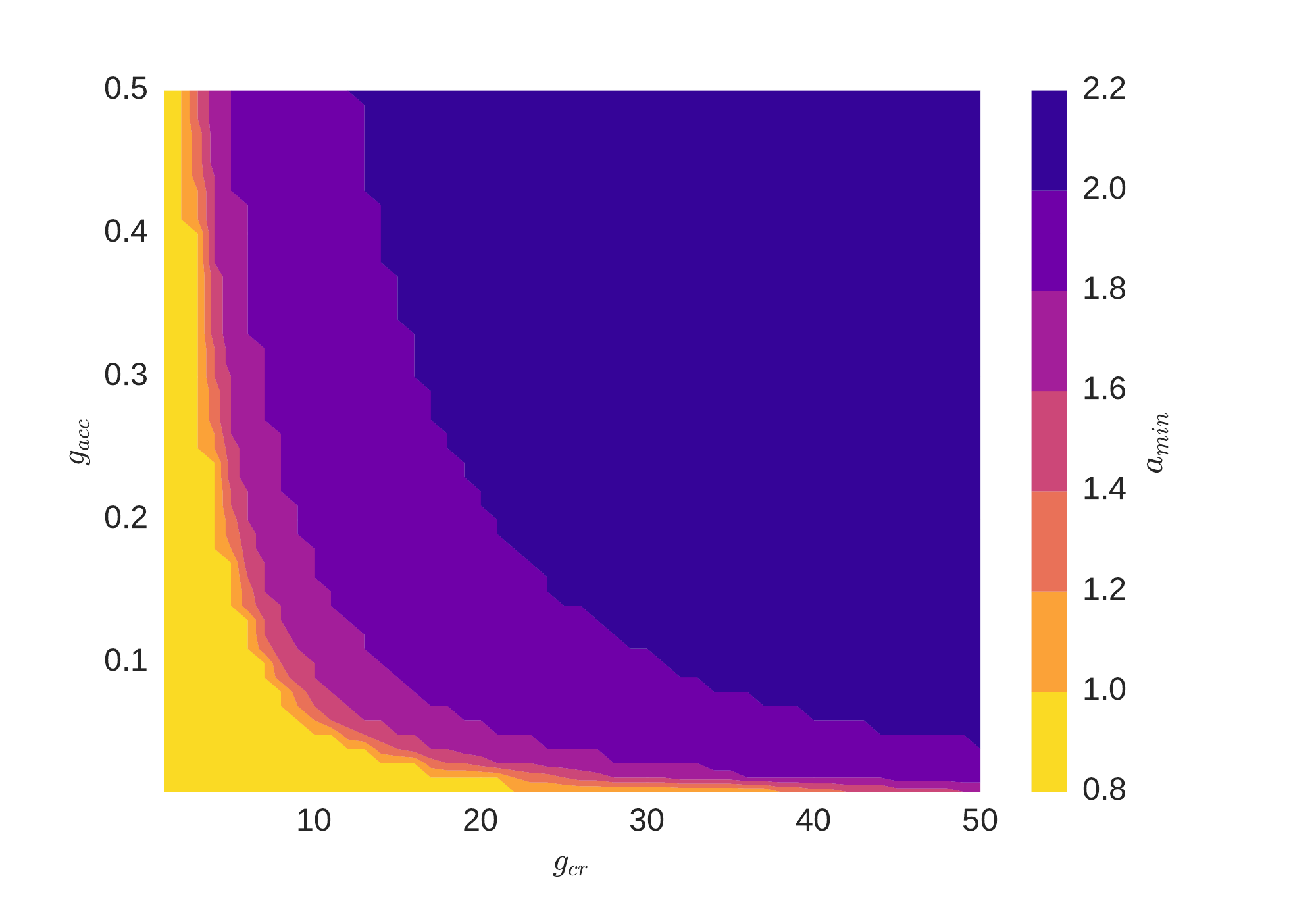}
    \includegraphics[width=0.49\textwidth]{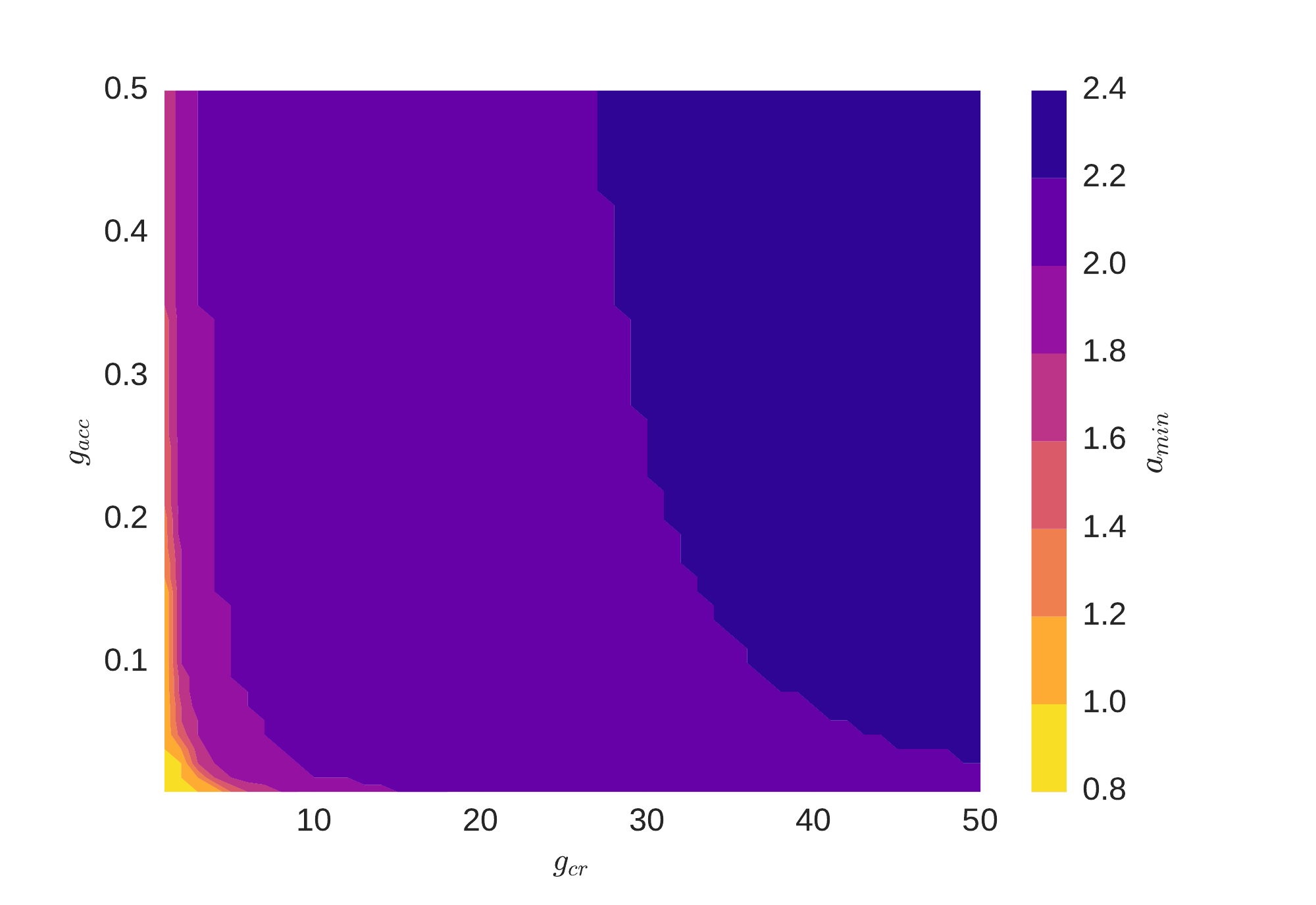}
\caption{Minimal spectral index of the average non-local contribution dependent on $g_{{\rm cr}}$ and $g_{{\rm acc}}$ to obtain a negligible contribution to the observed UHECR flux. Here, the limiting cases of a solar-like initial composition with $f_i=f_{\odot}$ (left plot) and a heavy initial composition with $f_i=f_{\bullet}$ (right plot) are shown.}
\label{MinSpecInd_nonLocal}
\end{figure}  

Hence, we can already draw the conclusion that the average non-local contribution needs to be subdominant above the ankle. As shown in Fig.\ \ref{MinSpecInd_nonLocal}, for the case of a heavy composition the parameter values of $g_{\rm cr}$ and $g_{\rm acc}$ need to be at the lower end of the possible range in order to make an initial spectral index $a<2$ possible. In the case of $f_i=f_{\odot}$, the range of possible values enlarges, but still the lower end of the parameter space is preferred. In the contrary, the previous Sect.\ has shown that at least a few, powerful local sources need a high value of $g_{\rm cr}$ to explain the highest energies observed. Thus we conclude, that we need to assume an exceptional parameter value for Centaurus A, M87, and/or Fornax A in order to explain the observational data.

\subsubsection{Cygnus A}
\label{Sec:CygnusA}

Cygnus A, with a CR luminosity $Q_{{\rm cr}}(L)=1.3\times 10^{45}\,g_{{\rm cr}}\,\text{erg/s}$ and a corresponding maximal rigidity $\hat{R}(L)=62.6\,g_{{\rm acc}}\,\sqrt{g_{{\rm cr}}}\,\text{EV}$ at a distance of $255\,\text{Mpc}$ \cite{2012A&A...544A..18V}, represents an ideal UHECR accelerator.\footnote{Note, that a distance of $(234\pm 16)\,\text{Mpc}$ is stated by the NASA/IPAC extragalactic database, however, we stick to the larger value given by the vV12 catalog in order to stay consistent in the use of its values.} It can accelerate protons to the highest energies observed, and nuclei to energies far beyond. Due to its large distance, its UHECR spectrum arriving at the observer is almost entirely shaped by the GZK process, unlike local FR-I galaxies, where the resulting spectrum is mostly determined by source properties. A slight residual effect of the source parameters is given through the strength of its GZK bump \cite{Hill:1983mk, 1988A&A...199....1B}, as shown in figure \ref{ContribOfCygA}.
\begin{figure}[tbh]
  \centering
    \includegraphics[width=0.7\textwidth]{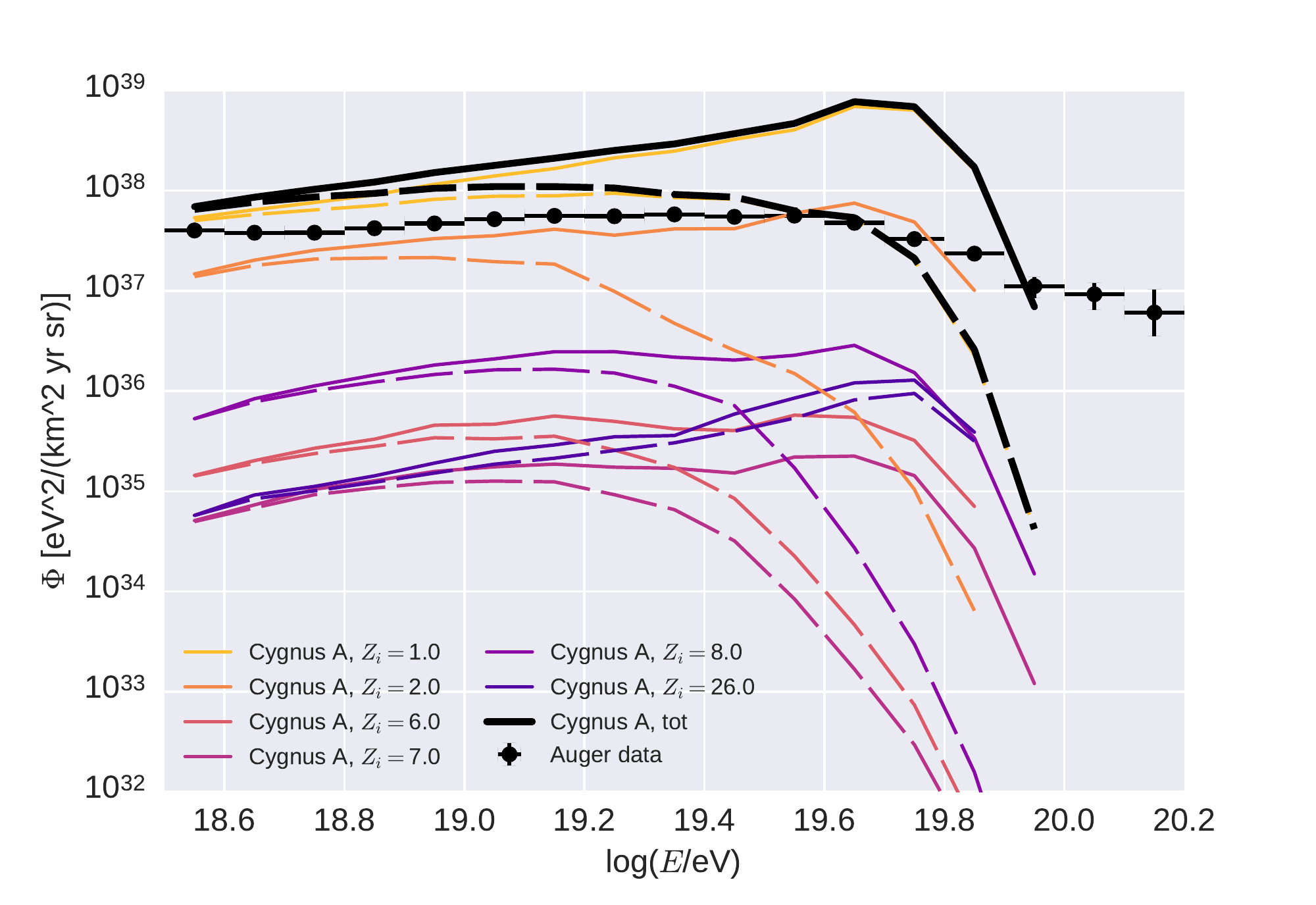}
\caption{Resulting flux from Cygnus A using $f_i=f_\odot$, $a=2$, $g_{{\rm cr}}=50$, $g_{{\rm acc}}=0.5$ (solid lines) and $f_i=f_\odot$, $a=2$, $g_{\rm cr}=50$, $g_{\rm acc}=0.05$ (dashed lines).}
\label{ContribOfCygA}
\end{figure} 
This shows that Cygnus A is in principle able to explain the observed UHECR flux up to a few tens of EeV, irrespective of its composition, where a better fit to the spectrum is obtained for a light composition scenario. It could therefore provide a source of light elements, balancing the required heavy composition of FR-I galaxies. In fact, the different morphology of FR-II galaxies like Cygnus A gives rise to assume not only a different composition, but also a different parameter set of $g_{{\rm cr}}^{\rm CygA}$ and $g_{{\rm acc}}^{\rm CygA}$. 

As we cannot perform a full 3D or even 4D simulation for Cygnus A due to the lack of a reliable EGMF structure out to its distance and other technical limitations (see Sect.~\ref{sec:SummConclu}), we use a simplified approach: We simulate Cygnus A as a point source in 3D neglecting deflections in the EGMF. In doing so, we are still able to normalize the flux --- as it will be introduced in Sect.\ \ref{AbsoluteNormalizationLocal} --- in the same manner as we do it for the local sources. Deflections by the EGMF are then estimated using a homogeneous random field with a given mean field strength and  coherence length (see Sect.\ \ref{sec:results} for further details).  As already shown for EGMF models assuming a high filling factor of ${>}\,1\,$nG fields throughout the universe \cite{AlvesBatista:2017vob}, this may allow even a full isotropization of UHECRs arriving from Cygnus A. Deflections inside the Galaxy can still be applied in the usual way.

By this treatment, we neglect the stochastic increase of UHECR propagation lengths due to random EGMF deflections, which will cause an increase of energy losses or photo-disintegration rates, and thus modify the observed spectrum and composition of UHECR from Cygnus A. To this adds the neglect of cosmological evolution effects in the 3D simulation. We discuss the potential impact on our results in Sect.~\ref{Case1}. 

We treat Cygnus A as the only point source outside 120\,Mpc distance, which is justified as is expected to provide more than half of all UHECRs observed. There are two other FR-II radio galaxies Pictor A and 3C353, both at about 125\,Mpc distance, which are expected to show very similar properties in spectrum and composition as Cygnus A and could make an individual contribution on a ${\sim}\,10\%$ level. We neglect them for simplicity. All other FR-II RGs can, with some restrictions, be considered to be properly treated by the continuous source function approach, and in fact FR-II RGs are included in the high luminosity part of MS07 RLF, and thus in the continuous source function.

\subsection{Simulation parameters}
\label{SimParameters}
As mentioned in Sect.\ \ref{Sec:SourceSample}, the free parameters of the model are optimized after the simulation by applying weights to the simulated candidates --- we describe this procedure in detail in Sect.\ \ref{AbsoluteNormalization}. Nevertheless, we need to consider the plausibility ranges of these parameters already during the simulation to keep the statistics of all simulated sources, energies and particle species in an optimal relation. 

All 3D simulations of individual sources are carried out up to the maximal rigidity of the source, $\hat{R} = g_{\rm acc} \sqrt{g_{\rm cr} Q_{\rm jet} / c}$ for the maximal parameters $g_{{\rm cr}}=50$ and $g_{{\rm acc}}=0.6$, where $Q_{\rm jet}$ is determined from the source luminosity $L_{1.1}$ as given in the vV12 catalog. We use a canonical spectral index $\tilde a =2$, thus the number of particles simulated per logarithmic energy bin follows an $E^{-1}$ power law. 
To obtain a sufficient number of candidates from each initial type of element, we simulate a similar number of H, He, CNO and Fe elements. 
Here, CNO is composed of 29\% ${}^{12}\text{C}$, 8\% ${}^{14}\text{N}$ and 63\% ${}^{16}\text{O}$, like the abundances of these elements in the solar system \cite{2009LanB...4B...44L}. 
In the case of 3D simulations that include CR particle interactions and deflections, the possible statistics is mostly determined by the available CPU time.\footnote{In our case, we used about $2\times 10^5$ CPU hours for the simulation of the total local source sample.} 
With respect to the available computational power, we aimed to reach sufficient statistics to obtained a statistical Poisson error per energy bin of less than $1$\% at low energies, i.e.\ up to $40\,\text{EeV}$. 
This way, the resulting statistical accuracy from the simulation becomes similar to the error of the observed UHECR flux. 
This translates to the requirement that about $10^4$ candidates per low energy bin need to arrive at the observer sphere.\footnote{To obtain these numbers in the high energy bins, we would have to increase the total number of simulated events by a factor of $10^2$ or more.} 
However, this statistical accuracy is neither needed nor realizable in less than $10^6$ CPU hours for each individual source. 
Due to the absolute normalization of the CR flux, as introduced in Sect.\ \ref{AbsoluteNormalizationLocal}, the low-luminous sources contribute to the result only in their total sum, so that in the end a comparable statistics to the powerful sources is obtained. 
Hence, it is a reasonable choice to simulate a total number $\tilde{N}_{i,j}$ of CR candidates per element $i$ and source $j$, that is proportional to its radio luminosity $L_{j}$, i.e. $\tilde{N}_{i,j}\propto L_{j}$. 

The simulation of CR candidates from Cygnus A is designed in the same manner as the simulation of the local sources, but disregarding the influence of the EGMF, which speeds up the simulations significantly. 
Here, we simulate the same total number for each element type $i$ and by using $\tilde{N}_{i,\rm CygA} = 6.1\cdot 10^{11}$ we obtain statistics that are comparable to ones from the powerful, local sources. 

For the continuous non-local source function (\ref{eq:SourceFunctionEvolution}), 1D simulations are carried out using a maximal rigidity of  $\hat{R}_{\rm nl}=100\,\text{EV}$ and a flat initial spectrum of $\tilde{a}_{nl}=1$. 

Although 1D simulations are so fast that CPU time is not a limiting factor, the computational time of subsequent analysis (the fitting procedure) strongly depends on size of the output data.
In order to obtain an appropriate analyzable output with sufficient statistics, we simulate about $10^6$ candidates in total with some more light than heavy elements, due to the dominant contribution of these elements according to the solar abundances. 
The final numbers of simulated particles together with the initial simulation parameters that characterize the energy distribution of the UHECR outflow at the sources are summarized in table \ref{InitSimPar}. Further details on the simulation program and the used parameters are given in Appendix \ref{detailsOfSim}.

\begin{table}[h!]
\centering
\caption{Initial simulation parameters.}
  \begin{tabular}{l c c c c c c c c c}
  \toprule
      & $\tilde{N}_H$ & $\tilde{N}_{He}$ & $\tilde{N}_C$ & $\tilde{N}_N$ & $\tilde{N}_O$ & $\tilde{N}_{Fe}$ & $\tilde{a}$ & $g_{\rm cr}$ & $g_{\rm acc}$  \\ 
   \midrule
     \textbf{local} & $1.8\cdot10^{11}$ & $1.3\cdot10^{11}$ & $3.7\cdot10^{10}$ & $1.1\cdot10^{10}$ & $8.0\cdot 10^{10}$ & $1.2\cdot10^{11}$ & $2$ & $50$ & $0.6$   \\ 
     \textbf{CygA} & $6.1\cdot10^{11}$ & $6.1\cdot10^{11}$ & $6.1\cdot10^{11}$ & $6.1\cdot10^{11}$ & $6.1\cdot10^{11}$ & $6.1\cdot10^{11}$ & $2$ & $50$ & $0.6$ \\
   \specialrule{0.7pt}{3pt}{6pt} 
  \end{tabular}
  \begin{tabular}{l c c c c c c c c}
     & $\tilde{N}_H$ & $\tilde{N}_{He}$ & $\tilde{N}_C$ & $\tilde{N}_N$ & $\tilde{N}_O$ & $\tilde{N}_{Fe}$ & $\tilde{a}_{nl}$ & $\hat{R}_{\rm nl}$ [EV] \\
   \midrule
     \textbf{non-local} & $5\cdot 10^5$ & $2.5\cdot 10^5$ & $8.3\cdot10^4$ & $7.1\cdot 10^4$ & $6.3\cdot 10^4$ & $2\cdot 10^4$ & $1$ & $100$ \\ 
   \bottomrule 
\end{tabular}
  \label{InitSimPar}
\end{table}

\subsection{Absolute normalization of the CR flux}
\label{AbsoluteNormalization}
As our model is absolutely normalized due to the connection between CR power and maximal rigidity of each source, we have no freedom to fit a spectrum obtained in ``arbitrary units'' to the measured CR spectrum.  
Instead, we need to transfer the absolute normalization to the simulation setup. 
In general, the task is to achieve absolute weights $W_k=dN_k/dS\,dt$ to each CR candidate $k$ which arrived at the observer with energy $E^*_k$ and charge $Z^*_k$ in the physical dimension of particles per square-cm per second. 
Here we need the information on each candidates initial rigidity $R_k=E_k/(eZ_k)$ at the source, which is provided in the CRPropa output.
Due to the different treatments of the local and non-local flux contribution, their absolute weights differ as shown in the following.

\subsubsection{Cosmic ray flux from local sources and Cygnus A}
\label{AbsoluteNormalizationLocal}

As rigidity is the natural measure for this problem, we use the simulated spectrum $d\tilde{N}/dR\propto R^{-\tilde{a}}$ dependent on the initial rigidity $R$ of the candidates and re-weight it to obtain
\be
{\diff N_i \over \diff R}=\nu_0\,f_i\,R^{-a}\,\exp(-R/\hat{R})
\label{ReWeightedSpectrum}
\ee
with an arbitrary normalization constant $\nu_0$. 
In order to derive a characteristic emission time scale, we use the CR power of the local source $j$, of the individual charge type $i$ given by 
\be
Q_{{\rm cr},i,j}=Q_{{\rm cr},j}\,{f_i\,Z_i \over \bar{Z}}\,,
\ee
as well as the CR energy of 
\be
E_{{\rm cr},i}= \nu_0\,f_i\,Z_i\,e\,\int_{\check{R}_0}^{\hat{R}_{\rm cut}}\text{d}R\,\,R^{1-a}\,\exp(-R/\hat{R})
\ee
that is provided by the re-weighted CR spectrum (\ref{ReWeightedSpectrum}). 
Here, $\check{R}_0=10\,\text{GV}$ denotes the minimal rigidity of the CRs according to the peak of the Galactic CR flux.
Hence, the CR emissivity of the local source $j$ with a given total CR power $Q_{\rm cr}$ yields
\be
{\diff N_{i,j} \over \diff R\,\diff t}={\diff N_{i,j} \over \diff R}\,{Q_{{\rm cr},i,j} \over E_{{\rm cr}}} = {f_i \over \bar{Z}}\,{Q_{{\rm cr},j}\,R^{-a}\,\exp(-R/\hat{R}) \over e\,\int_{\check{R}_0}^{\hat{R}_{\rm cut}}\text{d}R\,\,R^{1-a}\,\exp(-R/\hat{R})}
\ee
The surface of the observer\footnote{Note that negligible energy losses are supposed between the observer sphere in the simulation and the ``real observer'', i.e. the Earth.} is given by $S=4\pi\,r_{\rm obs}^2$. 
The mean rigidity transfer from the source to the observer sphere is determined by 
\be
\bar{R}_k={\int_{\check{R}_1}^{\hat{R}} \diff R\,\diff \tilde{N}/\diff R \over \diff \tilde{N}/\diff R_k}\,,
\ee
where $R_k$ denotes the \emph{initial} rigidity of an individual CR candidate that arrived at the observer, and $\check{R}_1=3EV/Z_i$ is the minimal initial rigidity of the simulated candidates.
Knowing the total number $\tilde{N}_{i,j}$ of simulated CRs of a given charge type $i$, we are able to determine the flux weight of an individual CR candidate that managed to arrive at the observer 
\be
W_{k,i,j} = {\bar{R}_k \over S\,\tilde{N}_{i,j}}\,{\diff N_{i,j} \over \diff R_k\,\diff t} = {f_i\,Q_{{\rm cr},j}\,(\hat{R}^{1-\tilde{a}}-\check{R}_1^{1-\tilde{a}}) \over 4\pi\,r_{{\rm obs}}^2\,\tilde{N}_{i,j}\,\bar{Z}\,e\,(1-\tilde{a})\, \left(G_a(\check{R}_0)-G_a(\hat{R}_{\rm cut})\right)}\,R_k^{\tilde{a}-a}\,\exp(-R_k/\hat{R})\,,
\label{localCRWeight}
\ee
with 
\be
G_a(x)= \hat{R}^{2-a}\,\Gamma\left(2-a,\, x/\hat{R} \right)\,
\ee
where $\Gamma(c,x)$ denotes the incomplete gamma function. 
Thus, the total, observed UHECR flux from all local sources yields
\be
W_{\kappa} = \sum_{k,i,j}\,\delta_{k\kappa}\,W_{k,i,j}.
\ee
Here, the Kronecker-delta accounts for the final binning of the observed flux in energy bins $E_\kappa$ (denoting the central energy of the bin) of equal logarithmic width $\Delta = \diff \log(E_\kappa)$, so that the linear width of the bin $E_\kappa$ is $\diff E_\kappa=\Delta E_\kappa$. 
Hence, $\delta_{k\kappa}=1$ for $E^*_k\in \diff E_\kappa$ and $\delta_{k\kappa}=0$ otherwise.
The absolute particle number flux density in the energy bin $E_\kappa$ (meaning in CGS units, particles per square-cm per erg per second) is given by 
\be
j_\kappa \equiv W_{\kappa}/\diff E_\kappa = \sum_{k,i,j}\, \delta_{k\kappa}\,{f_i\,Q_{{\rm cr},j}\,(\hat{R}^{1-\tilde{a}}-\check{R}_1^{1-\tilde{a}})\,R_k^{\tilde{a}-a}\,\exp(-R_k/\hat{R}) \over 4\pi\,r_{{\rm obs}}^2\,\tilde{N}_{i,j}\,\bar{Z}\,e\,(1-\tilde{a})\, \left(G_a(\check{R}_0)-G_a(\hat{R}_{\rm cut})\right)}\,.
\label{localParticleFlux}
\ee
It is shown in the Appendix \ref{TestNormLocal} that the previously derived absolute normalization is in good agreement with the theoretical expectations.

\subsubsection{Cosmic-ray flux from the non-local source function}
\label{AbsoluteNormalizationNonLocal}
For the case of 1D simulations of non-local sources, all CR candidates with a final energy $E^*_k\geq 3\,\text{EeV}$ are stored by the observer. 
Thus, the absolute normalization of the CR flux cannot be derived as before by using the rigidity transfer to a geometrical observer sphere with a certain surface. 
Rather, the production rate density Eq.\ (\ref{ProdRateDens1}) enables an absolute normalization, where we have to consider source evolution effects according to Eq. (\ref{eq:SourceFunctionEvolution}), fixing the evolution parameter to $m=3$ as it has only a minor influence on our results. 
We define an effective comoving volume element
\be
\Delta V_k \simeq 4\pi\,d_k^2\,\Delta d_i
\ee
with the propagation distance $d_k$ of the observed CR candidate and the thickness $\Delta d_i=\hat{d}/\tilde{N}_i$ of the spherical shell that is represented by each simulated candidate of type $i$. 
Here, the total number of simulated candidates of type $i$ is given by $\tilde{N}_i$, and $\hat{d}$ denotes to the maximal comoving distance that has been adopted in the 1D simulation.
Thus, the CR production rate yields
\be
\dot{P}_{{\rm cr},i,k} \equiv {\diff N_k \over \diff R_k \, \diff t} = \int \diff V\,\Psi_{i}(R_k,\,z) \simeq \Psi_{i,0}(R_k)\,(1+z)^{m-1}\,4\pi\,d_k^2\,\Delta d_i\,.
\label{CRProdRate}
\ee
Note that the result of the integral only holds if a homogeneous distribution of sources is used in the 1D simulation. 
Since the non-local candidates are simulated with a flat energy distribution, we obtain a mean rigidity transfer of 
\be
\bar{R}_k={\int_{\check{R}_1}^{\hat{R}_{\rm nl}} \diff R\,\diff\tilde{N}/\diff R \over \diff \tilde{N}/\diff R_k}=R_k\,\ln\left(\hat{R}_{\rm nl}/\check{R}_1 \right)\,,
\ee
so that the total CR production rate yields
\be
{\diff N_k \over \diff t} = \bar{R}_k\,\dot{P}_{{\rm cr},i,k}=R_k\,\ln\left(\hat{R}_{\rm nl}/\check{R}_1 \right)\,\Psi_{i,0}(R_k)\,(1+z)^{m-1}\,4\pi\,d_k^2\,\Delta d_i\,.
\ee
In three spatial dimensions, an isotropically emitting source with a total rate $\diff N/\diff t$ provides a particle flux of ${\diff N \over \diff t\,\diff A} = {\diff N \over \diff t}\,{1\over 4\pi\,d^2}$ at a distance $d$. 
The CR production rate (\ref{CRProdRate}) corresponds to $\diff N/\diff t$ for an isotropically emitting source at distance $d_k$ in three spatial dimensions, which can easily be shown by the transformation of $\dot{P}_{{\rm cr},i,k}$ into three spatial dimensions.\footnote{First $\dot{P}_{{\rm cr},i,k}$ needs to be integrated over the corresponding surface area, in order to obtain the total CR rate that is emitted in direction of the observer. Since, we expect the sources to emit CRs rather isotropically, we also need to divide by $4\pi\,d_k^2$, so that the surface area from the previous integration is canceled out.}
However, we still need to take into account that the cross section of an extended, spherical observer is reduced by a factor of $1/4$ by switching from a 1D to a 3D treatment, i.e. $(\diff N_k/ \diff t)_{3D}=(\diff N_k/ \diff t)/4$.
Hence, the flux contribution by a CR candidate with a propagation distance $d_k$ is given by
\be
W_{k,i} \equiv {\diff N_k \over \diff t\,\diff A_k} = \left({\diff N_k \over \diff t}\right)_{3D}\,{1\over 4\pi\,d_k^2} = {R_k\,\ln\left(\hat{R}_{\rm nl}/\check{R}_1 \right)\,\hat{d} \over 4\, \tilde{N}_i}\,\Psi_{i,0}(R_k)\,(1+z)^{m-1}
\ee
Like in the previous section, the total, observed UHECR flux from all non-local sources yields
\be
W_{\kappa} = \sum_{k,i}\, \delta_{k\kappa}\,W_{k,i}\,
\ee
where, the Kronecker-delta once again accounts for the final binning of the observed flux into energy bins $E_\kappa$ with a linear width of $\diff E_\kappa$. 
So, the absolute particle number flux density in the energy bin $E_\kappa$ (meaning in CGS units, particles per square-cm per erg per second) yields
\be
j_\kappa \equiv W_{\kappa}/dE_\kappa = \sum_{k,i}\, \delta_{k\kappa}\,{R_k\,\ln\left(\hat{R}_{\rm nl}/\check{R}_1 \right)\,\hat{d} \over 4\, \tilde{N}_i}\,\Psi_{i,0}(R_k)\,(1+z)^{m-1}\,.
\label{nonlocalParticleFlux}
\ee
A sanity check of the previously derived absolute normalization is shown in the Appendix \ref{TestNormNonLocal}. 

Of course, the total particle number flux density from local and nonlocal sources is given by the sum of Eq.\ (\ref{localParticleFlux}) and (\ref{nonlocalParticleFlux}). 
In addition, the normalization weights $W_{k,i,j}$ and $W_{k,i}$ from the local and nonlocal contribution, respectively, enable to determine the mean logarithm of the mass number $A_\eta$ within the energy bin $E_\eta$ according to
\be
\langle \ln(A_\eta) \rangle = {\sum_{k,i,j}\, \delta_{k\eta} \left(\ln(A^*_{k,j})\,W_{k,i,j}+\ln(A^*_k)\,W_{k,i} \right) \over \sum_{k,i,j}\, \delta_{k\eta} \left( W_{k,i,j}+W_{k,i} \right)} \,.
\ee

\subsection{Fit parameters of the model}
\label{Sec:FitPar}
As introduced in Sect.\ \ref{Sec:RG_physics}, the main free parameters of our model and their plausibility ranges are the spectral index $a\in [1.7, 2.2]$, the cosmic-ray load $g_{\rm cr} \in [1, 50]$, and the acceleration efficiency $g_{\rm acc} \in [0.01, 0.5]$. The abundances of nuclear species are connected, according to Sect.\ \ref{Sec:Lradio2Ljet2Rmax}, by the dependence $f_i = Z_i^q$, with an abundance parameter $q \in [0,2]$. Finally, the source-evolution parameter was fixed to $m=3$ as its influence is mainly important below the ankle \cite{1988A&A...199....1B}. 

In the derivation of the continuous source function, we assumed that these parameters are the same for all radio galaxies to allow an analytical solution. This, however, is a simplification and by no means implied. The scatter in the correlation plots showed by Willott et al.\ suggests that $g_{\rm cr}$ can vary from source to source within an order of magnitude centered at the best fit value for the sample. There is no reason to assume that a similar scatter does not apply to $g_{\rm acc}$ and $q$, as these parameters depend on individual source properties as jet speeds and the material in which the jet dissipates. Only for the spectral index $a$, we have good reasons to consider it as a parameter which is, in narrow ranges, determined by fundamental physics and thus universal. 

Hence, it would be in principle required for our treatment of individual sources to give each source its own parameter set. However, this would increase the number of free parameters far beyond the number of data constraints, making the problem ambiguous and ill-defined. Luckily, from the discussion in the previous section, we have good reasons to assume that only four sources, Centaurus A, M87, Fornax A, and Cygnus A, make a significant \textit{individual} contribution to the cosmic ray flux. All other individual sources can be given the same parameter values, and continuity implies that they should be identical to those chosen for the continuous source function. The contribution by latter needs to be subdominant, as already exposed in Sect.\ \ref{Sec:NonLocalContribution}, hence, there is no benefit in using different parameter values of $g_{\rm cr}$ \emph{and} $g_{\rm acc}$ in the case of these sources. Here, we only use an individual value of $g_{\rm cr}$, subsequently called $\bar{g}_{\rm cr}$, but stick to the same parameter value of $g_{\rm acc}$ that is applied for Centaurus A, M87, and Fornax A, respectively. 
Regarding the abundance parameter $q$, we assume from the discussion in Sect.\ \ref{Sec:Lradio2Ljet2Rmax} that $q=0$ for Cygnus A (meaning solar abundances), while for the FR-I galaxies Centaurus A, M87, and Fornax A, an enrichment of heavy elements is allowed by fitting $q$ within its plausibility limits. A general differentiation in the composition of FR-I and FR-II galaxies has no significant effect on the fit results, since only a few galaxies dominate the UHECR contribution of their source class --- as discussed in Sect.\ \ref{Sec:LocalSources} and \ref{Sec:CygnusA}. Thus, we are unable to expose whether all FR-I galaxies or just a few luminous, close-by ones emit UHECRs according to the heavy composition scenario. To keep the parameter set manageable, we will subsequently consider the two limiting cases where all of the other local and non-local radio galaxies emit either a heavy or a light composition. 
Thus, our model has a total number of seven free parameters, $a$, $\bar{g}_{\rm cr}$, $g_{\rm cr}^{\mu}$, $g_{\rm cr}^{\rm CygA}$, $g_{\rm acc}$, $g_{\rm acc}^{\rm CygA}$ and $q$, where the index $\mu$ refers to either Centaurus A, M87, Fornax A, or M87 and Fornax A combined.

\section{Fit Results}
\label{sec:results}
To obtain a best fit parameter set, we use the \texttt{minimize()} function provided by the python package \texttt{lmfit} \cite{newville_2014_11813}. This function minimizes the residual $\mathcal{R}$ of the weighted and binned results $j_\kappa$ and $\langle \ln(A_\eta) \rangle$ from the simulation and the 24 observational data points above $5\times 10^{18}\,\text{eV}$ from Auger \cite{ObservatoryMichaelUngerforthePierreAuger:2017fhr, PhysRevD.90.122005}. We perform the minimization algorithm 10 times using Powell's method \cite{Powell01011964} with a random set of initial weights within the allowed range, to increase the probability that the obtained minimum is indeed a global minimum. 
This is necessary as in particular for badly fitting models multiple local minima of $\mathcal{R}$ may occur in the parameter space. 
For well fitting models, however, the algorithm converges in most cases to the same minimum. 
As usual the goodness of the fit is quantified by the minimal reduced chi-square $\chi^2=\sum \mathcal{R}^2/n_{\rm dof}$, where $n_{\rm dof}$ refers to the corresponding degree of freedom of the fits.
The normalization of the observed $\langle \ln(A) \rangle$ data depends strongly on the used hadronic interaction model. 
There are three contemporary models that were either tuned to recent LHC data (QGSJetII-04 \cite{PhysRevD.83.014018}, Epos-LHC \cite{PhysRevLett.101.171101, PhysRevC.92.034906}) or found in good agreement with these measurements (Sibyll2.1 \cite{PhysRevD.80.094003}). 
The Auger collaboration processed the measured $X_{\rm max}$ data with these interaction models, exposing no difference in the spectral behavior, but a systematic shift in the normalization, where the QGSJetII-04 (Epos-LHC) model yields the lightest (heaviest) composition \cite{PhysRevD.90.122005}.
If there are no basic changes in the fit result, we display the results found for the Sibyll2.1 model unless noted otherwise.

In the case of a good fit to spectrum and composition, we further analyze the corresponding arrival directions of the best-fit models, but we do not include them into the fit. 
To estimate EGMF deflections for Cygnus A, we derive the rms deflection angle for a distance $d=255\,\text{Mpc}$ analytically \cite{2004PhRvD..70d3007S} by
\be
\theta_{k}\simeq 0.8\degree \,\bar{Z}_k\,\left( {\bar{E}_k \over 100\,\text{EeV}} \right)^{-1}\,\left( {d \over 10\,\text{Mpc}} \right)^{1/2}\,\left( {\lambda_{\rm c} \over 1\,\text{Mpc}} \right)^{1/2} \,\left( {B_{\rm rms} \over 1\,\text{nG}} \right)\quad,
\label{rms-deflection}
\ee
where we use an approximated mean candidate energy $\bar{E}_k=(E_k+E_k^*)/2$, an approximated mean candidate charge number $\bar{Z}_k=(Z_k+Z_k^*)/2$, and use as the rms strength $B_{\rm rms}=1.18\,\text{nG}$ of the D05 field. 
Further, a correlation length $0.1\,\text{Mpc}\leq\lambda_{\rm c}\leq 10\,\text{Mpc}$ is assumed, as for $\lambda_{\rm c}\ll 0.1\,\text{Mpc}$ we obtain negligible deflections, whereas in the case of $\lambda_{\rm c}\gg 10\,\text{Mpc}$ the UHECRs with $E_i\lesssim 10\,\text{EeV}$ from Cygnus A are completely isotropized. 
Then, the corresponding arrival directions $\vec a_k$ of the individual candidates from Cygnus A are drawn from the Fisher distribution \cite{Fisher295} that is also implemented in CRPropa3, using the source direction and the concentration parameter $\kappa = (81\degree / \theta_{k})^{2}$ for a distribution where 63\% of the candidates are deflected less than $\theta_k$ --- analogous to the standard deviation of the normal distribution. 

We limit the analysis of the arrival directions to the comparison of multipole moments, which is sufficient as mostly only upper limits are experimentally determined. For the detected dipole above $8\,$EeV, Auger provides also information on the direction \cite{1475-7516-2017-06-026,Aab:2017tyv}, but given the fact that its total amplitude is only about $7\%$, modelling the direction would require accuracies of the predicted arrival direction distribution significantly below this amplitude. Due to our approximate treatment of EGMF scattering from Cygnus A, and the neglect of the FR-II RGs Pictor A and 3C353, we do not expect to reach this accuracy in our present model. 
More sophisticated investigations are intended after the development of an extended EGMF structure as well as an efficient targeting method, as outlined in Sect.\ \ref{sec:SummConclu}. 

In addition, the resulting best-fit parameters also enable to draw further conclusions on the origin and the deflection of the observed UHECR candidates from local sources. 
Here, the corresponding best-fit weights $W_{k,i,j}^{\rm best}$ according to the absolute normalization of the local flux (\ref{localCRWeight}), have to be taken into account. 
Thus, the contribution of the local source $\mu$ is given by 
\be
N_\mu={\sum_{k,i,j}\,\delta_{\mu j} W_{k,i,j}^{\rm best} \over  \sum_{k,i,j}\, W_{k,i,j}^{\rm best}}
\ee
where the Kronecker delta $\delta_{\mu j}$ is used to ensure that the observed CR candidate originates from this source. 
Further, the mean deflection angle of the UHECR candidates from the source $\mu$ is determined by
\be 
\bar{\theta}_{{\rm def},\mu}={\sum_{k,i,j}\,\delta_{\mu j} W_{k,i,j}^{\rm best}\,\theta_{{\rm def},i,k} \over  \sum_{k,i,j}\, W_{k,i,j}^{\rm best}}\,.
\label{meanDefl}
\ee 
Here, the deflection angle of an individual candidate $i$ of type $k$ is estimated by
\be 
\theta_{{\rm def},i,k}=\arccos\left({\vec{p}_{i,k}\cdot \vec{p}^{\,*}_{i,k}\over |\vec{p}_{i,k}|\,|\vec{p}_{i,k}^{\, *}|} \right)
\label{minDefl}
\ee
where we only use its initial and final momentum $\vec{p}_{i,k}^{\,*}$ and $\vec{p}_{i,k}$, respectively. 
Note, that in general $\theta_{{\rm def},i,k}$ only determines the minimal deflection angle, since a total deflection $>\theta_{{\rm def},i,k}$ cannot be reconstructed\footnote{In principle, a detailed tracking of the candidates is possible with CRPropa. However, further developments on the propagation module are needed in order to handle the vast amount of simulated data.}. 
However, in the case that the minimal deflection angles of the candidates are predominantly small, most likely small-angle deflections take place and $\theta_{{\rm def},i,k}$ is also a good approximation for the total deflection angle.

\subsection{Case 1: All local sources equal}
\label{Case1}

First, we discuss the case that none of the powerful individual local sources offer deviating parameters, i.e. $g_{\rm cr}^\mu = \bar g_{\rm cr}$, and only Cygnus A provides solar abundances while for all other sources heavy elements get enhanced by fitting $q$. The resulting best-fit parameters, using the different hadronic interaction models, are summarized in table \ref{FitPar0}. 
\begin{table}[h!]
\centering
\caption{Best-fit parameters (no local exception).}
  \begin{tabular}{ l c c c c c c c}
  \toprule
              & $a$ & $\bar{g}_{\rm cr}$ & $g_{\rm cr}^{\rm CygA}$ & $\bar{g}_{\rm acc}$ & $g_{\rm acc}^{\rm CygA}$ & $q$ & $\chi^2$  \\ 
   \midrule
      \textbf{EPOS-LHC} & $1.7$ & $1.94$ & $22.17$ & $0.6$ & $0.11$ & $2$ & $5.4$  \\ 
     \textbf{QGSJetII-04} & $1.76$ & $2.23$ & $32.60$ & $0.6$ & $0.090$ & $1.84$ & $5.6$ \\
     \textbf{Sibyll2.1} & $1.83$ & $2.29$ & $42.51$ & $0.6$ & $0.085$ & $1.97$ & $5.7$ \\ 
   \bottomrule
\end{tabular}
  \label{FitPar0}
\end{table}
\begin{figure}[tbh]
  \centering
    \includegraphics[width=0.49\textwidth]{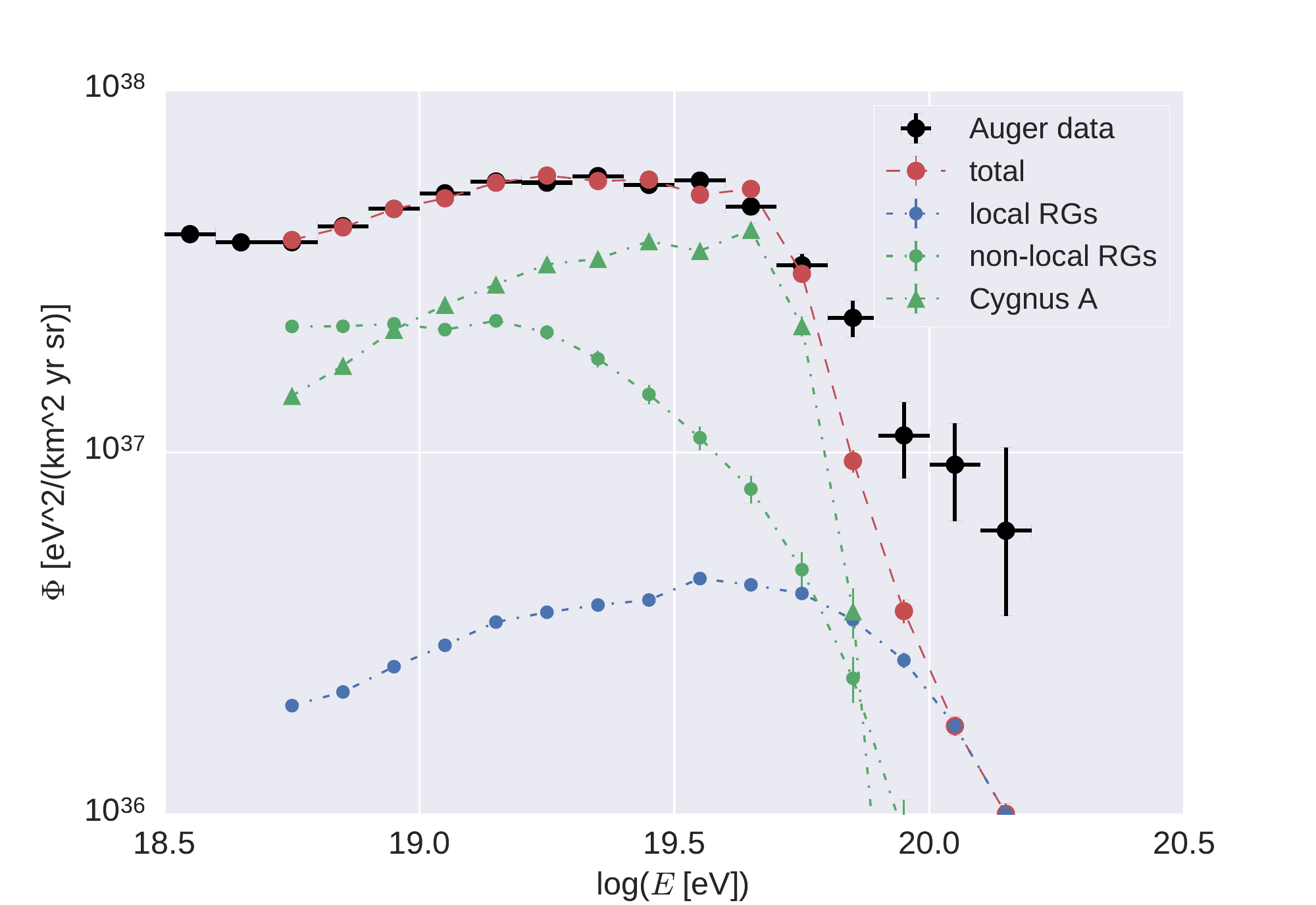}
    \includegraphics[width=0.49\textwidth]{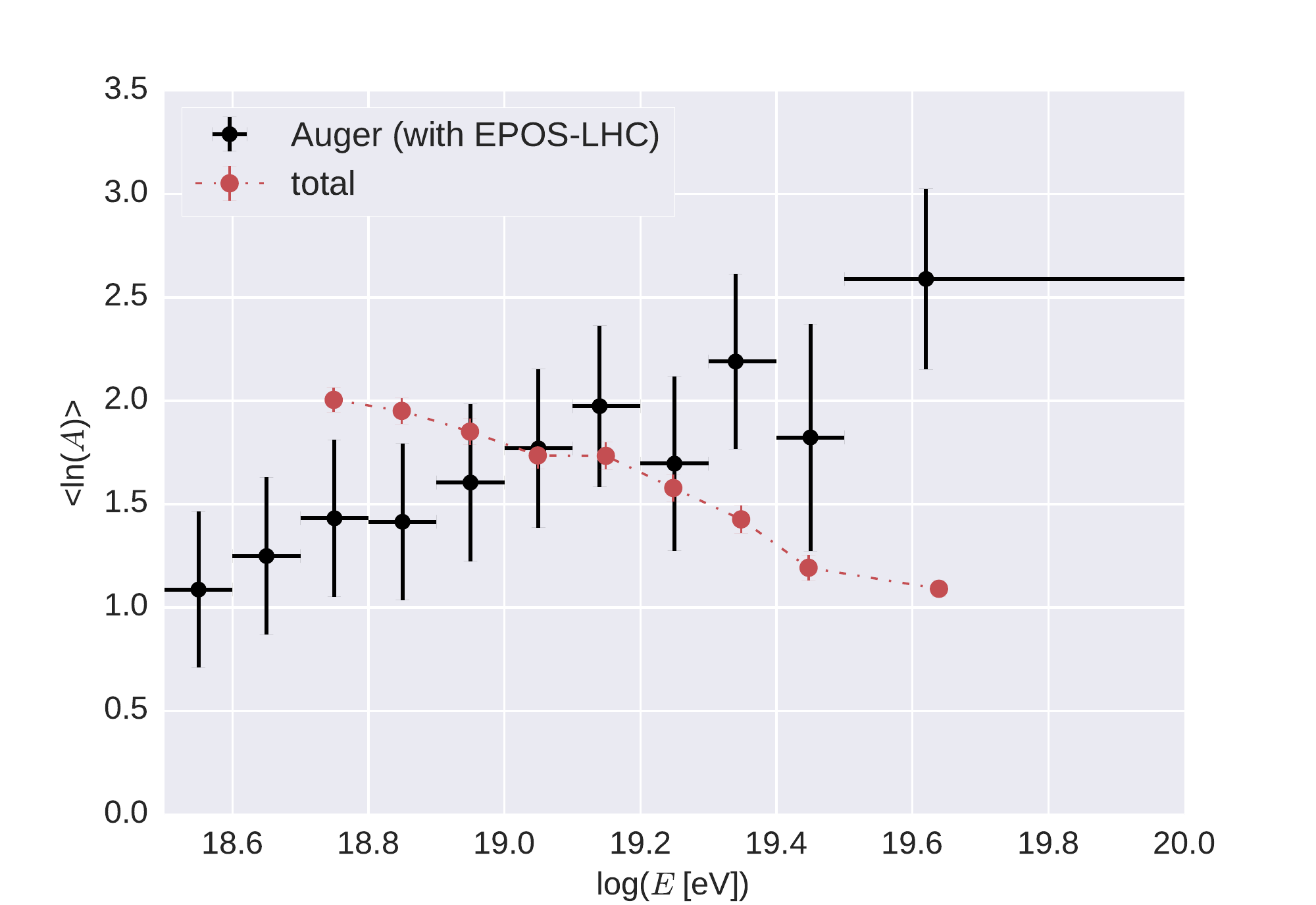}
\caption{Best-fit results to energy spectrum (left) and chemical composition (right) using the EPOS-LHC model and no local source exception.}
\label{EPOS_Fit0}
\end{figure}
As shown in Fig.\ \ref{EPOS_Fit0}, no satisfactory fit to the observational data is possible, even in the case of the EPOS-LHC model which leads to the heaviest observed composition. 
Apart from the decreasing heaviness with increasing energy, a distinct disagreement with the observations, also the CR flux at the highest energies cannot be explained. 
Hence, at least Centaurus A, M87, or Fornax A, need to obtain a higher value of $g_{\rm cr}$ than the average sources. 

We want to point out one additional aspect here which is relevant also for the following two cases. Our fit procedure finds for Cygnus A a rather low acceleration efficiency, $g_{\rm acc}^{\rm CygA}$. This is counter-intuitive, as from its jet power and radio structure one would expect that there can hardly be a more efficient accelerator among radio galaxies than Cygnus A. 
However, for a higher acceleration efficiency the spectrum of Cygnus A contains the GZK bump, as seen in Fig.~\ref{ContribOfCygA}, which is not present in the data. Here we need to point out again the several drawbacks of our approximate treatment of UHECR propagation from Cygnus A mentioned in Sect.~\ref{Sec:CygnusA}. Qualitatively, we expect that considering these effects will lead to a steeper and smoother spectrum of Cygnus A at the high energy end than obtained here, which would need to be compensated by choosing a larger value of $g_{\rm acc}^{\rm CygA}$. Quantifying this will require a precise simulation as outlined in Sect.~\ref{sec:SummConclu} for a follow-up work.

\subsection{Case 2: Powerful Centaurus A}
\label{Case2}
In this case we assume that candidates arriving from Centaurus A can be weighted with an individual parameter value of $g_{\rm cr}^{\rm CenA}$. We consider two limiting cases where (a) only Centaurus A has an enhanced heavy composition according to the parameter $q$ --- the \textit{light composition scenario}; and (b) all sources except Cygnus A eject an enhanced heavy composition --- the \textit{heavy composition scenario}. 
Cygnus A gets an individual parameter value of $g_{\rm cr}^{\rm CygA}$ and $g_{\rm acc}^{\rm CygA}$ as well, but all other sources (including Fornax A and M87) get an average parameter value of $g_{\rm cr}$ and $g_{\rm acc}$. 
The resulting best-fit parameters for both composition scenarios are summarized in table \ref{FitPar1} and \ref{FitPar2}. 
\begin{table}[h!]
\centering
\caption{Best fit parameters (light composition scenario with powerful Centaurus A).}
  \begin{tabular}{ l c c c c c c c c}
  \toprule
              & $a$ & $\bar{g}_{\rm cr}$ & $g_{\rm cr}^{\rm CenA}$ & $g_{\rm cr}^{\rm CygA}$ & $g_{\rm acc}$ & $g_{\rm acc}^{\rm CygA}$ & $q$ & $\chi^2$  \\ 
  \midrule
    \textbf{EPOS-LHC} & $1.85$ & $7.73$ & $41.54$ & $43.94$ & $0.127$ & $0.059$ & $2$ & $1.1$  \\ 
     \textbf{QGSJetII-04} & $1.82$ & $6.31$ & $21.20$ & $48.84$ & $0.220$ & $0.056$ & $2$ & $1.4$ \\
     \textbf{Sibyll2.1} & $1.83$ & $6.67$ & $24.77$ & $47.90$ & $0.19$ & $0.056$ & $2$ & $1.3$ \\
   \bottomrule
\end{tabular}
  \label{FitPar1}
\end{table}
\begin{table}[h!]
\centering
\caption{Best fit parameters (heavy composition scenario with powerful Centaurus A).}
  \begin{tabular}{ l c c c c c c c c}
   \toprule
       & $a$ & $\bar{g}_{\rm cr}$ & $g_{\rm cr}^{\rm CenA}$ & $g_{\rm cr}^{\rm CygA}$ & $g_{\rm acc}$ & $g_{\rm acc}^{\rm CygA}$ & $q$ & $\chi^2$  \\ 
   \midrule
      \textbf{EPOS-LHC} & $1.77$ & $1.02$ & $46.73$ & $30.48$ & $0.117$ & $0.068$ & $1.77$ & $1.3$  \\ 
     \textbf{QGSJetII-04} & $1.82$ & $1.06$ & $27.42$ & $49.34$ & $0.211$ & $0.057$ & $1.81$ & $1.8$ \\
     \textbf{Sibyll2.1} & $1.83$ & $1$ & $38.77$ & $50$ & $0.159$ & $0.057$ & $1.8$ & $1.6$ \\ 
   \bottomrule
\end{tabular}
  \label{FitPar2}
\end{table}
Independent of the preferred hadronic interaction model, we are able to explain the observational data of the energy spectrum and the chemical composition in the case of the light composition scenario, as shown in Fig.\ \ref{Sibyll_Fit} for Sibyll2.1. 
The heavy composition scenario leads to slightly less accurate fits, predominantly due to a too heavy composition around $10^{18.8}\,\text{eV}$, and the full range of possible values for the parameter $g_{\rm cr}$ is used in order to keep the average non-local contribution subdominant.
Concentrating on the light composition scenario, the significantly smaller parameter value of $\bar{g}_{\rm cr}$ compared to the values of Centaurus A or Cygnus A indicates that not every RG with a certain minimal radio luminosity has a high cosmic ray load. We could probably account for at most $15{-}25$\% of the RGs with a radio luminosity at $1.1\,\text{GHz}$ above $\check{L}\simeq 10^{38}\,\text{erg/s}$ to have parameters comparable to Centaurus A and Cygnus A, while the contribution of all other RGs then would have to be irrelevant (i.e., $g_{\rm cr} \ll \bar{g}_{\rm cr}$). The same would have to be the case for the abundance enhancement, which needs to be stronger in Centaurus A (or a small subset of sources) than in most other sources. Therefore, we would need to assume that both Cygnus A and Centaurus A, which are anyway unusual due to their proximity given their power, are also preferred as cosmic-ray accelerators. In this scenario, almost all UHECRs observed would originate in one of those two sources, and the isotropically ejected CR power above the ankle of these sources would be $Q^{\rm CenA}_{\rm uhecr}\simeq 2\times 10^{43}\,\text{erg/s}$ and $Q^{\rm CygA}_{\rm uhecr}\simeq 2\times 10^{47}\,\text{erg/s}$, weakly dependent on the chosen scenario or the hadronic interaction model. 
\begin{figure}[tbh]
  \centering
    \includegraphics[width=0.49\textwidth]{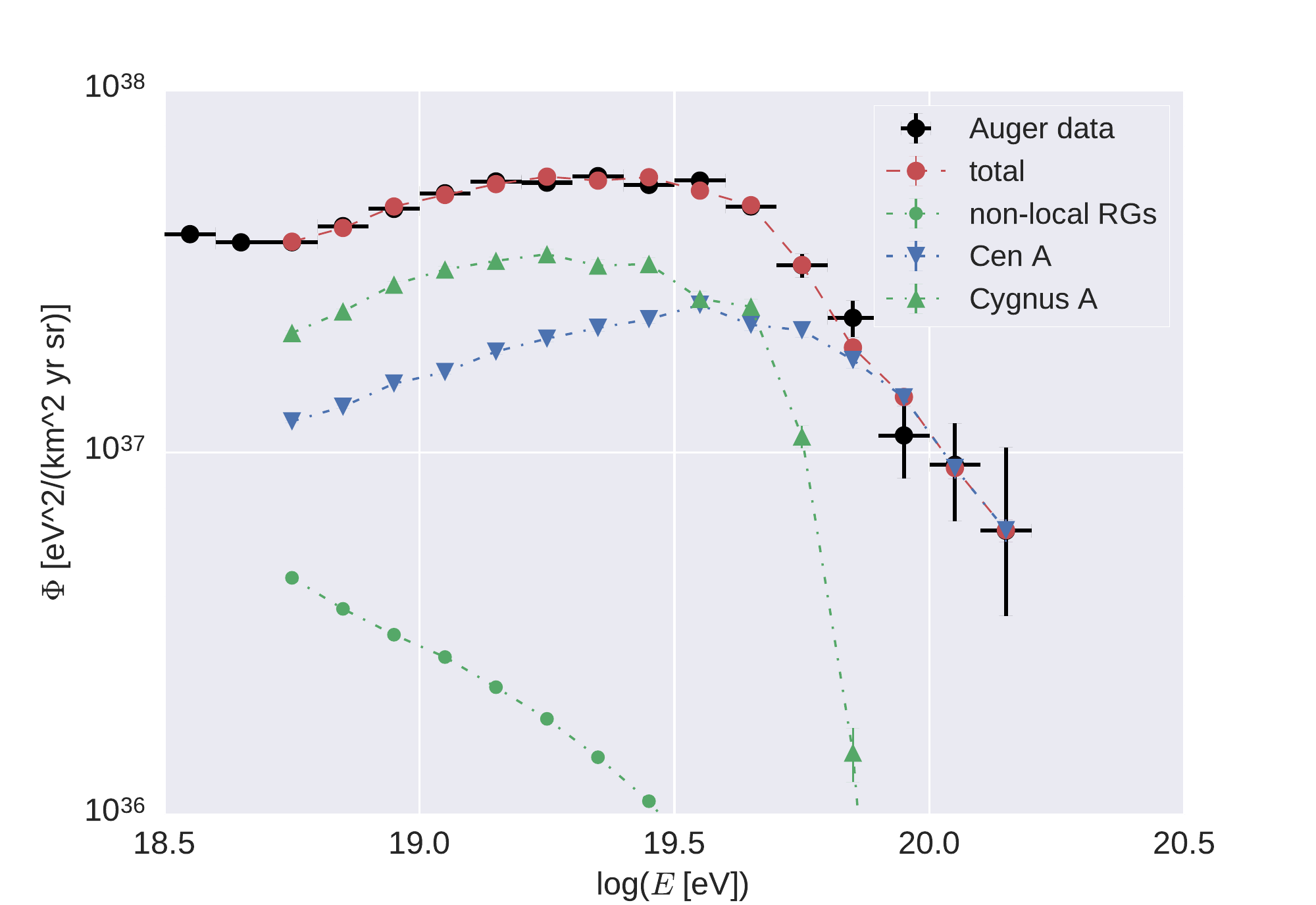}
    \includegraphics[width=0.49\textwidth]{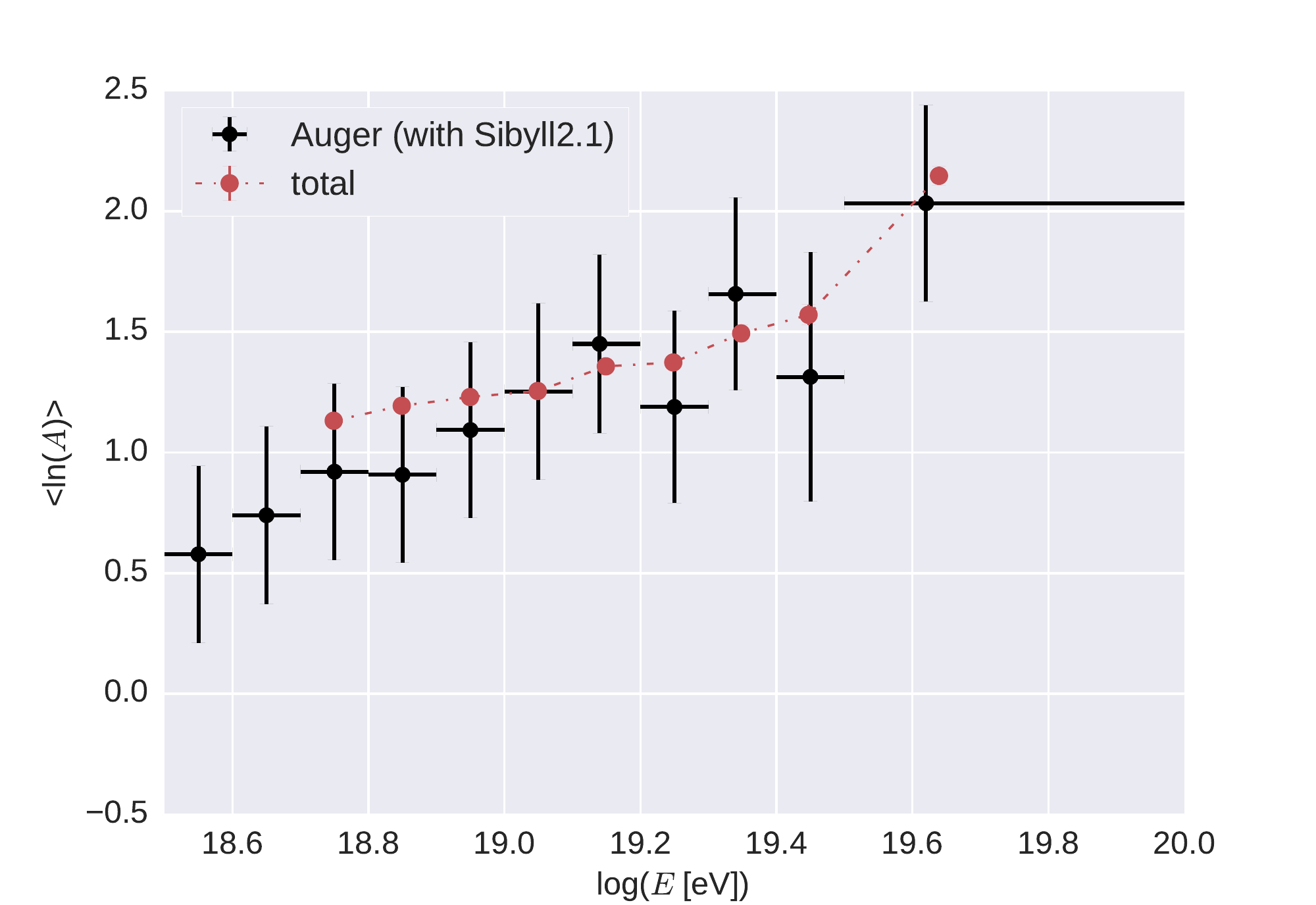}
\caption{Best-fit results to energy spectrum (left) and chemical composition (right) using Sibyll2.1 and the light composition scenario with powerful Centaurus A.}
\label{Sibyll_Fit}
\end{figure}
\begin{figure}[tbh]
  \centering 
    \includegraphics[width=0.49\textwidth]{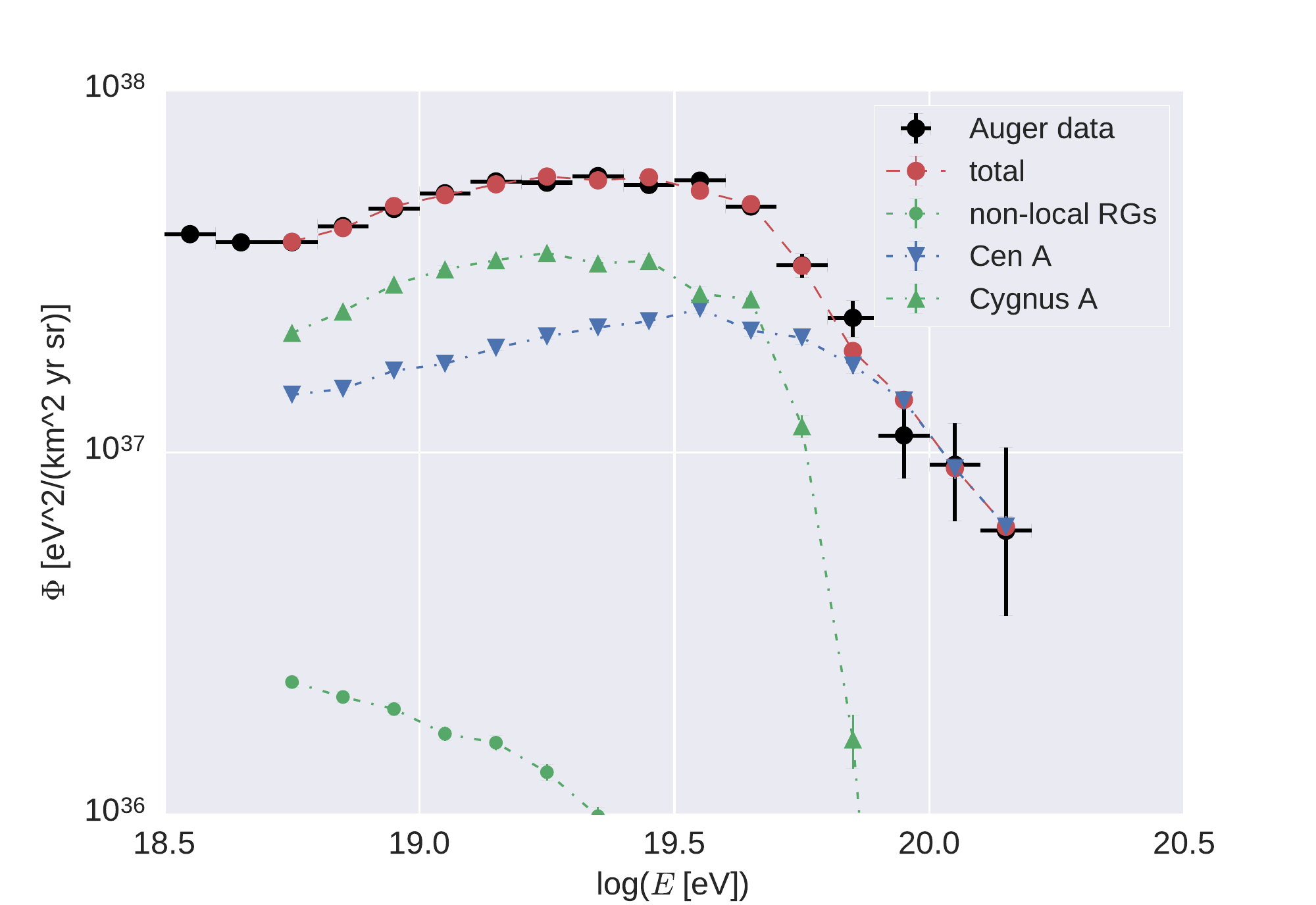}
    \includegraphics[width=0.49\textwidth]{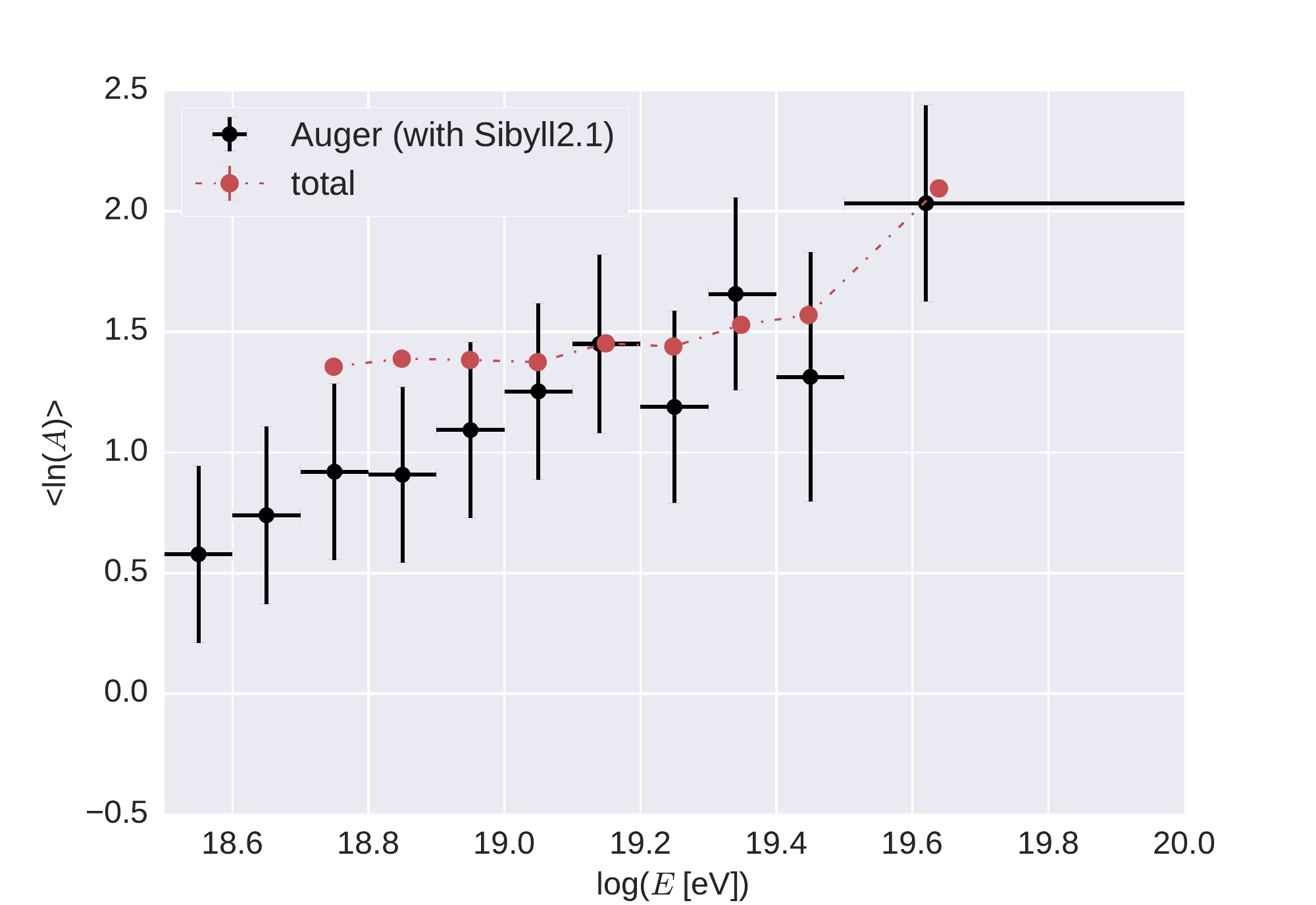}
\caption{Best-fit results to energy spectrum (left) and chemical composition (right) using Sibyll2.1 and the heavy composition scenario with powerful Centaurus A.}
\label{Sibyll_Fit2}
\end{figure}

The corresponding arrival directions for the different best-fit models or initial composition scenarios do not differ significantly. 
Figs.\ \ref{Sibyll_Skymap1}-\ref{Sibyll_Skymap2} show the resulting sky maps for the limiting EGMF deflection scenarios for Cygnus A (point source vs.\ complete isotropization) in the case of the Sibyll2.1 best-fit model for the light scenario. 
The angular power spectra of the arrival directions is analyzed for different deflection scenarios (see Fig.\ \ref{Sibyll_PowSpec}): Three different coherence lengths $\lambda_{\rm c}$ of a turbulent field with $B_{\rm rms}=1.18\,\text{nG}$, as well as the case of the complete isotropization (i.e.\ $\lambda_{\rm c}\gg 10\,\text{Mpc}$), and the complete isotropization where initial H and He candidates from Centaurus A are excluded. The latter assumption does not affect the total energy spectrum or the chemical composition, but lowers the level of anisotropy and improves consistency with the data set at $4{-}8\,$EeV.  
The motivation for this is that models for a particular heavy enhancement of Centaurus A would expect a pure heavy composition \cite{2010ApJ...720L.155G}. We also note that at $4{-}8\,$ EeV an impact of cosmic rays from the Galaxy or other extragalactic components cannot be excluded, and that at $E > 8\,$EeV our results are in good agreement with the data, for the case that UHECR from Cygnus A are completely isotropized.  
Further, an agreement with the observed arrival directions at energies $E > 8\,$EeV is obtained for 
\be
\label{rms-defl-condition}
\lambda_{\rm c}^{1/2}\, B_{\rm rms} \geq 6\,\text{Mpc}^{1/2}\,\,\text{nG},
\ee
i.e.\ a global coherence length $\lambda_{\rm c}=26\,\text{Mpc}$ in the case of a rms field strength of $B_{\rm rms}\geq 1.18\,\text{nG}$ (see Fig.\ \ref{AngPowSpec4_Skymap3}). We like to stress again that a more detailed inclusion of the EGMF effects on the propagation of UHECRs from Cygnus A is needed --- and intended in future investigations (see Sect.~\ref{sec:SummConclu}) --- to allow a stringent test against observational data. 

\begin{figure}[tbh]
  \centering
    \includegraphics[width=0.49\textwidth]{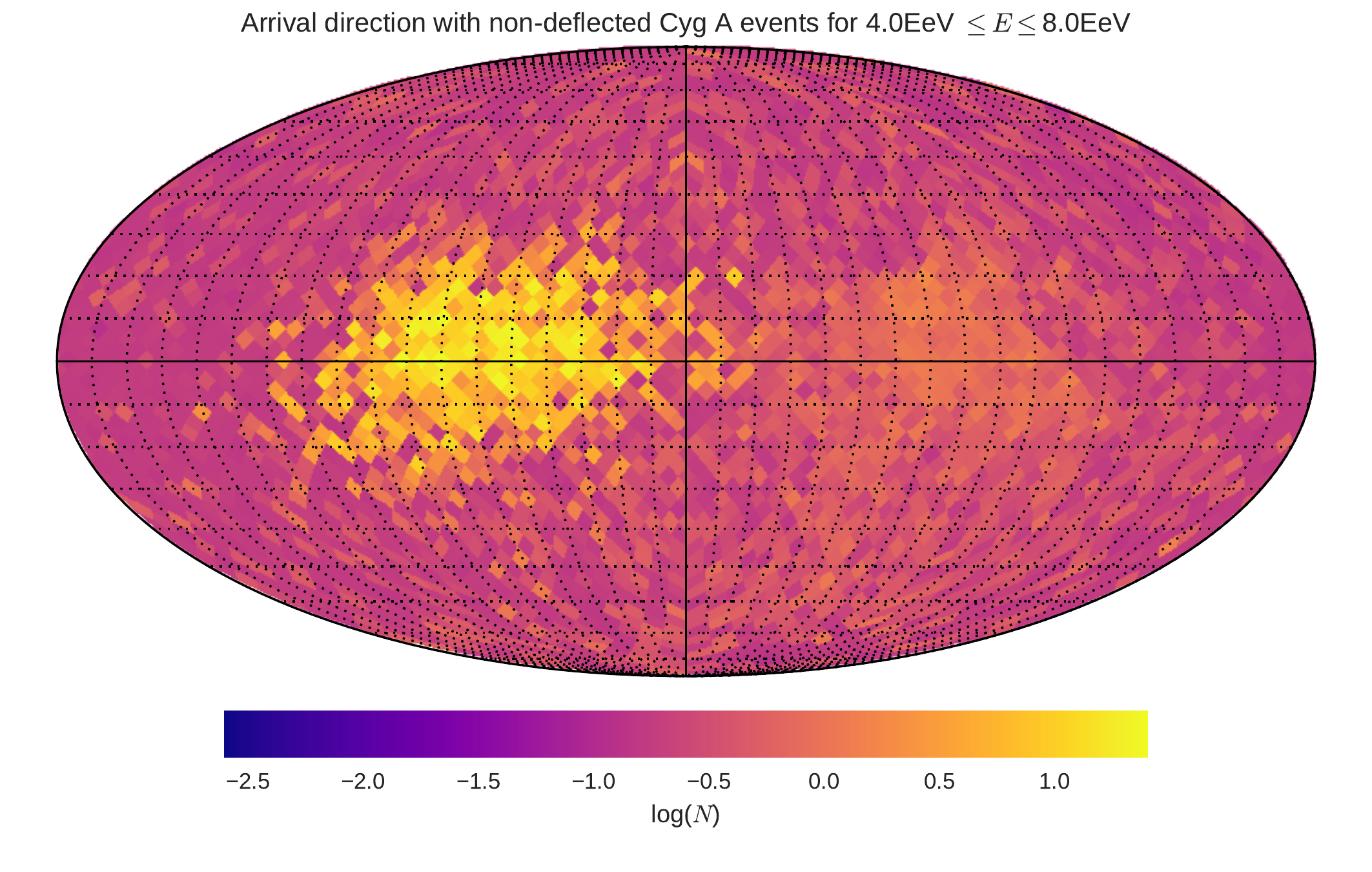}
    \includegraphics[width=0.49\textwidth]{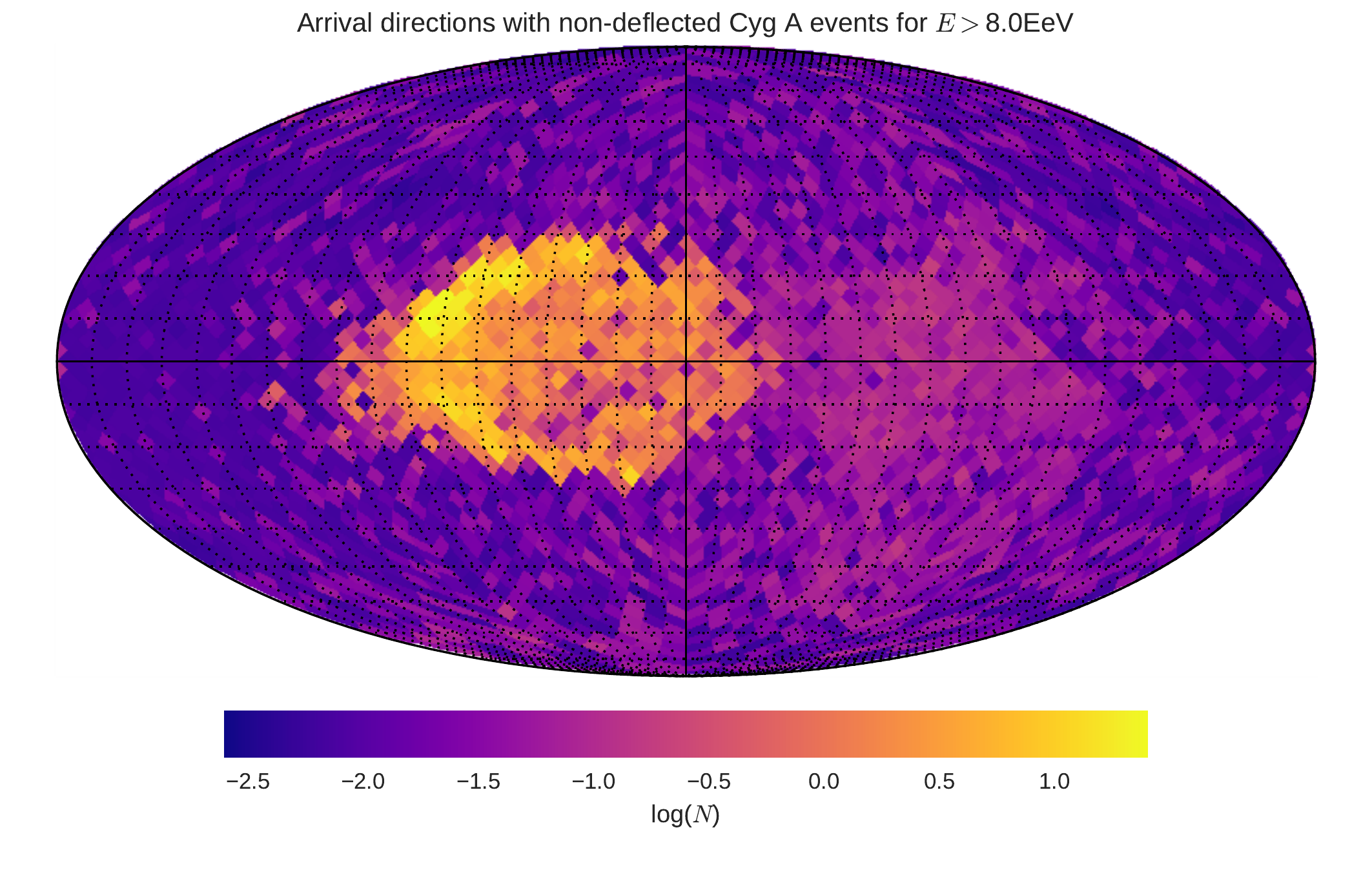}
\caption{Skymap with non-deflected Cygnus A events for $4\,\text{EeV}\leq E \leq 8\,\text{EeV}$ (left), and $E>8\,\text{EeV}$ (right) using Sibyll2.1 and the light composition scenario with a powerful Centaurus A.}
\label{Sibyll_Skymap1}
\end{figure} 
\begin{figure}[tbh]
  \centering
    \includegraphics[width=0.49\textwidth]{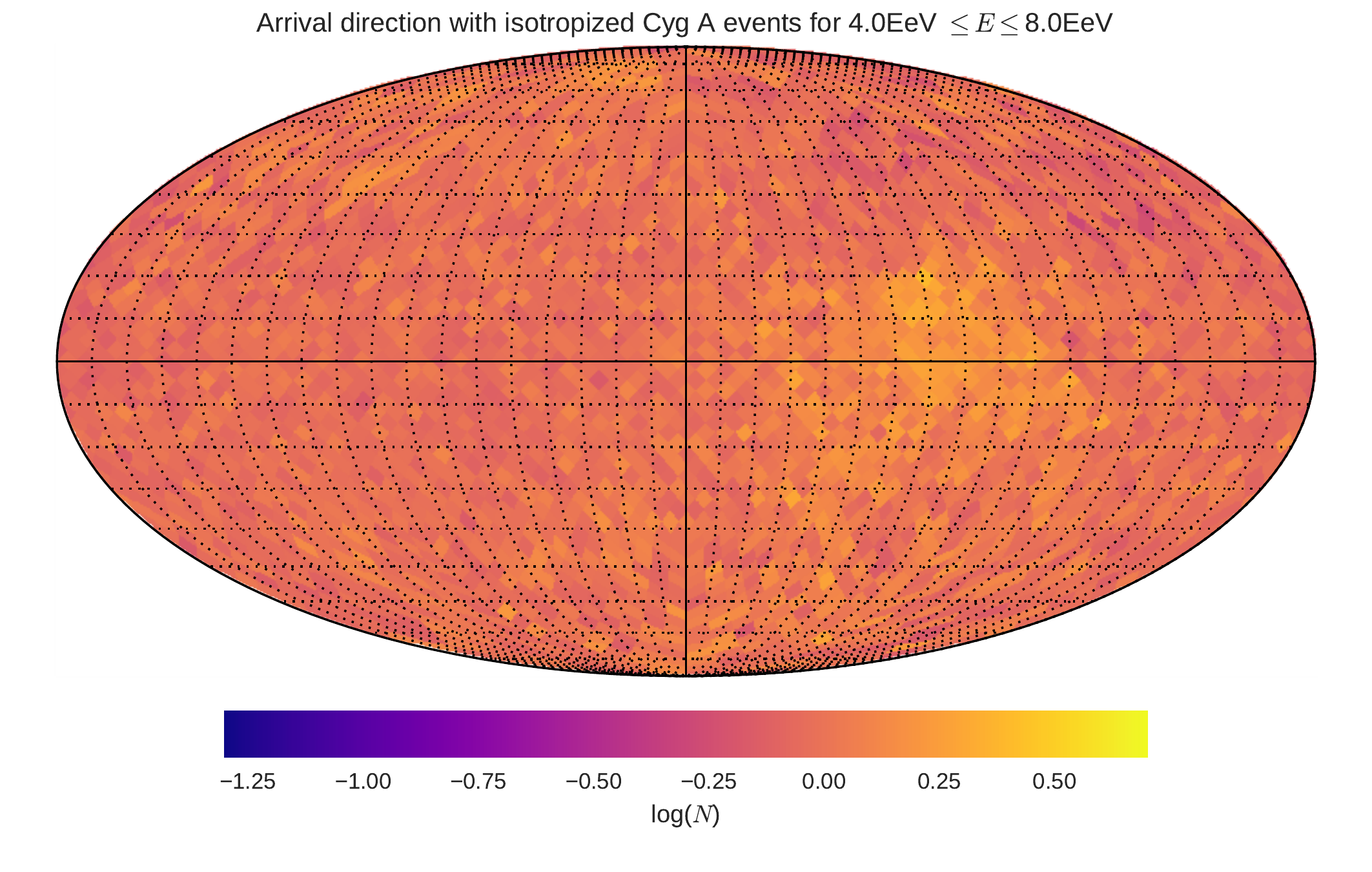}
    \includegraphics[width=0.49\textwidth]{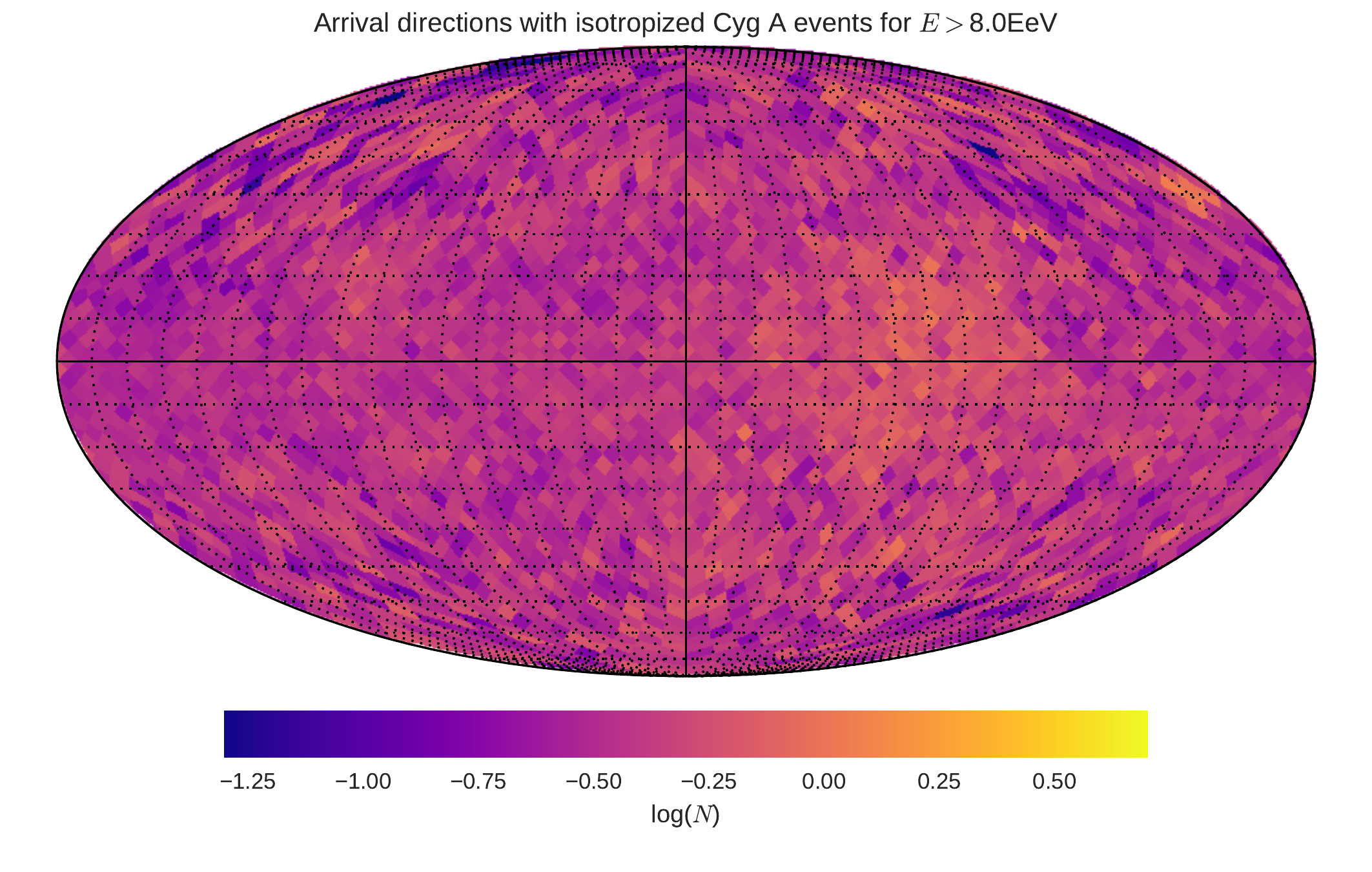}
\caption{Skymap with isotropized Cygnus A events for $4\,\text{EeV}\leq E \leq 8\,\text{EeV}$ (left), and $E>8\,\text{EeV}$ (right) using Sibyll2.1 and the light composition scenario with a powerful Centaurus A.}
\label{Sibyll_Skymap2}
\end{figure} 
\begin{figure}[tbh]
  \centering
    \includegraphics[width=0.49\textwidth]{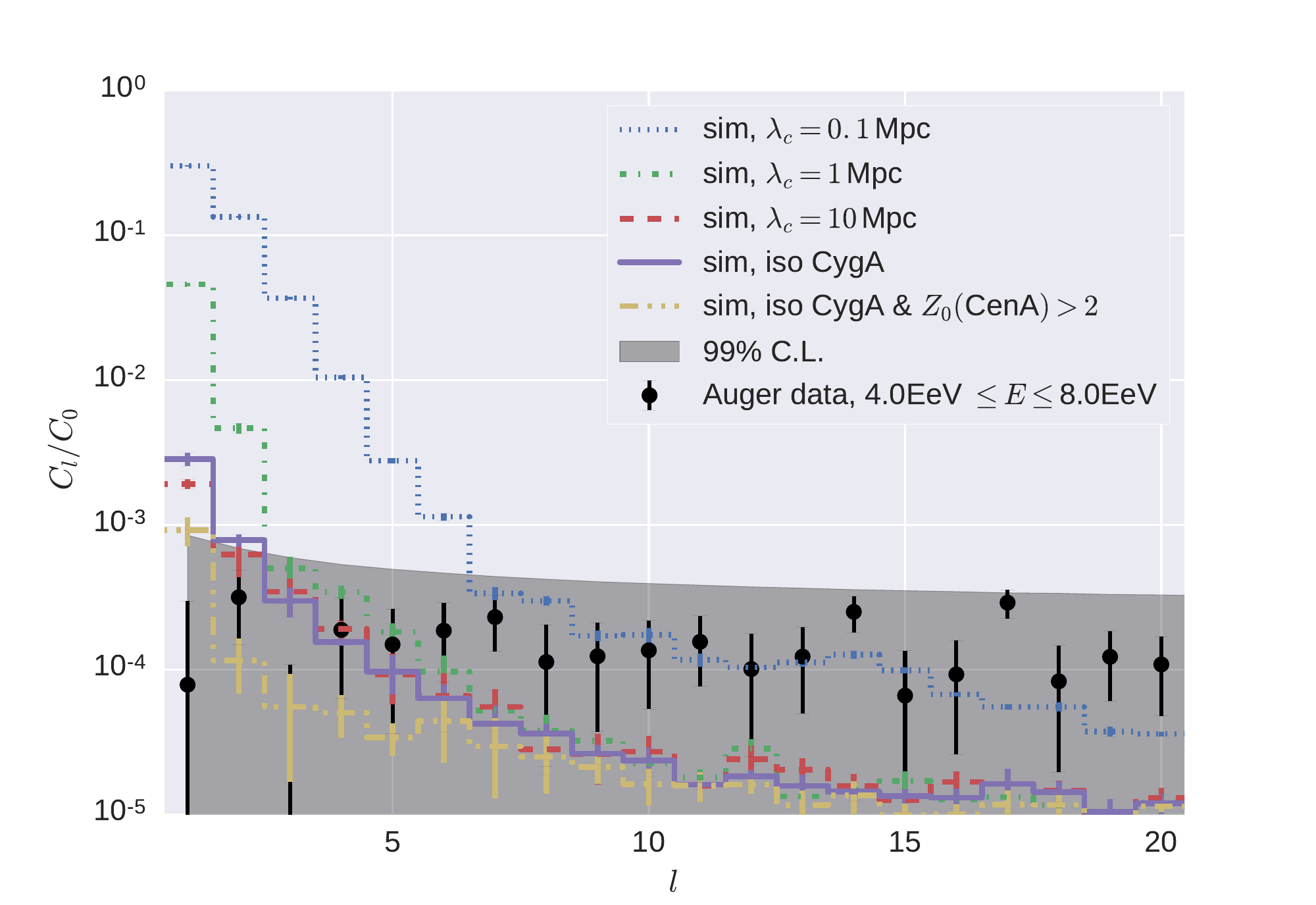}
    \includegraphics[width=0.49\textwidth]{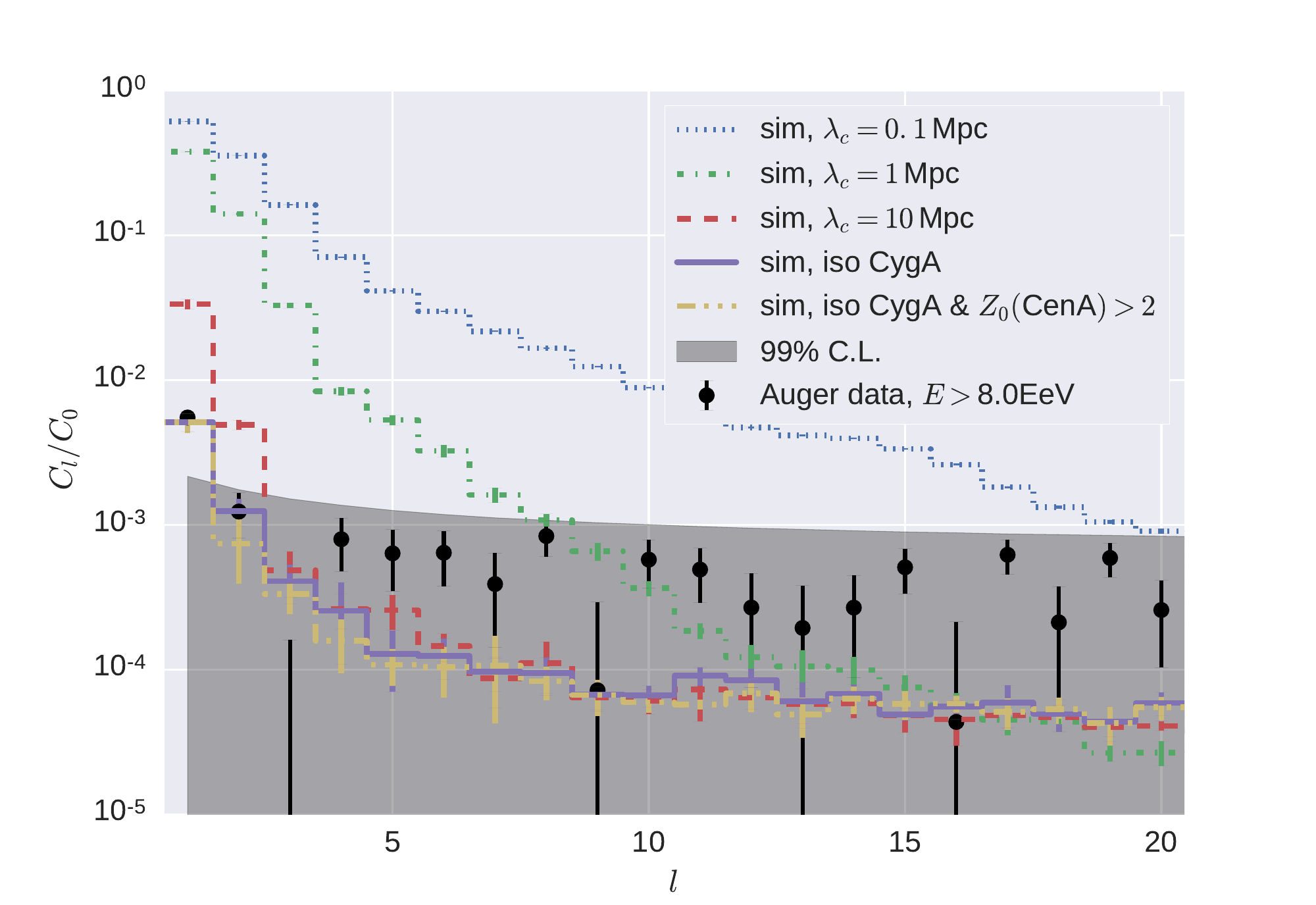}
\caption{Angular power spectrum with isotropized (solid and dash-dotted line) and non-deflected (dashed line) Cygnus A events for $4\,\text{EeV}\leq E \leq 8\,\text{EeV}$ (left), and $E>8\,\text{EeV}$ (right) using Sibyll2.1 and the light composition scenario with a powerful Centaurus A.}
\label{Sibyll_PowSpec}
\end{figure} 

Fig.\ \ref{Sibyll_Defl} shows the resulting mean deflection $\bar{\theta}_{\text{def},\mu}$ as well as $N_\mu$ of the individual local RG that contribute to the best-fit flux. 
It is seen that the dominant contribution from Centaurus A does not suffer significant deflections in the EGMF, rather it is the Galactic magnetic field which randomizes the arrival directions from Centaurus A to a large degree as they are mostly iron nuclei and thus their rigidity is small. Other radio galaxies which still have some significant contribution mostly show EGMF deflections up to about $10^\circ$. 
\begin{figure}[tbh]
  \centering
    \includegraphics[width=0.79\textwidth]{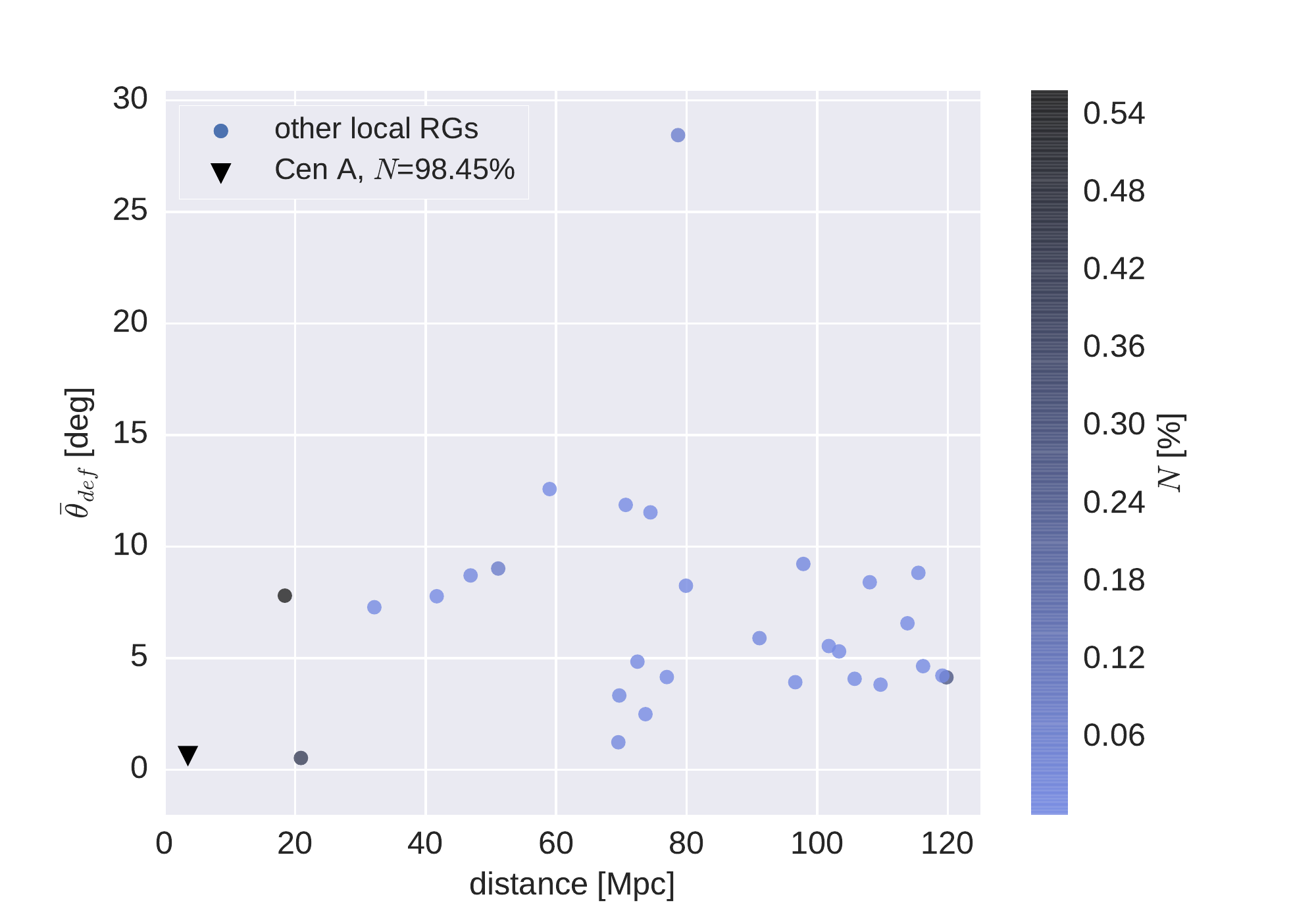}
\caption{The mean deflection angle due to the EGMF only of CR candidates from the individual local sources and its contribution (given by the color) to the best-fit CR flux. Here for the Sibyll2.1 model and the light composition scenario with a powerful Centaurus A.}
\label{Sibyll_Defl}
\end{figure}

\subsection{Case 3: Powerful M87 and Fornax A}
\label{Case3}
Finally, we discuss the case where M87 and Fornax A, instead of Centaurus A, have an individual parameter value of $g_{\rm cr}^{\rm M/F}$. 
The resulting best-fit parameters for the light and the heavy composition scenarios are summarized in table \ref{FitPar5} and \ref{FitPar6}, respectively. 
\begin{table}[h!]
\centering
\caption{Best-fit parameters (light composition scenario, powerful M87 and Fornax A).}
  \begin{tabular}{ l c c c c c c c c} 
  \toprule
              & $a$ & $\bar{g}_{\rm cr}$ & $g_{\rm cr}^{\rm M/F}$ & $g_{\rm cr}^{\rm CygA}$ & $g_{\rm acc}$ & $g_{\rm acc}^{\rm CygA}$ & $q$ & $\chi^2$  \\ 
  \midrule
    \textbf{EPOS-LHC} & $1.71$ & $12.83$ & $50$ & $32.04$ & $0.100$ & $0.065$ & $2$ & $4.8$  \\ 
     \textbf{QGSJetII-04} & $1.76$ & $9.83$ & $50$ & $41.61$ & $0.133$ & $0.061$ & $1.87$ & $2.8$ \\
     \textbf{Sibyll2.1} & $1.71$ & $10.02$ & $48.91$ & $34.05$ & $0.121$ & $0.064$ & $1.87$ & $3.1$ \\ 
   \bottomrule
\end{tabular}
  \label{FitPar5}
\end{table}
\begin{table}[h!]
\centering
\caption{Best-fit parameters (heavy composition scenario, powerful M87 and Fornax A).}
  \begin{tabular}{ l c c c c c c c c} 
  \toprule
              & $a$ & $\bar{g}_{\rm cr}$ & $g_{\rm cr}^{\rm M/F}$ & $g_{\rm cr}^{\rm CygA}$ & $g_{\rm acc}$ & $g_{\rm acc}^{\rm CygA}$ & $q$ & $\chi^2$  \\ 
   \midrule
    \textbf{EPOS-LHC} & $1.71$ & $2.17$ & $39.94$ & $27.70$ & $0.123$ & $0.078$ & $1.99$ & $2.7$  \\ 
     \textbf{QGSJetII-04} & $1.80$ & $2.11$ & $45.48$ & $44.20$ & $0.170$ & $0.065$ & $1.84$ & $2.4$ \\
     \textbf{Sibyll2.1} & $1.71$ & $2.14$ & $44.96$ & $31.50$ & $0.147$ & $0.072$ & $1.78$ & $2.4$ \\ 
   \bottomrule
\end{tabular}
  \label{FitPar6}
\end{table}
\begin{figure}[tbh]
  \centering
    \includegraphics[width=0.49\textwidth]{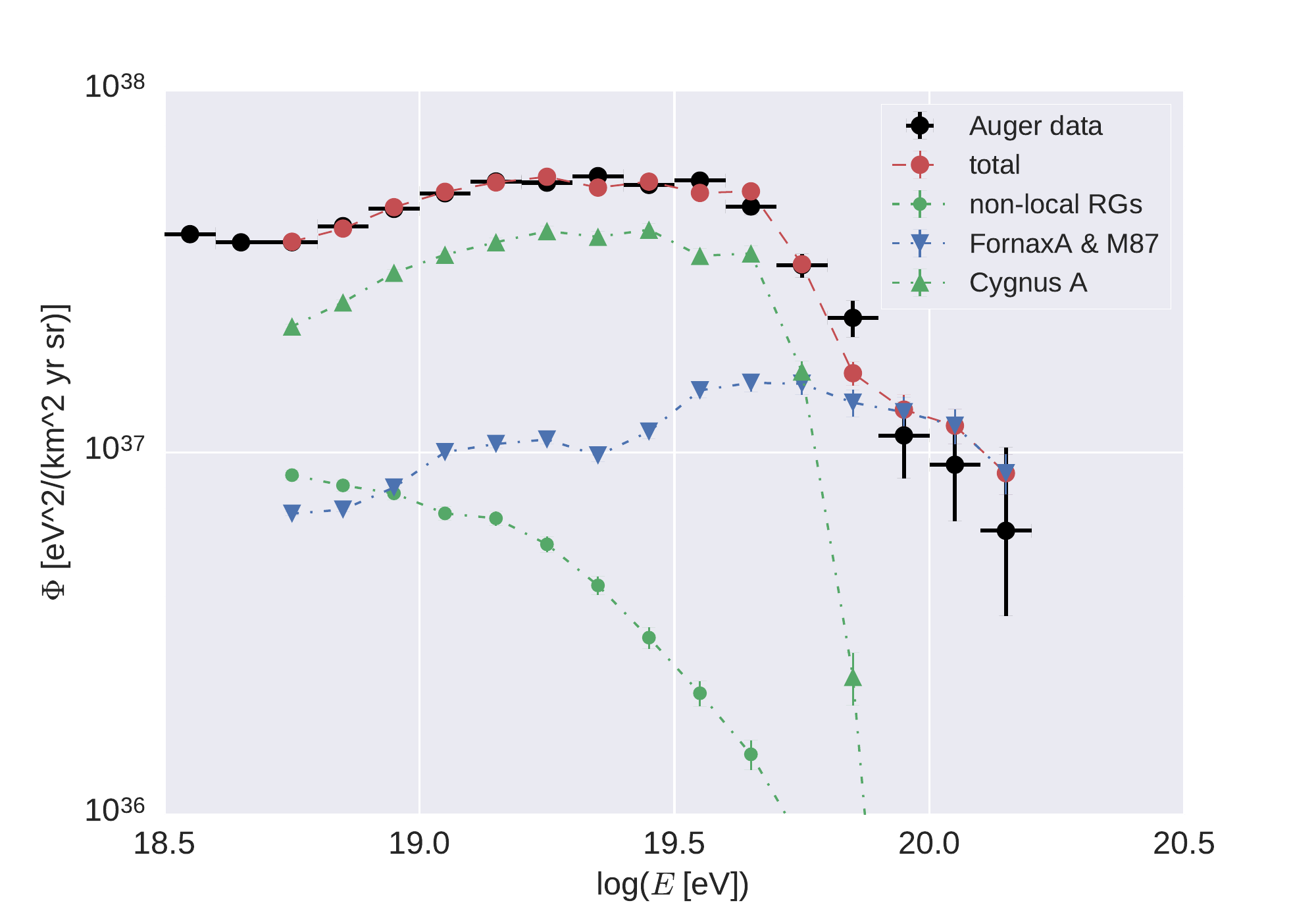}
    \includegraphics[width=0.49\textwidth]{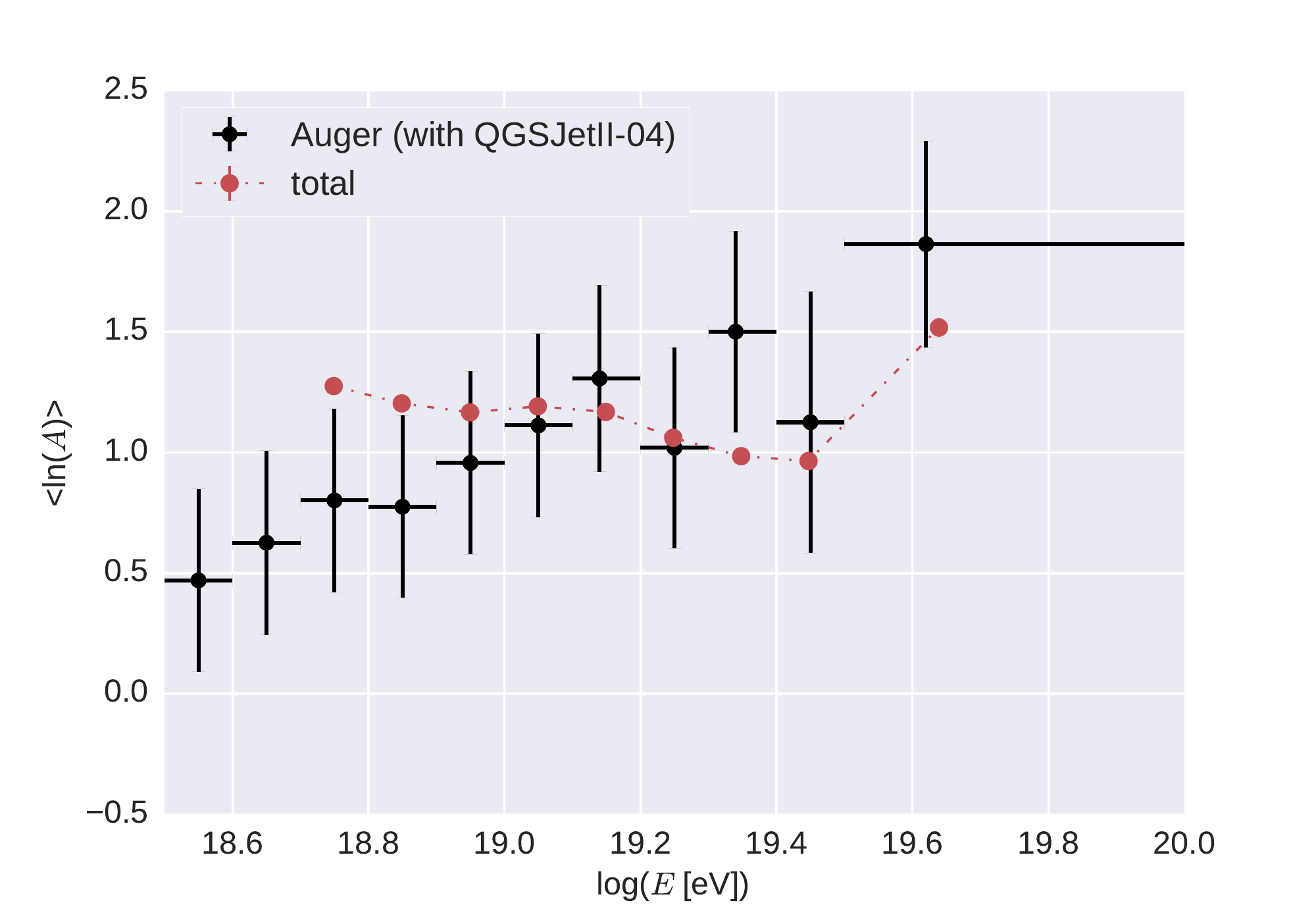}
\caption{Best-fit results to energy spectrum (left) and chemical composition (right) using QGSJetII-04 and the heavy composition scenario with a powerful M87 and Fornax A.}
\label{QGSJet_Fit4a}
\end{figure}
\begin{figure}[tbh]
  \centering
    \includegraphics[width=0.49\textwidth]{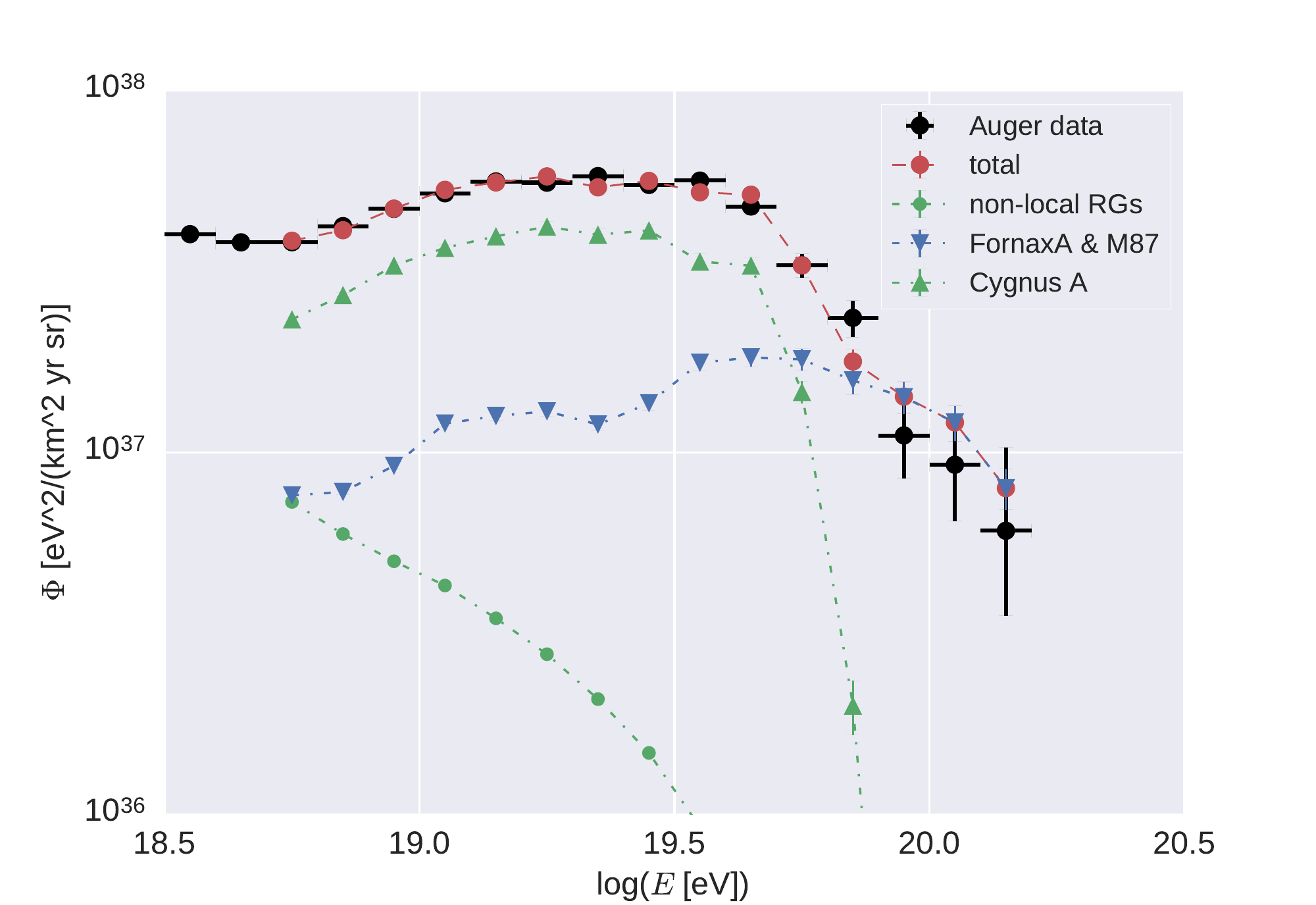}
    \includegraphics[width=0.49\textwidth]{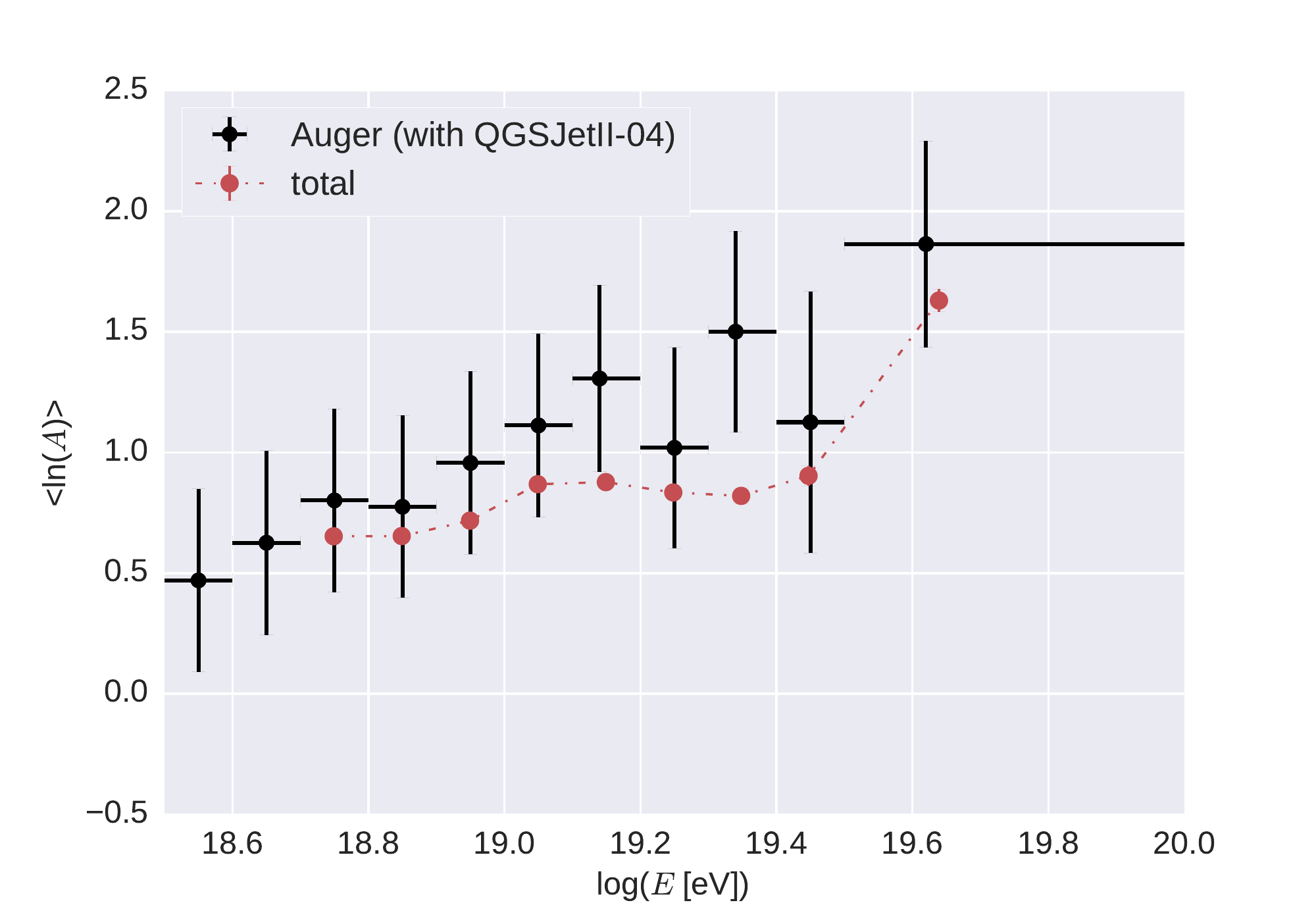}
\caption{Best-fit results to energy spectrum (left) and chemical composition (right) using QGSJetII-04 and the light composition scenario with a powerful M87 and Fornax A.}
\label{QGSJet_Fit4b}
\end{figure}
\begin{figure}[tbh]
  \centering
    \includegraphics[width=0.49\textwidth]{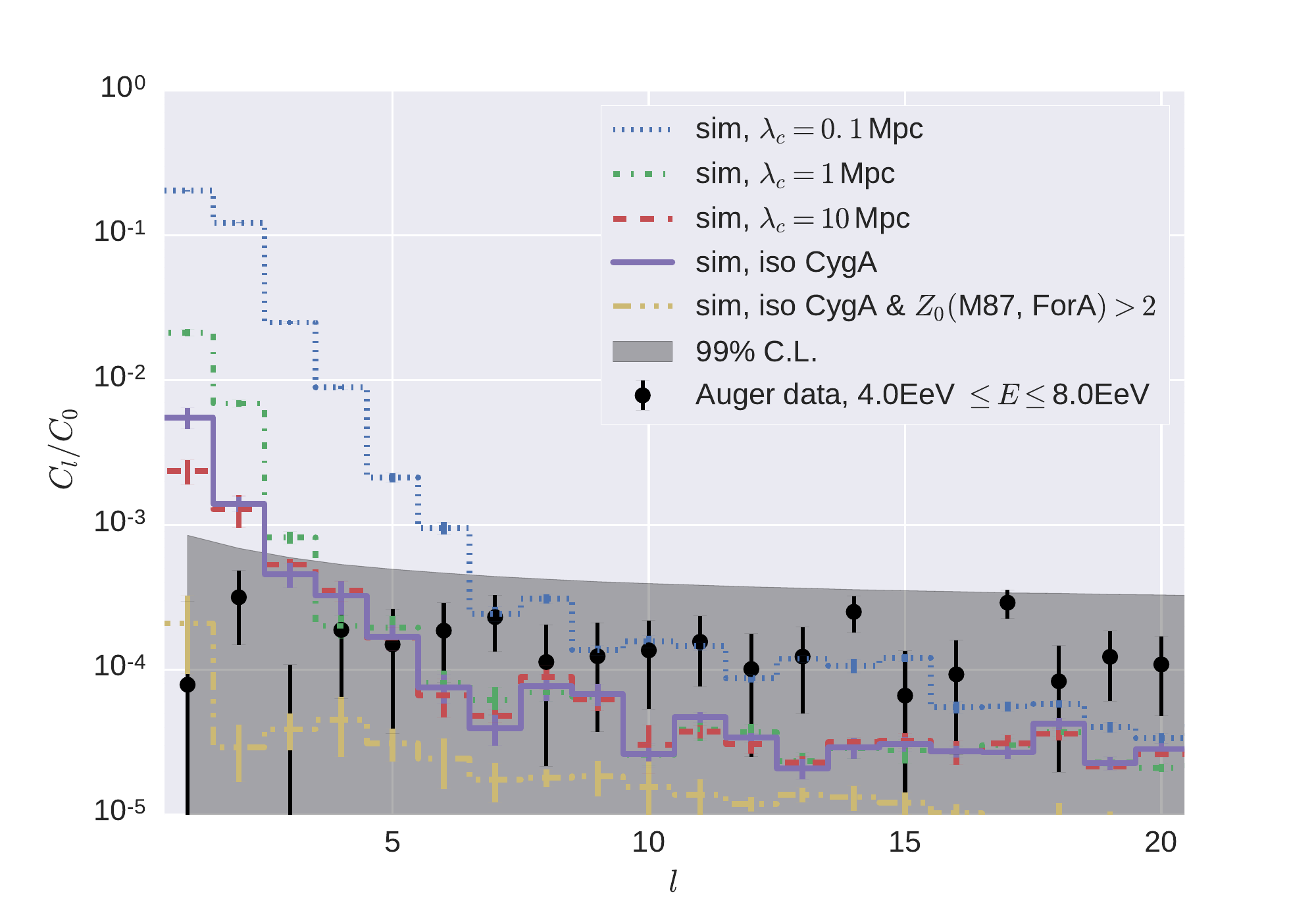}
    \includegraphics[width=0.49\textwidth]{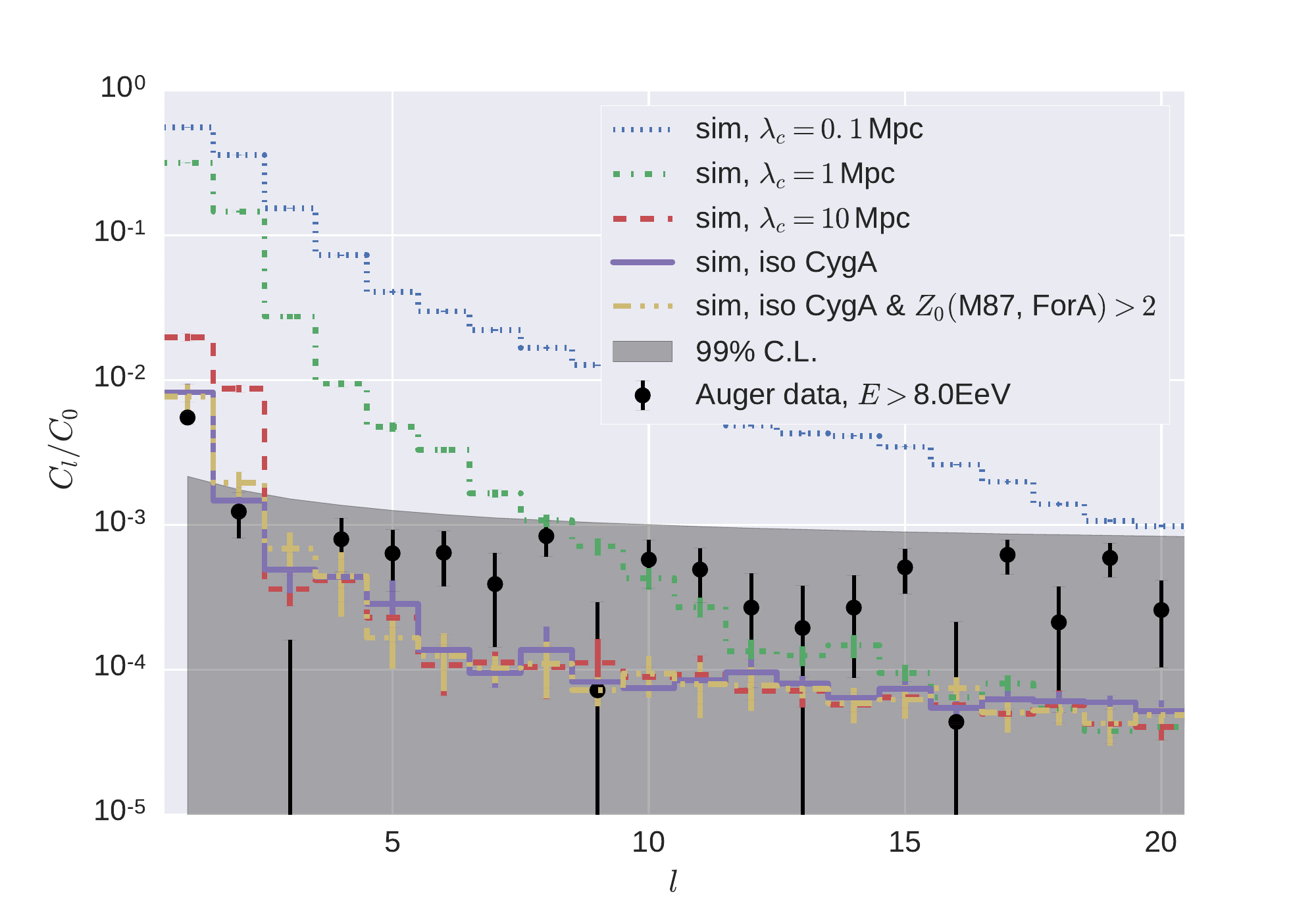}
\caption{Angular power spectrum of CR events with $4\leq E \leq 8\,\text{EeV}$ (left plot) and $E > 8\,\text{EeV}$ (right plot) for the light composition scenario with a powerful M87 and Fornax A using the QGSJetII-04 model. We show the cases for isotropized (solid and dash-dotted line) and non-deflected (dashed line) Cygnus A events}
\label{Sibyll_PowSpecComp_case4}
\end{figure} 

In general, the resulting fit is less accurate than in the case 2, as shown in Fig.\ \ref{QGSJet_Fit4a} and \ref{QGSJet_Fit4b}: 
Dependent on the chosen initial composition scenario the resulting composition at the observer is either too light or too heavy (with a rather decreasing heaviness with increasing energy). 
The resulting CR flux shows a better agreement with the observations, although, the data point at $10^{19.85}\,\text{eV}$ is missed.
Even though the fit-procedure has already chosen the maximal possible value of $g_{\rm cr}^{\rm M/F}$, M87 and Fornax A together are still too weak to explain the CR flux below the three highest data points, as Centaurus A could. 
Hence, there is obviously no need in generating multiple realizations of the distribution function $\Phi$ of the unknown local sources.

Aside from the fit accuracy the resulting spectra expose features that refer to the transition from helium to CNO at $\sim 10^{18.9}\,\text{eV}$ and from CNO to iron at $\sim 10^{19.4}\,\text{eV}$ which have not been seen in the previous case due to the different $q$ value (note that for this single source spectrum we did use exponential cut-offs for all nuclear components, so these features are not mathematical artifacts). 
Thus, for $q>1.9$ the observed CR spectra become completely dominated by the initial iron ejecta of the sources.
The corresponding arrival directions of the best-fit model yields a similar but slightly worse result --- with respect to the observed dipole --- compared to the case of a powerful Centaurus A (case 2): The resulting dipole moment for CR energies $>8\,\text{EeV}$ is slightly above the observed one --- for completely isotropized Cygnus A events --- but at CR energies $<8\,\text{EeV}$ we obtain a very good agreement with isotropy when we exclude the initially light elements (H, He) from M87 and Fornax A (see Fig.\ \ref{Sibyll_PowSpecComp_case4}). 

The previously discussed disagreements with the data in terms of energy spectrum and composition only get worse when we suggest to provide either M87 or Fornax A with an individual parameter value of $g_{\rm cr}$. 
Here, a good value for $g_{\rm cr}$ within its plausibility limits cannot be found, and the fit-procedure tries to compensate this by increasing $g_{\rm acc}$ to the maximal value. 
This leads to a wrong spectral behavior of the CR flux at the highest data points, as well as a stronger contribution by the average non-local contribution at the lowest energies.
Since Fornax A and M87 possess an almost equal radio luminosity and distance to Earth, energy spectrum and chemical composition expose the same deficits independent of which source is assumed to be powerful. 
The main difference between both scenarios is shown in the resulting arrival directions of the CRs and its corresponding angular power spectrum, which finds its explanation that the two sources have different positions and are thus UHECR trajectories are differently influenced by EGMF and GMF structures. 
As shown in Fig.\ \ref{Sibyll_PowSpecComp}, a powerful M87 source is in the case of the heavy composition scenario and completely isotropized Cygnus A events able to explain the observed dipole above $8\,\text{EeV}$, whereas the scenario with a powerful Fornax A source is not.
\begin{figure}[tb]
  \centering
    \includegraphics[width=0.49\textwidth]{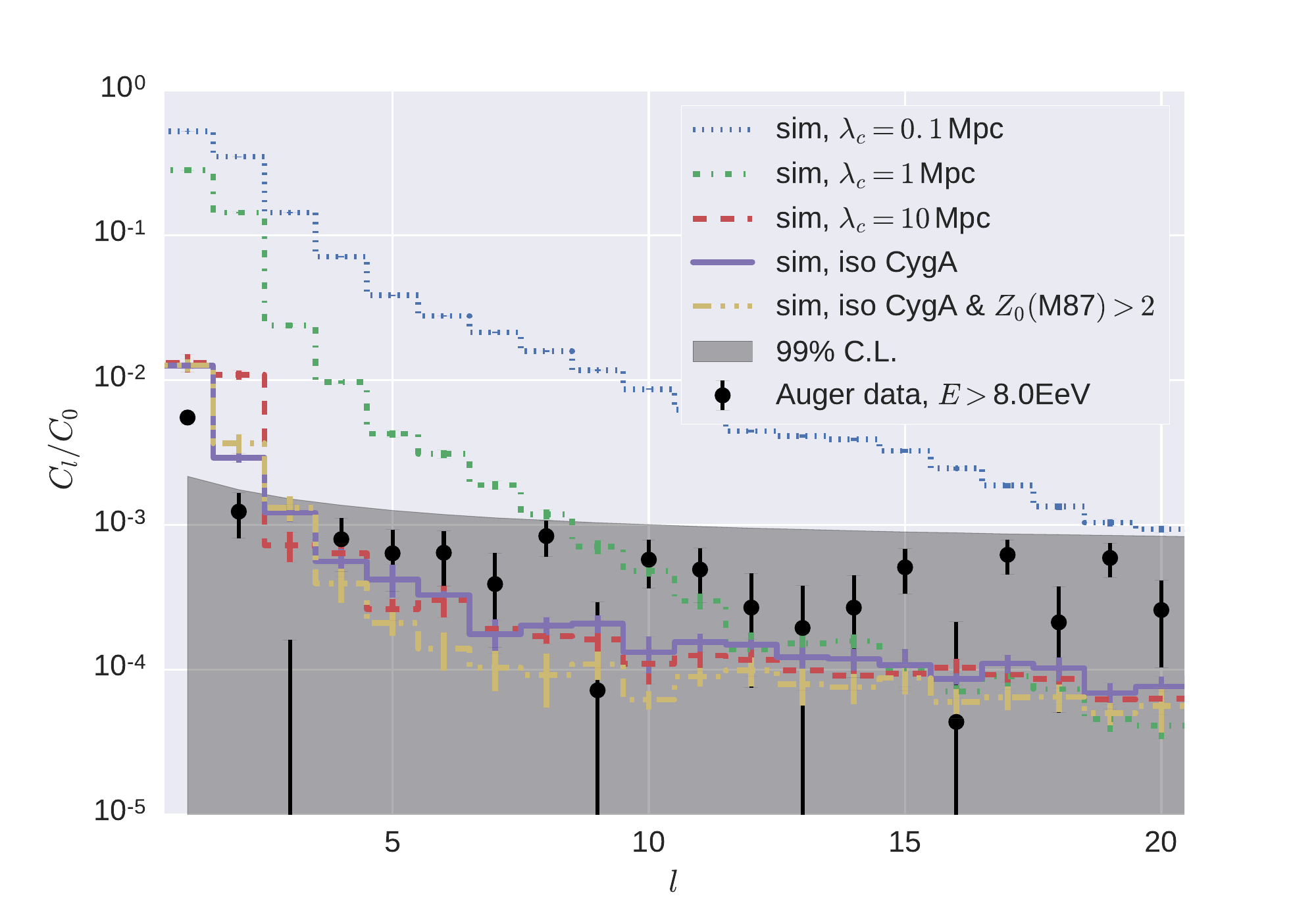}
    \includegraphics[width=0.49\textwidth]{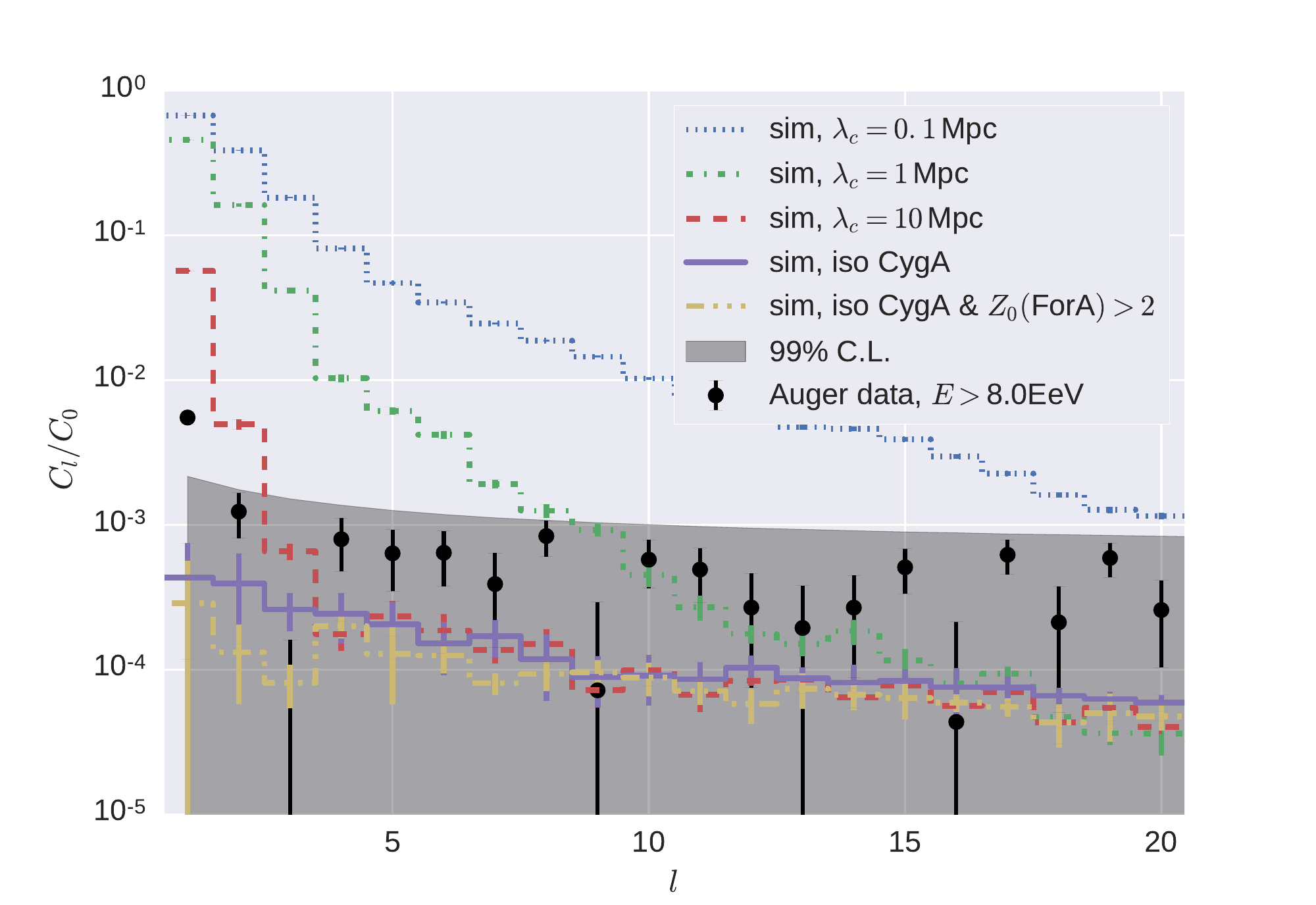}
\caption{Angular power spectrum with isotropized (solid line) and non-deflected (dashed line) Cygnus A events for $E>8\,\text{EeV}$ using Sibyll2.1 and the light composition scenario. The left plot shows the case of a powerful CR source by M87 and the right plot the case of a powerful CR source by Fornax A.}
\label{Sibyll_PowSpecComp}
\end{figure} 

Certainly one could discuss whether the results could be improved by further loosening parameter constraints, e.g., allowing $\bar{g}_{\rm acc}$ and $g_{\rm acc}^{\rm M/F}$ to be fitted independently, but in total we consider the scenario that M87 and/or Fornax A are the main contributor(s) of the cosmic ray flux beyond the GZK cutoff for Cygnus A as disfavored. Certainly, both can make a significant individual contribution which might become important when comparing UHECR results in the northern and southern hemisphere (like we discuss also for Cygnus A, see Sect.\ \ref{Case1} and \ref{discussion} --- note that M87 is a northern source while Centaurs A and Fornax A are in the south), but we do not consider an accurate explanation of the Auger data plausible without a dominant contribution by Centaurus A.

\section{Discussion}
\label{discussion}

Radio galaxies are highly individual sources; both their intrinsic luminosities and their radio brightness as seen from Earth spread over many orders of magnitude. Given the physical constraints explained in Sect.\ \ref{Sec:RG_physics} regarding their potential role as UHECR accelerators, these highly individual properties should reflect on total power and spectrum of their UHECR emission. Nevertheless, we present here results from a simplified model describing UHECR emission from radio galaxies, which involves maximally seven free parameters, adjustable in narrow ranges. In spite of the simplifications we used, we think we can make three tentative conclusions regarding the possibility of radio sources to be the sources of UHECR, which we outline and discuss as follows.

The most solid conclusion we can draw is that an average contribution of all radio galaxies cannot explain both spectrum and composition of UHECR. This is a result of what we called the Lovelace-Hillas relation, which sets the maximal rigidity cosmic rays can reach in the acceleration process in relation to the radio luminosity of the source. Here of course, we need to note that our continuous source function has been derived by applying average values of cosmic-ray load, acceleration efficiency, and composition to all radio galaxies. We disregarded both the spread and potential correlations of these parameters between themselves or to observed properties of the radio galaxies. What would happen if we release this constraint? Clearly, the main effect would be it would smear out our continuous source function to something which is no longer a sharply broken power law, but a smooth function with slowly steepening spectrum. How strong this effect can be is difficult to say on a qualitative base, but as only the flat end of this function would be able to provide good fits to the observed UHECR spectrum, this assumption would not evade our second conclusion.

The second conclusion is, that we need some outstanding, individual sources which explain the cosmic ray flux above the ankle. Here, two sources recommend themselves: Cygnus A, the by far brightest radio galaxy in the sky, and Centaurus A, the second-brightest radio galaxy and also the one nearest to Earth. The fact that we have such bright sources in our vicinity is already a cosmic coincidence, but additionally we need to assume that they have a cosmic ray load significantly above the average of the bulk of radio galaxies. While this can easily be argued for Cygnus A, which is the by far the most powerful radio galaxy in our cosmic environment, it is difficult to argue for Centaurus A --- this is just a regular radio galaxy, and is only distinguished by its proximity to Earth. Clearly, we do not need to assume that Centaurus A is the \textit{only} local source with such special properties, but it would need to be one of maximally 25\% of the sources --- a number which stems from assuming the extreme scenario that all other radio galaxies fall in two strict classes, in which one class has the cosmic-ray load as Centaurus A, and the other not producing any UHECR. Among these 25\%, there could be the other two exceptionally bright local radio galaxies, M87 and Fornax A, but we showed that they cannot solve the issue alone --- Centaurus A is needed. 

The third conclusion is that Cygnus A needs to provide a predominantly light (meaning: near to solar) composition, while Centaurus A would need to have a heavy composition, with an iron fraction comparable to protons at a given cosmic-ray energy. Only this way, the increase of heaviness seen in the Auger data can be explained. We argued already that if the injection process into first-order Fermi acceleration is purely rigidity dependent (as the acceleration process itself), and since FR-II galaxies like Cygnus A can be expected to feed their powerful jets exclusively from the accretion disk of the supermassive black hole, which again are expected to have abundances not far from solar, one would indeed expect a light composition from Cygnus A. On the other hand, for FR-I galaxies where the much weaker jets dissipate inside the galactic environment, one would assume that some of them mix much stronger with heavy material present in their ISM. Indeed, such scenarios have been proposed for Centaurus A based on observations of the jet environment \cite{2010ApJ...720L.155G}. 

A necessary condition for all this to work is that UHECRs from Cygnus A are largely isotropized by the EGMF. 
We estimate that for $\lambda_{\rm c}^{1/2}\, B_{\rm rms} = 6\,\text{Mpc}^{1/2}\,\,\text{nG}$ all anisotropy constraints at energies ${>}\,8\,$EeV are satisfied, whereas the deflection maps which would arise for CRs with energies ${>}\,40\,$EeV show a clear anisotropy (Figs.~\ref{Sibyll_Skymap3} and \ref{AngPowSpec4_Skymap3}). From comparing with observation maps from Auger and TA, one sees that a large part of the ``Cygnus A hot spot'' lies in the ``white spot'' (i.e., the unobservable part of the sky) of Auger, while a large part of the ``Centaurus A hot spot'' lies in the ``white spot'' of TA. Given the expected differences in the composition of UHECR from both sources, this could provide the first physics-based explanation for the lighter composition in the northern hemisphere as reported by TA \cite{ABBASI201549}. Apart from this, one may also expect differences between the spectra measured at different declination ranges of the sky, for which tentative evidence has been delivered recently \cite{2018arXiv180107820A}. However, a detailed, quantitative prediction of such effect is difficult within the current limitations of our approach.

\begin{figure}[b]
  \centering
    \includegraphics[width=0.49\textwidth]{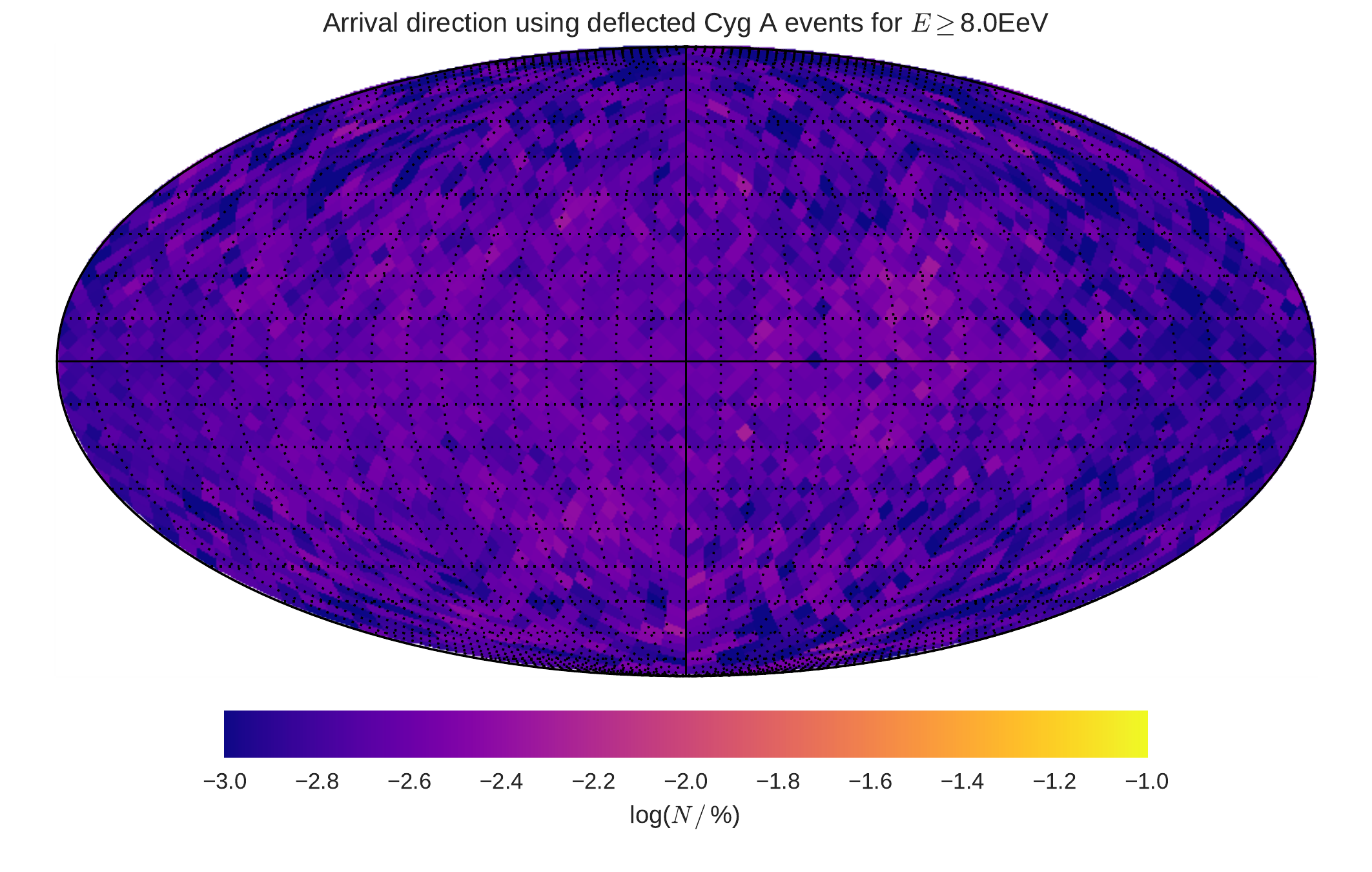}
    \includegraphics[width=0.49\textwidth]{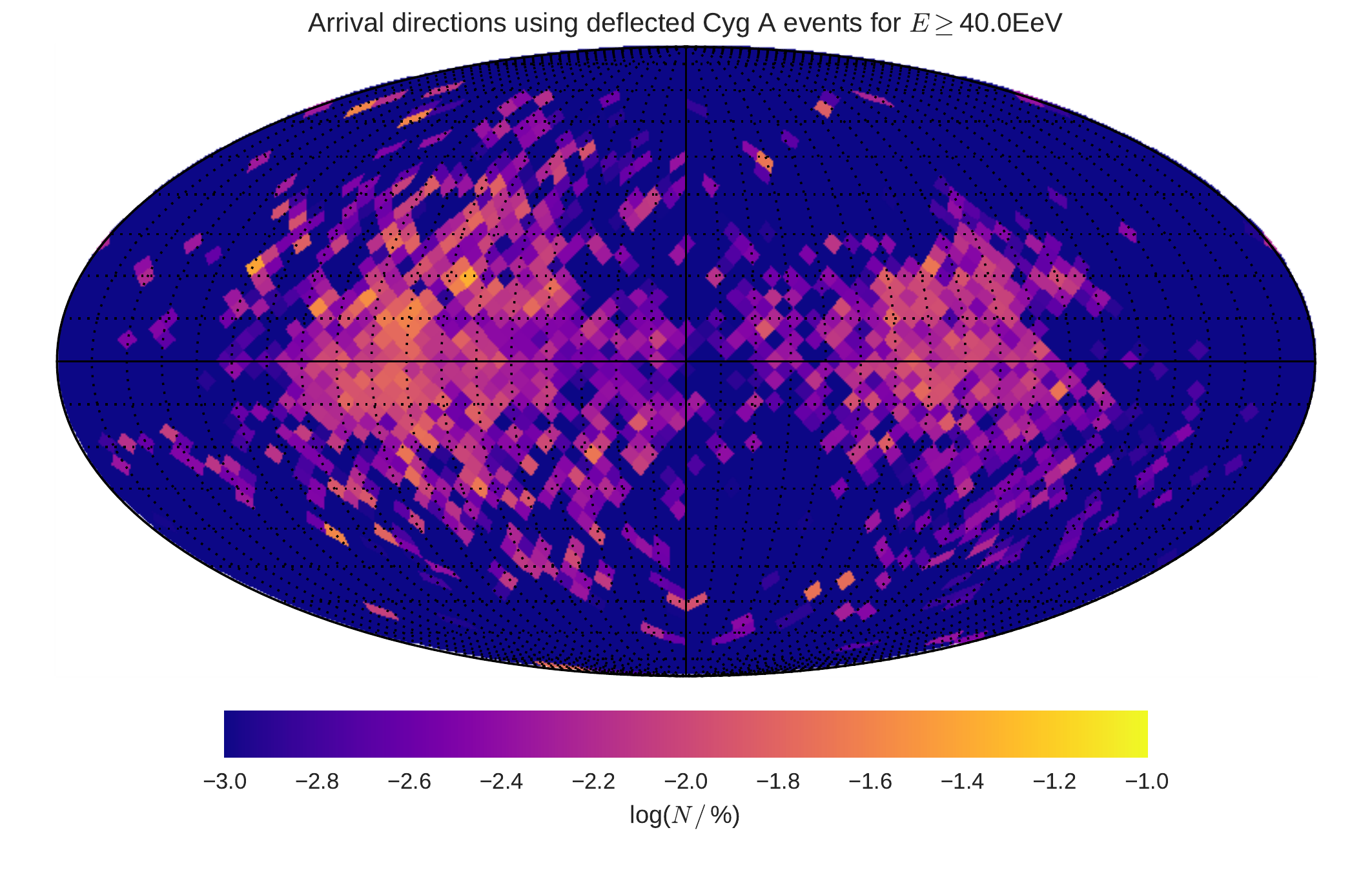}
\caption{Skymap with deflected Cygnus A events according to $\lambda_{\rm c}^{1/2}\, B_{\rm rms} = 6\,\text{Mpc}^{1/2}\,\,\text{nG}$ for two different energy regimes: $E>8\,\text{EeV}$ (left), and $E>40\,\text{EeV}$ (right). Here, Sibyll2.1 and the light composition scenario with a powerful Centaurus A is used.}
\label{Sibyll_Skymap3}
\end{figure} 
\begin{figure}[tb]
  \centering
    \includegraphics[width=0.7\textwidth]{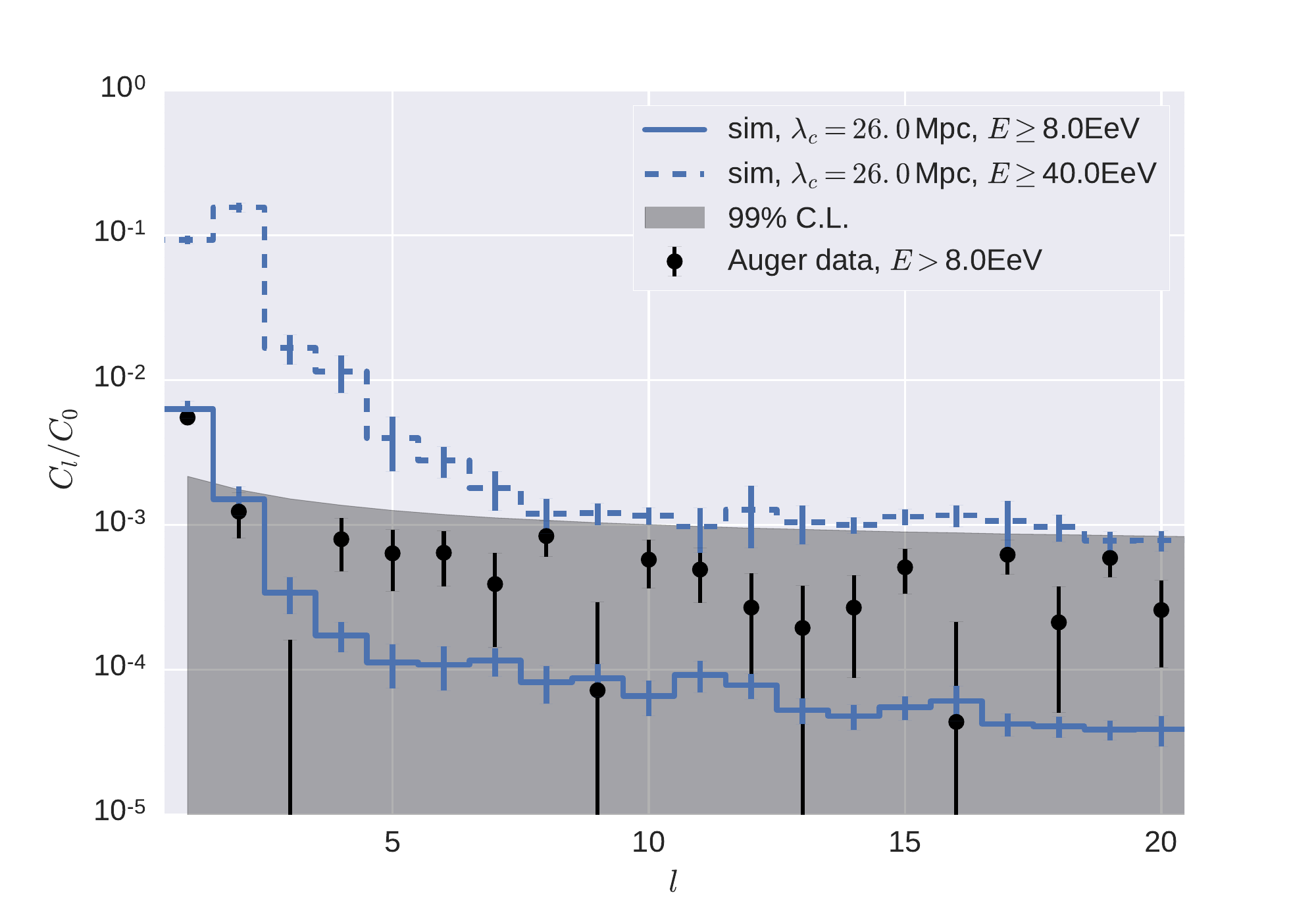}
\caption{Angular power spectrum corresponding to Fig.\ \ref{Sibyll_Skymap3} for $E>8\,\text{EeV}$ (solid line), and $E>40\,\text{EeV}$ (dashed line).}
\label{AngPowSpec4_Skymap3}
\end{figure} 

Thus, for the simplified EGMF scenario discussed we expect that TA would see significant anisotropy in their sky maps at energies ${>}\,40\,$EeV, which would to a large part be attributed to the contribution of Cygnus A. 
At energies ${>}\,57\,$EeV, TA already reported on a cluster of events now known as the ``TA hot spot'' \cite{Abbasi:2014lda}, which, however, is not aligned with the direction of Cygnus A. 
This does not contradict our expectation as the light elements from Cygnus A hardly contribute at these energies due to the GZK cutoff. 
Further, the Virgo cluster at about $20\,$Mpc distance including the strong RG M87 in its center, is not far from the ``TA hot spot'' when its extension of roughly $15^\circ$ diameter in the sky is taken into account. 

Recently, Auger reported on an indication of an intermediate-scale anisotropy at energies $\gtrsim\,40\,$EeV, where tentative evidence for a partial origin of UHECR from starburst galaxies \cite{Aab:2018chp} has been claimed. We note, however, that the largest \textit{observed} anisotropy presented in this paper coincides with the position of Centaurus A, so exactly where we would expect it in our model. 
Additionally, due to the large deflections by the Galactic and extragalactic magnetic fields that we expected for \textit{all} our source candidates (north and south), our predictions do not contradict the lack of event clustering reported by Auger \cite{Abreu:2013kif}, which is given assuming that UHECR events arrive within 30 degrees of their sources.

In any case, even if the Auger result and potentially also the TA hot spot are taken as a hint that other sources contribute to the UHECR flux, the current results from both observatories do not exclude that the \textit{dominant} contribution above the ankle still comes from radio galaxies. This would keep most of our conclusions valid, and released from the necessity to explain all UHECR, we could even relax our need to assign particularly high values for the cosmic-ray load $g_{\rm cr}$ to Cygnus A and Centaurus A, making our model even more plausible. A critical, observational test would be easiest in the northern sky, where the RG scenario predicts Cygnus A and potentially M87 as dominant sources, while the starburst scenario predicts a dominant contribution from M82. As all three sources are well distinguished in their direction in the sky, sufficient statistics of TA events in the regime $40{-}60\,$EeV and a proper consideration of EGMF and GMF deflections could allow to make distinct conclusions.

For the time being, we just note that we manage with our simplified 7-parameter model (a) to provide an excellent fit to the UHECR spectrum, (b) to reproduce the heaviness of chemical composition and their trend as a function of energy, (c) to match the strength of the dipole in arrival directions, and (d) to remain consistent with the limits on anisotropy for higher multipoles --- all measured by the Pierre Auger Observatory. If we take this success as a reason to assume that radio galaxies, with the properties as assumed in our best-fit scenario, are indeed the dominant sources of UHECR, two interesting corollaries would follow. 

First, the contribution of the rest of the radio galaxies (i.e, those except Cygnus A and Centaurus A), although irrelevant as CR sources at ultra-high energies,  would have a spectrum so steep that towards lower energies it has the potential to reach the level of the measured cosmic-ray flux at a break energy around $10^{17}\,$eV. Thus, it would have a shape which perfectly matches the feature known as the ``second knee''. It would thus add radio galaxies to the various scenarios which predict a strong contribution of extragalactic cosmic rays \textit{below} the ankle \cite{Thoudam:2016syr}. We did not check this scenario further as it is out of the scope of this paper, but it is an intriguing possibility to follow.

Second, as we have shown that indeed the predictions of the continuous source scenario are consistent with the sum of individual sources in our local universe, we need to conclude that if we are able to determine the average UHECR flux over a very large volume, it should match the one predicted by the continuous source function. But as its contribution lies below the UHECR flux \textit{measured at Earth} (at least in the energy range above the ankle), it means in turn that this local UHECR flux is roughly one order of magnitude \textit{higher} than the average UHECR flux throughout the universe. One immediate conclusion would be that predicted cosmogenic neutrino fluxes (e.g. \cite{2001PhRvD..64i3010E,Kotera:2010yn,Roulet:2012rv}) would be overestimated by an order of magnitude, as they all rely on the assumption that our measured UHECR flux is typical for the universe. Furthermore, it would also reduce the slight tension that predicted cosmogenic photon fluxes show in the case of models with a strong source evolution \cite{vanVliet:2017obm}.

Finally, we want to note two conclusions which affect the modeling of the UHECR origin in general.
Firstly, our model is based on what we call the Lovelace-Hillas relation, which puts the highest rigidity of accelerated particles in relation to the radio luminosity of the source, and which has been used to argue that UHECR sources must be very numerous transients. The simple reason why we manage to remain within the isotropy limits although we propose that essentially all measured UHECR come from just two steady sources, is that one of these sources is far away enough that extragalactic magnetic fields can reasonable isotropize its contribution \cite{AlvesBatista:2017vob}, and the other one is heavy enough that the Galactic magnetic field is able to do the job.  

Secondly, we obtain all our fits for a power-law spectrum approximately $\propto E^{-1.8}$, thus for an index which is quite reasonable and expected from first-order Fermi acceleration at weakly relativistic shocks. This questions the necessity to assume extremely hard injection spectra (e.g.\ \cite{Aab:2016zth}) to get a good fit of the data. Our result is even slightly softer (closer to theoretical expectations for first-order Fermi acceleration) than the recent results found by Wittkowski and Kampert \cite{2017arXiv171005617W} from 4D simulations with CRPropa3, which point in the same direction.

\section{Conclusions and Outlook}
\label{sec:SummConclu}
We examined the possibility that radio galaxies are the source of the observed ultra-high energy cosmic rays (UHECRs). We performed 3D simulations using the CRPropa3 code \cite{1475-7516-2016-05-038} together with the extragalactic magnetic field obtained in constrained structure-formation simulations by Dolag et al.\ \cite{2005JCAP...01..009D}, which represents the structure of our local universe out to a distance of $120\,\text{Mpc}$. Within this volume, we used individual radio galaxies taken from the catalog of van Velzen et al.\ \cite{2012A&A...544A..18V}, for which we derived a parametrized model for their UHECR emission based on their radio luminosity. For the contribution outside this volume, we derived a continuous source function for UHECRs from radio galaxies using the same physical relations as for the individual sources, and a radio-luminosity function provided by Mauch and Sadler \cite{2007MNRAS.375..931M}, to determine the measured UHECR flux by 1D simulations with CRPropa3. To complete the model, we needed further to include the ultra-luminous radio galaxy Cygnus A, which is expected from its radio brightness to provide about half of all observed UHECRs within our model. As our setup did not allow a detailed propagation calculations for Cygnus A, it was simulated as a point source neglecting scattering in the extragalactic magnetic field. In the final analysis of the results, we then added a scattering pattern of UHECR events arriving from Cygnus A derived from an analytical random field approximation. As we use a physical model to link UHECR emission power and spectrum to the radio luminosity of the sources, our model is absolutely normalized in flux for all UHECR components, and we present an approach how to implement this absolute normalization into a Monte Carlo simulation.

The results of our simulations were then fitted to data published by the Pierre Auger Observatory for spectrum, composition, and arrival-direction distribution of UHECRs, using a re-weighting mechanism common for fitting Monte Carlo results with the residual minimizer contained in the \texttt{lmfit} python package. To keep the parameter space manageable, we constrained our fits to several plausible scenarios, in which usually the parameters for all sources except a few were set equal.  Adopting this necessarily simplified approach, we could obtain two main conclusions:

\begin{itemize}
\item[(1)] The continuous source function of UHECRs from radio galaxies cannot explain the spectrum and composition of cosmic rays above the ``ankle'' of the spectrum at ${\sim}\,5\,{\times}\,10^{18}\,$eV. However, we found from our fits that this component, which represents a broken power law with a break at about $3\,{\times}\,10^{17}\,$eV might have the potential to explain part of the cosmic-ray spectrum between the ``second knee'' (at ${\sim}\,10^{17}\,$eV) and the ankle. This requires further investigation.
\item[(2)] The spectrum and composition above the ankle can be explained by the contribution of the two brightest radio galaxies seen from Earth, Cygnus A at a distance of about $250$\,Mpc, and Centaurus A at a distance of about $4\,$Mpc. Here we have to assume that (a) both sources have a baryonic load in their jets which is significantly above the average of radio galaxies; (b) that Cygnus A provides a predominately light composition, not far from solar abundances of accelerated cosmic rays compared at the same rigidity, while Centaurus A provides a heavy composition, in which the abundance of iron nuclei is enhanced roughly to the level of hydrogen.     
\end{itemize}
A requirement for this scenario not to violate isotropy constraints is that extragalactic magnetic fields are strong enough to largely isotropize UHECR arriving from Cygnus A, which we found possible for extreme EGMF scenarios \cite{AlvesBatista:2017vob}. We did check several other scenarios, for example for the radio galaxies M87 and/or Fornax A being the dominant local source(s) instead of Centaurus A, but found none of them more plausible than assuming a dominant contribution of Centaurus A. We deliver a critical discussion of the assumptions required for this scenario to work, and also mention some intriguing corollaries which might follow from it, in particular the possibility of differences in UHECR spectrum and composition between the northern and southern hemispheres near the highest observed energies, and a possible inhomogeneity of the UHECR density throughout the universe, with the locally measured flux being up to one order of magnitude above average. 

Eventually, we presented a physics-based, absolutely normalized, 7-parameter model which is able to explain all currently-known features of UHECR. We consider this result a clear demonstration that radio galaxies still present very good candidates for an explanation of the origin of UHECR. To further substantiate this claim, we need to overcome several simplifications which we applied to obtain our result within the technical limitations present. In particular, we see the need to set up a simulation and optimization procedure which fully accounts for the astrophysical properties of the model, which means (i) performing 3D, or better 4D simulations at least out to the distance of Cygnus A (250\,Mpc), ideally out to the magnetic horizon at about 500\,Mpc \cite{Stanev:2000fb}, (ii) investigating several possible scenarios of magnetic-field structure out to this distance, and (iii) applying a Bayesian parameter inference in high-dimensional parameter spaces, which would allow to treat at least all radio galaxies which make a measurable individual contribution, with an own parameter set. While by the time this work was started, such a setup still seemed illusionary, it will become possible in the near future thanks to several developments:
\begin{itemize}
\item[(a)] The availability of detailed reconstructions of the matter distribution of the local universe out to 400\,Mpc distance and beyond \cite{2015JCAP...01..036J, 2015JCAP...06..015L}, which could be used for a phenomenological, parametrized construction of extragalactic magnetic fields with various field strengths and volume filling factors as demonstrated by Alves Batista et al. \cite{AlvesBatista:2017vob}.
\item[(b)] The development of efficient targeting methods in CRPropa which could speed up 3D and 4D simulations by factors up to $10^7$ \cite{JascheVlietRachenInprep} compared to the still common $4\pi$ emission.
\item[(c)] The development and computational implementation of very efficient Bayesian Markov Chain Monte Carlo methods in the last years (e.g. PyMC3 \cite{2016ascl.soft10016S}), where in particular Hamiltonian MCMC sampling promises to optimally explore high-dimensional parameter spaces \cite{2012arXiv1206.1901N}. 
\end{itemize} 
The next, straightforward step, however, will be to test the model developed here as a possible extragalactic explanation of cosmic rays below the ankle, for which we expect to be able to present results very soon.

\acknowledgments
This work most notably benefits from the development of CRPropa3 and in particular from the support by David Walz, Gero M\"uller and Nils Nierstenh\"ofer. 
We are grateful to G\"unter Sigl, Peter Biermann, David Wittkowski and the unknown referee for very useful comments that helped to improve the original version of the paper.
Some of the results in this paper have been derived using the software packages Numpy \cite{vanDerWalt2011}, Pandas \cite{mckinney-proc-scipy-2010}, Matplotlib \cite{Hunter:2007} and HEALPix/ healpy \cite{2005ApJ...622..759G}. 
Further, we acknowledge support from the MERCUR project St-2014-0040 (RAPP Center) and the research department of plasmas with complex interactions (Bochum). AvV acknowledges financial support from the NWO Astroparticle Physics grant WARP.
\appendix

\section{Validation of the flux normalization}
\label{TestNorm}
Within this work the radio luminosities are used to normalize the observed UHECR flux from the simulations. 
To validate the normalization suggested in Sect.\ \ref{AbsoluteNormalization} we first made a simplified simulation with a well known theoretical expectations to show that the local flux contribution according to Eq.\ (\ref{localParticleFlux}) provides the expected behavior. 
Afterward, the non-local flux contribution according to Eq.\ (\ref{nonlocalParticleFlux}) is determined in the local region with $d\leq120\,\text{Mpc}$ and compared with the previously validated local contribution. 
\subsection{Local contribution}
\label{TestNormLocal}
A three-dimensional simulation without any particle interactions or deflections is performed using a random subset of the luminous sources of the vV12 catalog. 
Here, the source $j$ provides a radio luminosity $L_j$ and a corresponding CR power $Q_{{\rm cr},j}$ between $\check{R}_0$ and $\hat{R}$ according to Eq.\ \ref{CRluminosity}. 
Within the UHE regime between $\check{R}_1$ and $\hat{R}$ we obtain
\be
Q_{\rm uhecr, i,j}={f_i\,Z_i\,Q_{{\rm cr},j} \over \bar{Z}}\,{G_a(\check{R}_1)-G_a(\hat{R}_{\rm cut}) \over G_a(\check{R}_0)-G_a(\hat{R}_{\rm cut})}
\label{UHECRpower}
\ee
with 
\be
G_a(x)=-\int\text{d}R\,\,R^{1-a}\,\exp(-R/\hat{R}) = \hat{R}^{2-a}\,\Gamma\left(2-a, x/\hat{R} \right)\,
\ee
where $\Gamma(c,x)$ denotes the incomplete gamma function. 
The first term in Eq.\ (\ref{UHECRpower}) accounts for the distribution of the total CR power between the different types of nuclei $i$ with respect to the initial abundances $f_i$. 
The second term determines the proportion of the CR power that goes into UHECRs with a spectral behavior according to the re-weighted energy spectrum under the assumption that the spectral behavior does not change between $\check{R}_0$ and $\check{R}_1$. 
Thus, the theoretically expected total energy flux $J^{{\rm (th)}}_{i,j}$ for an extended, spherical observer with a radius $r_{{\rm obs}}$, surface $S_{{\rm obs}}$ and cross-section $\sigma_{{\rm obs}}$ at a distance $d_j\gg r_{{\rm obs}}$ yields
\be
J^{{\rm (th)}}_{i,j} = {Q_{\rm uhecr, i,j} \over 4\pi\, d_j}\,{\sigma_{{\rm obs}} \over S_{{\rm obs}}} \simeq {1 \over 16\pi\,d_j}\,{f_i\,Z_i\,Q_{{\rm cr},j} \over \bar{Z}}\,{G_a(\check{R}_1)-G_a(\hat{R}_{\rm cut}) \over G_a(\check{R}_0)-G_a(\hat{R}_{\rm cut})}\,.
\ee
It should be kept in mind that the used cross-section $\sigma_{{\rm obs}}=\pi\,r_{{\rm obs}}^2$ is only valid for parallel incident CRs, which is almost the case for $d_j\gg r_{{\rm obs}}$. The total energy flux $J^{{\rm (sim)}}_{i,j}$ due to the normalization (\ref{localParticleFlux}) is given by
\be
J^{{\rm (sim)}}_{i,j} = \sum_{k,i,j}\, E^*_k\,W_{k,i,j} =  \sum_{k,i,j}\, {E^*_k\,f_i\,Q_{{\rm cr},j}\,(\hat{R}^{1-\tilde{a}}-\check{R}_1^{1-\tilde{a}}) \,R_k^{\tilde{a}-a}\,\exp(-R_k/\hat{R}) \over 4\pi\,r_{{\rm obs}}^2\,\tilde{N}_{i,j}\,\bar{Z}\,e\,(1-\tilde{a})\, \left(G_a(\check{R}_0)-G_a(\hat{R}_{\rm cut})\right)}\,.
\ee
Using 100 random sets of the parameters $a$, $f_i$, $g_{\rm cr}$, $g_{\rm acc}$ the corresponding relative error 
\be 
\Delta_{i,j}=100\times{|J^{{\rm (sim)}}_{i,j}-J^{{\rm (th)}}_{i,j}| \over J^{{\rm (th)}}_{i,j} }
\ee
for a single source as well as the total relative error $\sum_i\Delta_{i,j}$ for multiple sources is shown in the Fig.\ \ref{Test1}. 
\begin{figure}[h!]
  \centering
    \includegraphics[width=0.49\textwidth]{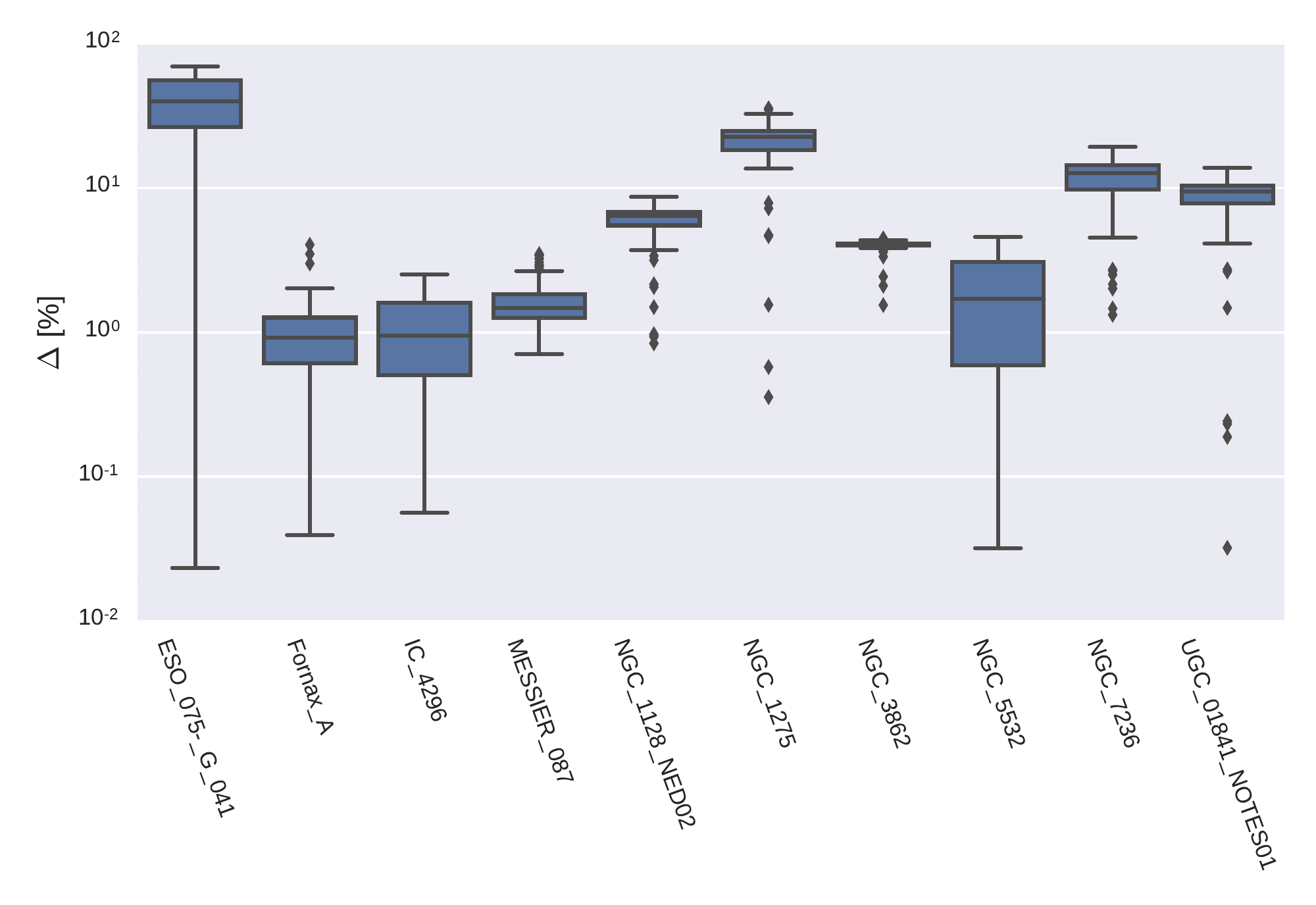}
    \includegraphics[width=0.49\textwidth]{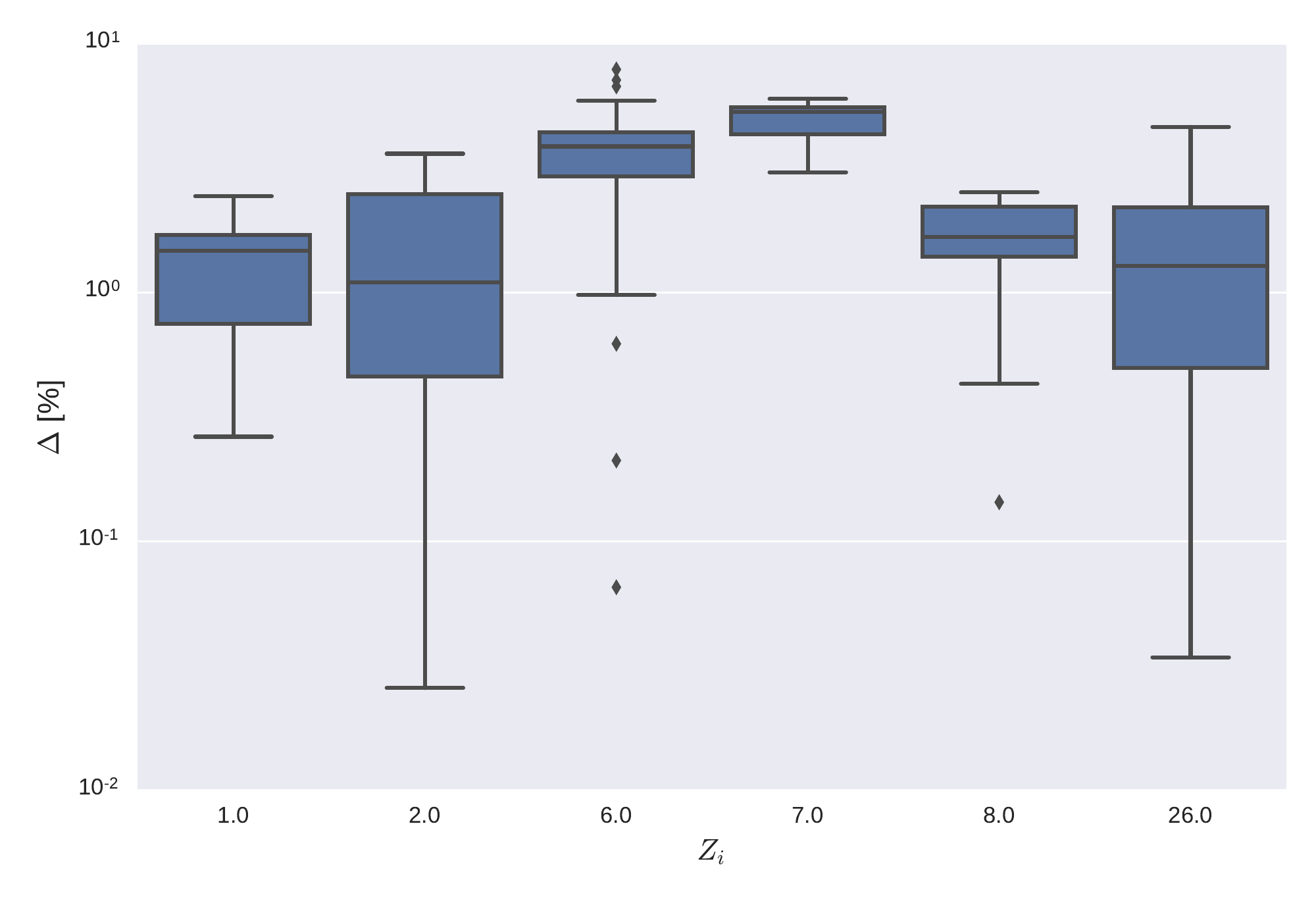}
\caption{Difference between the simulated and the theoretically expected CR flux using 100 random sets of parameters: (i) Relative error $\Delta=\sum_i\Delta_{i,j}$ dependent on the local source (left plot); and (ii) relative error $\Delta=\Delta_{i,\rm Fornax\,A}$ using only candidates from Fornax A dependent on the initial nucleus type $i$ (right plot). The boxes display $50\%$ of the data and the mean value is indicated by a black, horizontal line. Except for outliers, the width of rest of the distribution is indicated by the error bars.}
\label{Test1}
\end{figure} 
Here, the error increases in the case of smaller statistics but predominantly stay below a few percentage level yielding a good validation of the normalization of the local contribution. 
\subsection{Non-local contribution}
\label{TestNormNonLocal}
The contribution of non-local sources is only taken into account for candidates with a propagation distance $>120\,\text{Mpc}$. 
Nevertheless, we simulated CR candidates in this 1D approach with $0<d\leq120\,\text{Mpc}$ in order to compare their resulting flux with the local flux contribution from the 3D approach. 
Since we cannot account for deflections in 1D and the cosmic variance can cause a local excess or lack of CRs, we do not expect the results to be match exactly, but they should still agree within in a reasonable range. 
In order to avoid differences by the different types of cutoff --- exponential in the 3D case and by a Heaviside function in the 1D case --- we use $g_{\rm cr}=50,\,g_{\rm acc}=0.6$ so that $\hat{R}_{\rm cut} =\hat{R}$ and both spectra have a sharp cutoff. 
Fig.\ \ref{Test2a} displays the resulting energy flux from candidates that originate between $5\,\text{Mpc}$ and $120\,\text{Mpc}$ using the 3D approach and a subsequent normalization with Eq.\ (\ref{localParticleFlux}) and the 1D approach normalized with Eq.\ (\ref{nonlocalParticleFlux}), respectively. Here, the shown energy flux 
\be
J^{(1D)}_{i,\kappa}= {E_\kappa \over dE_\kappa}\,\sum_{k}\, \delta_{k\kappa}\,W_{k,i}
\ee
from the 1D simulation and 
\be
J^{(3D)}_{i,\kappa}= {E_\kappa \over dE_\kappa}\,\sum_{k,j}\, \delta_{k\kappa}\,W_{k,i,j}
\ee
from the 3D simulation of the different initial nuclei type $i$ in the energy bin $E_\kappa$ have the same spectral behavior as well as a similar absolute magnitude. 

\begin{figure}[h!]
  \centering
    \includegraphics[width=0.69\textwidth]{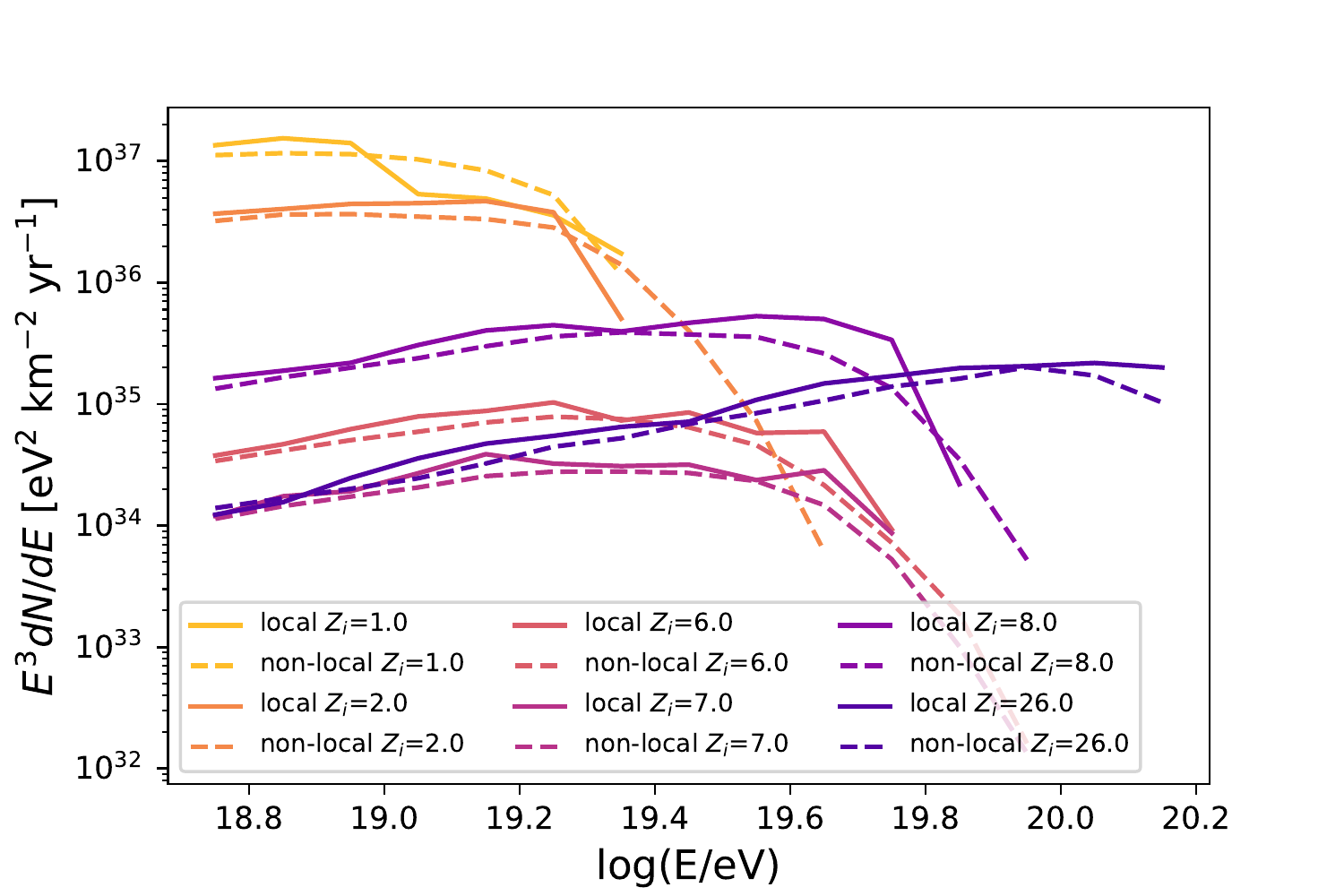}
\caption{Energy flux $J^{(1D)}_{i,\kappa}$ and $J^{(3D)}_{i,\kappa}$ resulting from the 1D and the 3D approach, respectively, by candidates that originate at a distance between $5\,\text{Mpc}$ and $120\,\text{Mpc}$. Here, the canonical parameter values of $a=1.8$, $g_{\rm cr}=50$, $g_{\rm acc}=0.6$ and $f_i=f_{\odot}$ are used.}
\label{Test2a}
\end{figure} 

The selected distance range is chosen such that the contribution by Centaurus A, at a distance of $3.6\,\text{Mpc}$, gets excluded, otherwise $J^{(3D)}_{i,\kappa}$ would significantly exceed $J^{(1D)}_{i,\kappa}$, as due to the close distance to Earth and the high radio luminosity it is most likely that Centaurus A causes a local excess of UHECRs.
To ensure, that the relative error 
\be
\Delta_{i,\kappa}=100\times{|J^{(1D)}_{i,\kappa}-J^{(3D)}_{i,\kappa}| \over J^{(3D)}_{i,\kappa}}
\ee
is independent of the chosen parameters $a$, $f_i$ and $m$, we draw 100 random sets of these parameters and determine the corresponding $\Delta_{i,\kappa}$. 
In addition, we also change the spatial region the candidates need to originate between $\check{d}\in[5,\,30]\,\text{Mpc}$ and $\hat{d}\in[100,\,120]\,\text{Mpc}$, to account for cosmic variance. 
As shown in the Fig.\ \ref{Test2b} the 1D flux hardly differs from the 3D flux by more than a few hundred percent and the mean is well below. So we conclude that the normalization (\ref{nonlocalParticleFlux}) yields the correct CR flux for the non-local source contribution. 
\begin{figure}[h!]
  \centering
    \includegraphics[width=0.49\textwidth]{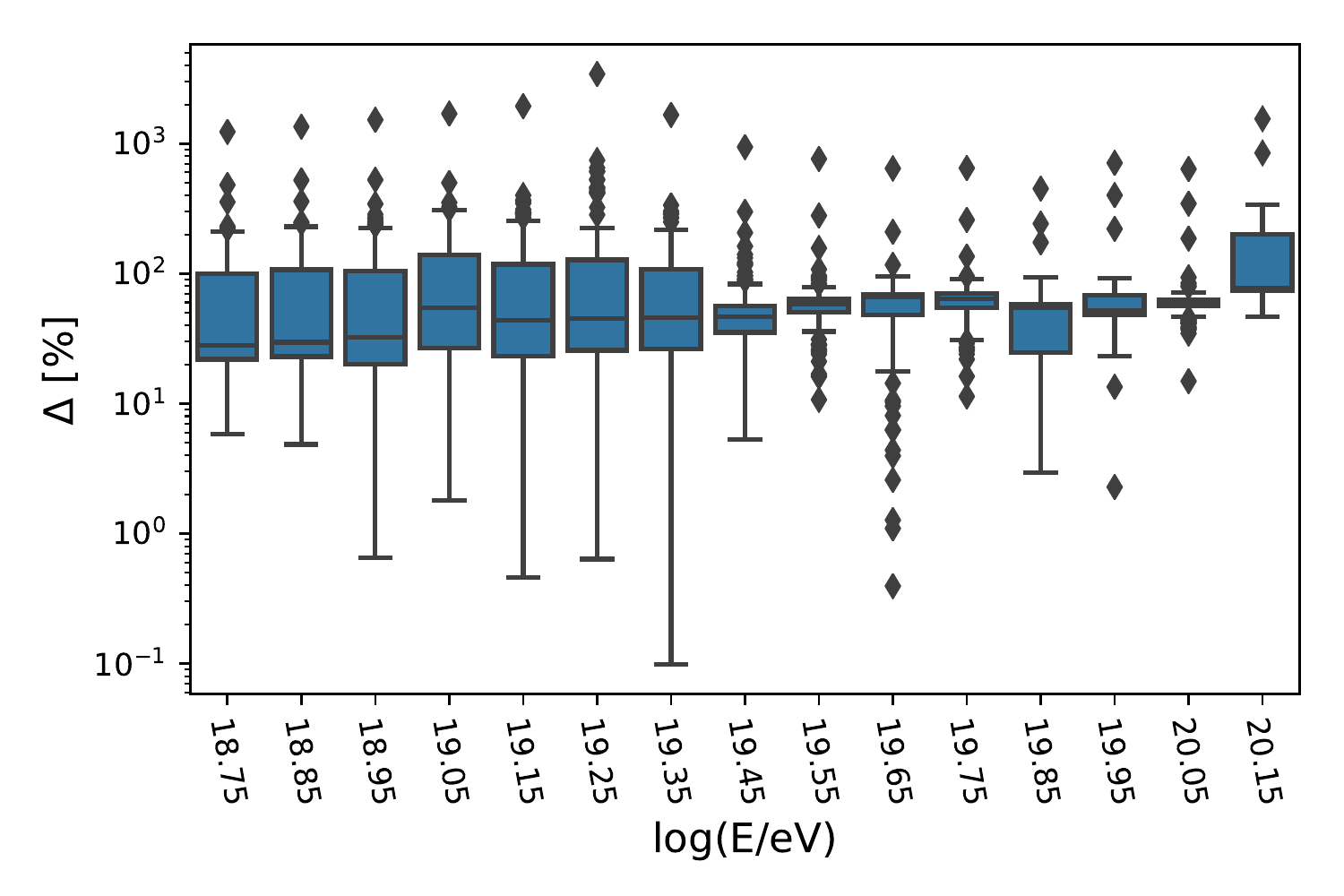}
    \includegraphics[width=0.49\textwidth]{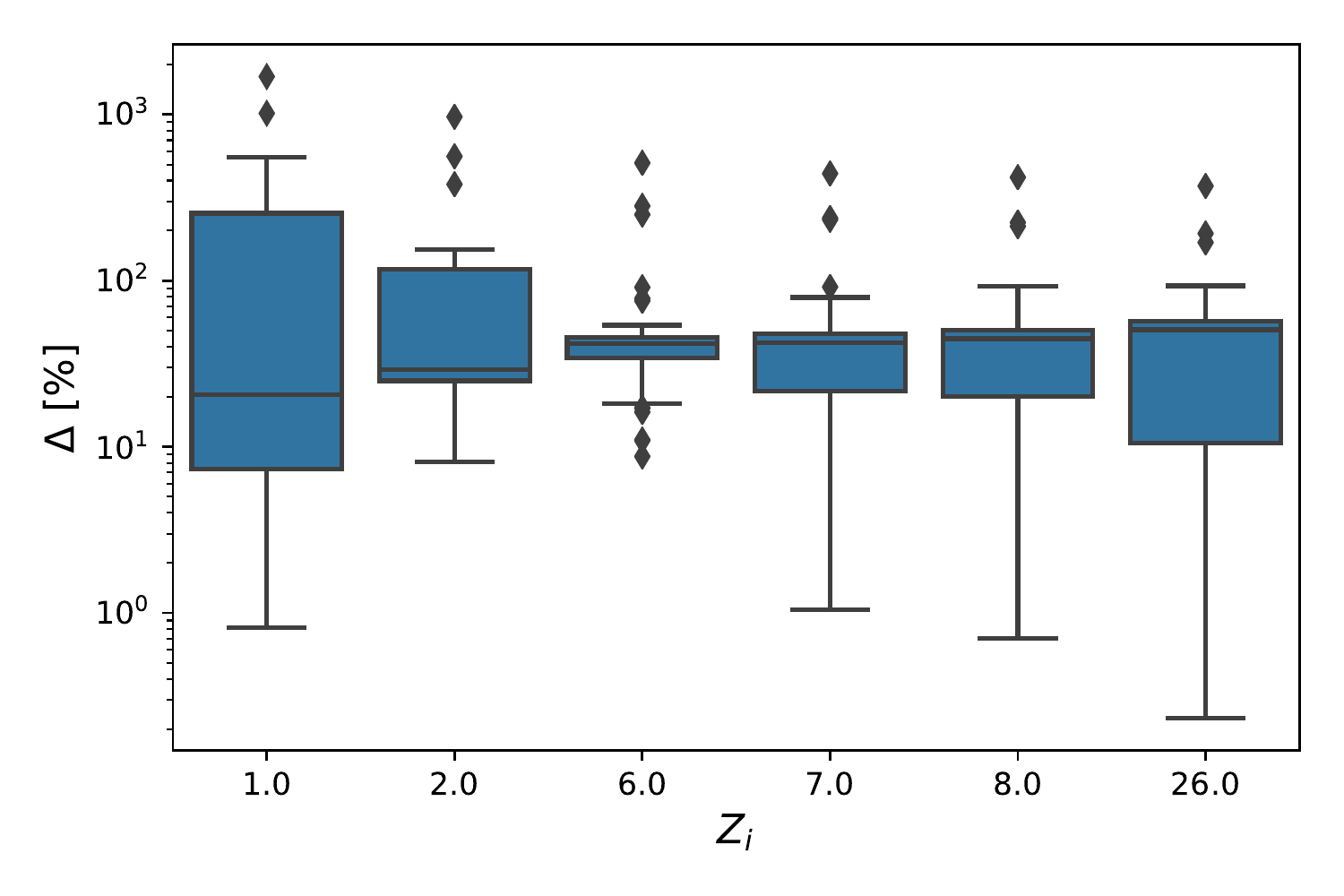}
\caption{Difference of the 1D and the 3D flux normalization in the local regime using 100 random sets of parameters: (i) Relative error $\Delta=\sum_i\Delta_{i,\kappa}$ dependent on the energy bin $E_\kappa$ (left plot); and (ii) relative error $\Delta=\sum_\kappa\Delta_{i,\kappa}$ dependent on the initial nucleus type $i$ (right plot). See Fig.~\ref{Test1} for symbol conventions.}
\label{Test2b}
\end{figure} 
\section{Further details of the simulation setup}
\label{detailsOfSim}
The results shown in the Sect.\ \ref{sec:results} are based on the following simulation setup: 
\begin{enumerate}
 \item[(i)] \emph{The observer} is located at $\vec{r}_{{\rm obs}}=(118.34,\, 117.69,\, 119.2)\,\text{Mpc}$ with a radius of $0.1\,\text{Mpc}$. 
 \item[(ii)] \emph{The sources} are located according to the catalog of RGs from van Velzen et al.\ \cite{2012A&A...544A..18V} with a maximal distance to the observer of $120\,\text{Mpc}$. 
 In total, $N_{\rm RG}=838=121+100+617$ different source positions are used within $120\,\text{Mpc}$ where all RGs are considered as a point source.  
 In addition, we take the average CR contribution by the non-local sources beyond $120\,\text{Mpc}$ into account as well the the impact of an extraordinarily powerful radio source, like Cygnus A, at a distance of $255\,\text{Mpc}$. 
 \item[(iii)] \emph{The extragalactic propagation} in the case of the local sources is performed with the \texttt{DeflectionCK()} module from CRPropa3, where the equation of motion is solved by a Runge-Kutta integration method. 
 Thereby, we used a local accuracy of $10^{-3}$ between the fourth and the fifth order of the algorithm, a minimal step of $1\,\text{kpc}$ and a maximal step of $1\,\text{Mpc}$. 
 Further, we use the magnetic field data from Dolag et al.\ \cite{2005JCAP...01..009D}, which is stored in a multi-resolution grid according to the \texttt{QuimbyMagneticField()} module \cite{2016JCAP...08..025M}. 
 These EGMF data are intended to become part of the CRPropa program and are currently available at \url{https://forge.physik.rwth-aachen.de/public/quimby/mhd/}.
 In addition, we account for nuclear decay as well as electron pair production, photo-pion production and photo-disintegration due to the cosmic microwave background (CMB) and the UV/optical/IR background (IRB). 
 
 To decrease the necessary CPU time for the propagation, we use the \texttt{EmissionMap()} module from CRPropa3. 
 Here, three prior simulations are made, where each simulation generates a so-called emission map for the individual sources and the initial candidate types with a certain angular resolution in the spherical coordinates $\theta$ and $\phi$. 
 Each emission map provides information on the number of candidates that hit the observer sphere dependent on its initial momentum. 
 By using this emission map (in the subsequent simulation) a CR candidate will immediately be rejected from the propagation, if its initial momentum has not provided any hit according to the map. 
 In order to optimize the resolution of the final emission map, we run three simulation with a gradually decreasing observer size, in which each simulation uses the previously generated emission map and generates a new one with a higher resolution. 
 The final emission map of each source and candidate type has an angular resolution of 10$\degree$ in $\theta$ and $\phi$.
 
 In the case of the non-local sources, no realistic EGMF structure is available, so that the \texttt{SimplePropagation()} module is used, where CR candidates keep their initial direction of momentum. 
 CR candidates from the single non-average source at $255\,\text{Mpc}$ are simulated in three spatial dimensions like the local sources, whereof each source emits UHECR candidates isotropically. 
 However, the average non-local contribution is simulated in one spatial dimension, where the sources are homogeneously distributed in 1D according to the \texttt{SourceUniform1D} module. 
 Here, all CR candidates will reach the observer, if their energy does not drop below the minimal energy of $3\,\text{EeV}$ that is used for all simulations. 
 Further, the effects due to the redshift evolution of the photon target are taken into account. 
 \item[(iv)] \emph{The galactic propagation} is performed with the \texttt{PARSEC} code \cite{2014APh....54..110B} in order to obtain the arrival directions of the UHECR candidates at the Earth. 
 Here, only the deflection by the galactic magnetic field from the JF12 model, its regular and turbulent component, is taken into account. 
 The magnetic lensing according to the \texttt{PARSEC} code is already part of the CRPropa program. 
 \end{enumerate}

\bibliographystyle{JHEP} 
\addcontentsline{toc}{section}{Bibliography}
\bibliography{references}

\end{document}